\documentclass[11pt]{article}
\usepackage{graphicx,graphics,color}
\usepackage{amsmath,amssymb, bm}
\usepackage{amsthm}
\usepackage{mathrsfs}
\usepackage{mathabx}
\usepackage{algorithm}
\usepackage{algpseudocode}
\usepackage{setspace}
\usepackage[utf8]{inputenc}
\usepackage{natbib}%
\usepackage[dvipsnames]{xcolor}
\usepackage{multirow}
\usepackage{caption}
\usepackage{subcaption}
\usepackage{bbm}
 \usepackage{parskip}
\usepackage{comment}
\usepackage{listings}
\usepackage{color}
\usepackage{fancybox}
\usepackage[dvipsnames]{xcolor}
\definecolor{dkgreen}{rgb}{0,0.6,0}
\definecolor{gray}{rgb}{0.5,0.5,0.5}
\definecolor{mauve}{rgb}{0.58,0,0.82}
\definecolor{blue}{RGB}{22,138,173}
\definecolor{red}{RGB}{213,94,0}
\definecolor{yellow}{RGB}{240,228,66}
\definecolor{green}{RGB}{0,158,115}

\usepackage[colorlinks=true,linkcolor=blue,citecolor=blue]{hyperref}%

\usepackage[margin=1.0 in, top = 1.0in]{geometry}

\newcommand{\R}{{\rm I}\kern-0.18em{\rm R}}
\newcommand{\h}{{\rm I}\kern-0.18em{\rm H}}
\newcommand{\K}{{\rm I}\kern-0.18em{\rm K}}
\newcommand{\p}{{\rm I}\kern-0.18em{\rm P}}
\newcommand{\E}{{\rm I}\kern-0.18em{\rm E}}
\newcommand{\Z}{{\rm Z}\kern-0.18em{\rm Z}}
\newcommand{\1}{{\rm 1}\kern-0.24em{\rm I}}
\newcommand{\N}{{\rm I}\kern-0.18em{\rm N}}

\newcommand{\argmin}{\mathop{\mathrm{argmin}}}

\def\@begintheorem#1#2{\trivlist \item[\hskip \labelsep{\bf #1\ #2.}]\sl}
\def\@opargbegintheorem#1#2#3{\trivlist
      \item[\hskip \labelsep{\bf #1\ #2\ (#3).}]\sl}

\newtheorem{theorem}{Theorem}[section]

\newtheorem{lemma}[theorem]{Lemma}
\newtheorem{assumption}{Assumption}

\newcommand{\indep}{\perp \!\!\! \perp}
\DeclareMathOperator*{\plim}{plim}

\title{Dynamic Biases of Static Panel Data Estimators}

\author{Sylvia Klosin\footnote{I am grateful to Isaiah Andrews, Victor Chernozhukov, Anna Mikusheva, and Whitney Newey for their guidance and support. I thank Alberto Abadie, Karl Aspelund, Tamma Carleton, Lindsey Currier, Vitor Hadad, Chris Hansen, Kelsey Jack, Sara Johns, Clair Lazar Reich, Ishan Nath, Benjamin Olken, Dev Patel, Deborah Plana, Jacquelyn Pless, Ashesh Rambachan, Rahul Singh, Sophie Sun, Jenny Wang and seminar participants at MIT for helpful discussions. I acknowledge generous support from the Jerry A. Hausman Graduate Dissertation Fellowship and the NSF GRFP. This draft is a work in progress and comments are
welcome; all errors are my own. This paper has been publicly circulated as my job market paper
since October 2024 via my personal website: https://klosins.github.io/.
Department of Economics, MIT, 77 Massachusetts Avenue, Cambridge, MA 02139. Email: klosins@mit.edu}}

\date{ \today} 

\begin{document}
  \doublespacing
\maketitle
\vspace{-.75cm}
\begin{center}
\href{https://klosins.github.io/Klosin_JMP.pdf}{%
{\color{blue}\Ovalbox{Please click here for the latest version}}
}
\end{center}

\vspace{-.0cm} 
\begin{abstract}
This paper identifies an important bias — termed dynamic bias — in fixed effects panel estimators that arises when dynamic feedback is ignored in the estimating equation. Dynamic feedback occurs if past outcomes impact current outcomes, a feature of many settings ranging from economic growth to agricultural and labor markets. When estimating equations omit past outcomes, dynamic bias can lead to significantly inaccurate treatment effect estimates, even with randomly assigned treatments. This dynamic bias in simulations is larger than Nickell bias. I show that dynamic bias stems from the estimation of fixed effects, as their estimation generates confounding in the data. To recover consistent treatment effects, I develop a flexible estimator that provides fixed-T bias correction. I apply this approach to study the impact of temperature shocks on GDP, a canonical example where economic theory points to an important feedback from past to future outcomes. Accounting for dynamic bias lowers the estimated effects of higher yearly temperatures on GDP growth by 10\% and GDP levels by 120\%.
\end{abstract}


\newpage

\section{Introduction}

Treatment effects are often estimated with fixed effects panel models \citep{currie2020technology}. These models accounted for 19\% of the empirical articles published in the American Economic Review from 2010 to 2012 \citep{de2020two}.\footnote{Fixed effects panel models are especially important in economic contexts where randomizing treatment is not feasible. For example, in environmental economics, it is impossible to randomize exposure to floods or temperature shocks to estimate their effects. Instead, researchers use observational data and typically assume that, conditional on location, the treatment is random.}  It is common for these fixed-effects models to be static, meaning the models do not control for past outcomes and, therefore, do not account for dynamics.  Static models are frequently used even when economic theory suggests a dynamic relationship between past and current outcomes. For example, past agricultural yields impact future yields through soil health and market demand \citep{griliches1963sources}, human capital formation is dynamic \citep{cunha2007technology}, past labor market states impacts future states \citep{blanchard1988beyond}, past GDP impacts current GDP \citep{solow1956contribution}, and migration flows are functions of historical migration chains \citep{massey1993theories}. Despite this theoretical expectation, empirical papers studying these outcomes often run fixed effects analysis without controlling for past outcomes.\footnote{Examples include  \cite{annan2015federal}, \cite{burke2015global}, \cite{cho2017effects}, \cite{jessoe2018climate}, \cite{drabo2015natural}, \cite{mahajan2020taken}, \cite{missirian2017asylum}, \cite{graff2018temperature} \cite{garg2020temperature}.}

 One reason why researchers often use static models is because they are concerned that adding past outcomes as controls can \textit{cause} estimation problems. This concern is rooted in the understanding that panel models with both fixed effects and past outcomes are subject to Nickell bias \citep{nickell1981biases}. Nickell bias arises from the estimation of fixed effects, in particular the failure of strict exogeneity conditions when dynamics are present, leading to biased estimates. Another reason why researchers use static models is because they think it is unnecessary to control for past outcomes when treatment is random. For example, if treatment is an exogenous temperature shock, it is commonly assumed that treatment is random conditional on fixed effects. Consequently, researchers believe they can obtain unbiased treatment effect estimates without controlling for past outcomes. Random treatment assignment ensures that excluding past outcomes does not lead to omitted variable bias. However, this paper shows that a different bias arises due to the omission of these dynamics and the fixed effect estimation.

The main contribution of this paper is two-fold. First, I identify ``dynamic bias'', a bias that arises when static fixed effects panel models are used in settings with dynamics in outcomes. Dynamic bias occurs because past treatment is related to past outcomes through the outcome equation. Therefore, de-meaning or differencing to account for fixed effects generates confounding. This generated confounding leads to biased treatment effect estimates even if the treatment is random. Treatment effects are even more biased if treatment is related to past outcomes and therefore endogenous. A contribution of this paper is to explicitly characterize the resulting dynamic bias. Using analytical derivations, simulations, and applied examples, I demonstrate that dynamic bias is often substantial. The dynamic bias, caused by \textit{omitting} past outcomes from the model, is often much larger than the Nickell bias caused by \textit{including} past outcomes in the model.\footnote{In simulations, no matter how correlated past outcomes are with current outcomes, dynamic bias is larger than Nickell bias.} Therefore, when researchers avoid controlling for past outcomes because they are worried about introducing bias, they may in fact be making the bias on their treatment effect estimates \textit{larger}.

The second contribution of the paper is to develop a new estimator that corrects for both dynamic bias and Nickell bias, which works even when the number of time periods $T$ is fixed. I create a novel bias correction, ``dynamic biases correction'' (DBC), by deriving a formula for the asymptotic\footnote{Under asymptotics where the number units $N \rightarrow \infty$ but number of time periods $T$ is fixed.} bias term following \cite{kiviet1995bias}. 
The explicit formula can be derived because, given the model, we know exactly how the Nickell bias is generated: the errors become correlated with regressors due to the demeaning required for fixed effects estimation. Therefore, I analytically derive an expression for the demeaned errors and use this expression to calculate how the demeaned errors correlate with the demeaned regressors, leading to an explicit formula for the bias. The bias correction is achieved by subtracting an estimate of the bias from the original estimated coefficients.

The DBC correction for treatment effects is the first analytic bias correction that accommodates endogenous treatment related to past outcomes. Allowing for endogenous treatment is important in many economic settings because selection into treatment is often a function of past outcomes.\footnote{\cite{marx2022parallel} and \cite{ghanem2022selection} show how many economic models lead to treatment selection that depends on past outcomes. For example, selection into environmental policy treatment is often a function of past environmental conditions. Policies that target air pollution in particular areas are implemented because of past pollution rates \citep{chay2005does}. As another example, regional deforestation protection in Brazil is based on past deforestation in the region \citep{harding2021commodity, assunccao2023optimal}.} My correction also works when the treatment is exogenous, e.g. randomly assigned. Additionally, my correction is the first analytical correction not to impose homogeneity in treatment effects, even when treatment is endogenous. A large econometrics literature highlights problems that arise when homogeneity in treatment effects is assumed incorrectly -- and how important it is to allow for treatment effects to vary depending on the treatment group in panel data.\footnote{Discussed in \cite{sun2020estimating}, \cite{callaway2021difference}, \cite{goodman2021difference}, \cite{de2024difference} .}

The exact bias correction approach has some appealing characteristics in comparison to alternative corrections, which are based on instrumental variables \citep{holtz1988estimating, arellano1991some}. The instrumental variable methods are based on using further outcome lags as instruments for outcome lags. The correct choice of instrument is often unclear and can lead to problems caused by weak instruments.\footnote{Problems caused by weak instruments are discussed by \cite{andrews2019weak, mikusheva2021many, mikusheva2024weak}.} In simulations, I find that my analytical solution keeps standard errors as small as the original linear regressions and maintains proper coverage, as compared to instrumental variable methods which lead to larger standard errors.


The biases and proposed estimator are illustrated through Monte Carlo simulations and empirical replications. I generate simulation data where treatment is a function of past outomes, as well as data where treatment is random conditional on fixed effects. Even in the simulation with random treatment, treatment effect estimates from models omitting past outcomes are biased significantly more than models that control for past outcomes. 
In Monte Carlo simulations, the DBC estimator is unbiased and has smaller standard errors as compared to the Arellano-Bond-based alternative. To validate my results with real data, I use data from \cite{dell2012temperature}. This paper studies the ``contemporary causal effect of temperature on the development process'' by using a yearly panel of countries with GDP and temperature information \citep{dell2012temperature}. The treatment variable of interest is temperature, which is taken to be random, conditional on the country. Controlling for past outcomes significantly changes the results both when the outcome is GDP growth (10\% change) and GDP levels (120\% change).\footnote{The p-value for the GDP growth result is .06, so it is significant at the 10\% level while the GDP level result is significant at the 5\% level.}\footnote {Both GDP growth and levels are used as outcomes in the literature that studies the effect of temperature on economic outcomes \citep{newell2021gdp, nath2024much}.}

The rest of the paper proceeds as follows. Section \ref{sec:literature_review} discusses related work in
more detail. I provide empirical motivation and an overview of the biases discussed in this paper in Section \ref{section:applied_motivation}. Section \ref{sec:theoretical_results} presents theoretical results. Section \ref{subsection:estimation} introduces the DBC estimator of treatment effects. Section \ref{sec:simulation_study} conducts a simulation study to illustrate the asymptotic properties of the estimators I propose. Section
\ref{sec:emperical_examples} provides an empirical example illustrating how correcting for dynamic bias impacts treatment effect estimation. Section \ref{sec:conclusion}
 concludes.

\subsection{Literature review}
\label{sec:literature_review}

This paper contributes to the literature on dynamic panel estimation with fixed effects. This literature has a long history, beginning with \cite{griliches1967distributed}\footnote{\cite{griliches1967distributed} discussed how time series regression parameters are estimated with bias when intercepts are included. This phenomenon was also studied by \cite{nerlove1971further}. } and other researchers, who investigated how dynamics in outcomes lead to violations of the strict exogeneity of errors assumption, which is necessary for unbiased ordinary least square error (OLS) coefficient estimation including fixed effect estimation. \cite{nickell1981biases} derived an explicit formula for the bias of OLS parameters when past outcomes and unit fixed effects are included in the regression, assuming all other regressors are strictly exogenous. Nickell bias can be thought of as a specific type of incidental parameter bias \citep{neyman1948consistent}.\footnote{Incidental parameter bias arises in fixed effects panel models with the number of time periods $T$ is small relative to the number of units $N$. This bias occurs because each unit fixed effect is estimated only using a few observations, leading to biased estimates of the fixed effects.}  The primary focus of the literature has been on investigating how including past outcomes in the regression leads to bias, particularly with regard to the coefficient on the past outcomes. However, in applied work the statistical object of interest is often the treatment effect estimate, rather than the dynamic process itself, and do not include past outcomes as controls.  This paper is therefore uniquely contributing to the literature by focusing on the coefficient on the treatment effect rather than the coefficient on the past outcome, which is treated as a nuisance parameter. It is also the first to focus on parameter estimation when the past outcome is not included in the OLS regression model. By studying treatment effects in models that exclude past outcomes, I am able to characterize dynamic bias. This characterization reveals that dynamic bias can be much larger than the previously studied Nickell bias. In all simulation specifications, dynamic bias is larger than Nickell bias, even when treatment is random.

To correct for dynamic bias, I provide a bias-corrected (DBC) estimator. Given that Nickell bias is often much smaller than dynamic bias, my bias correction procedure calls for first controlling for past outcomes, and then correcting the resulting Nickell bias.  In settings with fixed $T$, there are two main approaches for dealing with Nickell bias. The first and most well-known approach is that of \cite{holtz1988estimating} and \cite{arellano1991some}, which is based on instrumental variables. The second approach is an analytical method that I build upon, following the works of \cite{kiviet1995bias}, \cite{juodis2015iterative}, and \cite{breitung2022bias}.

The instrumental variables (IV) approach is based on using past outcomes as instruments for endogenous regressors. Its goal is to use further lags of the outcomes, or treatments, as instruments for current differenced outcomes and treatments. These instruments may be weak, and the more time periods available the more possible instruments, which can lead to problems associated with many weak instruments \citep{mikusheva2024weak}.\footnote{The weak instrument issue is partially addressed by \cite{blundell1998initial}, who proposed a system GMM solution by including both first differences and levels of past outcomes as instruments. However, this approach still suffers from the weak instrument problem when the variance of individual effects is greater than the variance of the errors (see \cite{bun2010weak}).}  In practice, estimates obtained using instrumental variables are quite sensitive to the choice of instruments, making instrument selection a daunting task for applied researchers.\footnote{See Section \eqref{sec:emperical_bias_correction} for application to \cite{dell2012temperature}. Depending on the instruments used, point estimates for both treatment and past outcomes flip signs.}

Instead of using instrumental variables, I follow \cite{kiviet1995bias} and analytically correct the bias. The analytical bias correction avoids problems associated with instruments while still providing a correction in the fixed-T setting. The downside of the analytical bias correction method is that it requires analytical work that is outcome and treatment model-specific, which IV methods do not. The past analytical bias literature provided corrections for a specific class of models.  Both \cite{kiviet1995bias} and \cite{breitung2022bias} focus on corrections for models with endogeneity from past outcomes and do not accommodate endogenous treatment. \cite{breitung2022bias} extends the work of \cite{kiviet1995bias} by allowing for multiple lags of the outcome variable and for incorporating the bias correction into a GMM framework.  I build off \cite{breitung2022bias} and also use a Generalized method of moments (GMM) framework for the DBC. Instead of GMM, \cite{juodis2015iterative} provides an iterative analytical bias correction for Vector Autoregressive (VAR) systems. VAR models also do not accommodate endogenous treatments; endogenous treatments in time period $t$ impact outcomes in time period $t$. Therefore to allow for endogenous treatments I extend the analytical bias corrections to  structural VARs (SVARs). This extension requires that I make a structural assumption to avoid simultaneity problems: treatment in a time period $t$ impacts the outcome in time period $t$, but the outcome in time period $t$ doesn't impact treatment in time period $t$. I allow past outcomes, like those from time period $t-1$, to impact treatment in time period $t$. This paper is the first to provide an analytical correction for settings with endogenous treatments. This analytical correction is also the first to allow for interaction terms between endogenous variables and exogenous variables, allowing treatment effects to vary based on observable characteristics.

I work in the fixed-T setting to help applied researchers sidestep the issue of having to guess whether they have ``enough'' time periods for a correction that yields valid inference. Although asymptotic normality is only guaranteed in a fixed-T setting with instrumental or analytical bias approaches, there are other approaches for de-biasing Nickell bias as long as T is allowed to grow. One correction for Nickell bias in the large T asymptotic setting is the jackknife approach (e.g., \cite{dhaene2015split}), which involves taking samples of the panel data with different lengths of T to calculate the bias. This hands-off approach is easy to implement but results in larger standard errors than the analytical method \citep{fernandez2016individual}. Fixed-T analytical corrections work with flexible linear models but do not allow for non-linear non-separable  models (e.g., logit models). Marginal effects are not identified in fixed-T settings in non-separable models \citep{chernozhukov2013average}. Many applied papers in environmental economics seek to estimate treatment effects, a type of marginal effect, so I impose flexible linearity to allow for their estimation.

This paper demonstrates the importance of explicitly including past outcomes in the regression model for panel data settings where there is a relationship between past and current outcomes,\footnote{A method for testing whether past outcomes influence current outcomes, as opposed to merely exhibiting autocorrelation in the model's errors, is discussed in \cite{chamberlain1982multivariate}.}  Currently, applied researchers often employ various alternative methods to account for outcome history without incorporating past outcomes into their models.  Common approaches include adding time trends to the models or employing factor model-based methods.\footnote{For examples of applied research using factor models to control for outcome history, see \cite{damm2024beyond}; for examples using time trends, see \cite{annan2015federal}.} However, these methods do not control for dynamics in outcomes—that is, the influence of past outcomes on current outcomes, meaning the dynamics bias persists in these estimates.

Time trends are often introduced into models by adding unit-specific trends, which are constructed by interacting unit-specific dummy variables with a polynomial of the time variable. While these trends capture a general time-specific pattern for each unit, they fail to account for the influence of past outcomes on current outcomes, such as in autoregressive processes. In other words, time trends alone do not capture the dynamic relationships in the data where past values directly affect present outcomes. This is also seen empirically in the applied example in Section \ref{sec:emperical_examples}. Another prevalent method for controlling for outcome history is the use of factor models. Factor model-based approaches, such as synthetic controls and synthetic differences-in-differences, \citep{arkhangelsky2021synthetic, abadie2010synthetic} rely on low-rank factor assumptions that do not accommodate unit-specific dynamics.\footnote{In the context of panel data, a low-rank factor model aims to capture cross-sectional correlations among different units (e.g., individuals, firms, or countries) by assuming that these correlations can be explained by a limited number of common factors. Importantly, such models focus primarily on capturing variation between units rather than time-series dynamics within each unit.} Consequently, to properly account for the dynamics where past outcomes influence current outcomes, researchers must directly include past outcomes in their models.\footnote{In bio-statistics, the g-formula is also used to control for outcome history, but it does not accommodate selection on unobserved fixed effects \citep{robins1986new, naimi2017introduction}.}

In the applied literature, when the correlation between past and current outcomes is large, another approach for dealing with dynamics is transforming outcomes to reduce the correlation. For example, GDP levels are correlated highly over time, with a correlation parameter of .95, the correlation of the difference transformed GDP (GDP growth) is smaller and closer to .3. However, transforming the outcomes through this approach still creates bias in treatment effect estimates. This is discussed in greater detail in the next section.

\section{Preview of Results}
\label{section:applied_motivation}
Before presenting the theoretical results of the paper, I preview the main empirical results of this paper. I give brief intuition for the origins of dynamic bias. I also compare dynamic bias and Nickell bias in simulations both when treatment is random and when it is endogenous (related to past outcomes). I then discuss how commonly used transformations of the outcome variable interact with these biases.

\subsection{Random Treatment}
My applied example is based on the literature that studies the effect of temperature on GDP. The units are countries and the time periods are years. Suppose that the true model is the simple model given in Equation \eqref{eq:true_model}: temperature, $\text{Temp}_{i,t}$, in every time period is an independent and identically distributed (i.i.d.) random shock conditional on country.\footnote{$\text{Temp}_{i,t} = c_i + u_{i,t}$ where $c_i$ is a country fixed effect and $u_{i,t}$ is a random shock.} In a fixed effects model, treatment is therefore as good as random conditional on the fixed effects. Additionally, past GDP is included in the true model, as \cite{solow1956contribution} explains that past GDP impacts current GDP.\footnote{One way that past GDP affects current GDP through its impact on capital accumulation. Higher GDP in the past implies higher savings and investment, leading to a larger capital stock in the current period. Since the capital stock is an input in the production function, a larger capital stock results in higher current GDP.} 
\begin{equation}
\label{eq:true_model}
\begin{aligned}
     \text{True Model with Random Treatment: } \quad  \text{GDP}_{i,t} = a_i + \tau_0 \text{Temp}_{i,t} +  \rho_{10} \text{GDP}_{i,t - 1} + \epsilon_{i,t}.
\end{aligned}
\end{equation}

\subsubsection{Models}

In this section, I introduce two models that could be used to estimate the treatment effect $\tau_0$. Researchers often estimate these models using OLS. A necessary condition for OLS to be unbiased is that treatment in time period $t$, $\text{Temp}_{i,t}$, is uncorrelated with the model error $e_{i,t}$ in time period $t$.

\begin{enumerate}

      \item[] 
      \begin{equation}
      \label{eq:dynamic_model_intro}
  \text{Dynamic Model: } \quad  \text{GDP}_{i,t} = a_i + \tau_0 \text{Temp}_{i,t}  + \rho_{10} \text{GDP}_{i,t -1} + \epsilon_{i,t}.
\end{equation}

The model error here $\epsilon_{i,t}$ is simply the original true model error, so by construction we know that it is uncorrelated with treatment.  

    \item[] \begin{equation}
  \text{Static Model: } \quad  \text{GDP}_{i,t} = a_i + \tau_0 \text{Temp}_{i,t}  + \underbrace{e_{i,t}}_{ \rho_{10} \text{GDP}_{i,t -1} + \epsilon_{i,t}}.
  \end{equation}

  In the Static Model, researchers do not control for $\text{GDP}_{i,t-1}$. However, $\text{GDP}_{i,t-1}$ is part of the true model given Equation \ref{eq:true_model}, therefore the $\text{GDP}_{i,t-1}$ term appears in the new model error along with the original error. Therefore, $e_{i,t} :=  \rho_{10} \text{GDP}_{i,t -1} + \epsilon_{i,t} $. Still, $\text{Temp}_{i,t}$ remains uncorrelated with the error term $e_{i,t}$. This is because the temperature is random, so it is not related to previous outcomes $\text{GDP}_{i,t-1}$ or model errors $\epsilon_{i,t}$ that appear in $e_{i,t}$.

\end{enumerate}

\subsubsection{Cause of Bias}
Researchers often estimate these models because they see that treatment in time period $t$, $\text{Temp}_{i,t}$, is uncorrelated with model error in time period $t$. However,researchers still have to estimate fixed effects $a_i$ by either using dummies, the within transformation, or first differences.\footnote{Running a  regression with within-transformed data leads to same coefficients as running a regression with unit dummies \citep{wooldridge2010econometric}. Running first-differences also generates bias of a similar form.} These transformations create new variables that are functions of data in multiple time periods. For example, consider the within transformation, 

\begin{equation}
    \widetilde{\text{Temp}}_{i,t} = {\text{Temp}}_{i,t} - \frac{1}{T} \sum_{s = 1}^T {\text{Temp}}_{i,s}. 
\end{equation}

This new treatment variable $\widetilde{\text{Temp}}_{i,t}$ is now a function of $\text{Temp}_{i,s}$ in all time periods. Because of this, unbiased OLS estimation requires not only that treatment in time period $t$, $\text{Temp}_{i,t}$, is uncorrelated with model error in time period $t$, but also that regressors in \textit{all} time periods are uncorrelated with model error, which is known as strict exogenieity. The fixed effect estimation turns past outcomes into generated confounders if they are correlated with \textit{either} the outcome or the treatment.\footnote{In classic cross sectional causal inference we think of variables as confounders if they are correlated with \textit{both} the outcome and treatment. } Both models, the Static Model and Dynamic Model,  lead to biased treatment effects. 

\begin{enumerate}

    \item In the Dynamic Model, treatment in all time periods is uncorrelated with model error because the past outcome, $\text{GDP}_{i,t}$,  is controlled for. However $\text{GDP}_{i,t}$ itself in other time periods is correlated with model error, leading to the well studied \textbf{Nickell bias}. Then because the coefficient on $\text{GDP}_{i,t-1}$ is estimated with bias, this spills over and biases the treatment estimate. Nickell bias is discussed in detail in Section \ref{sec:nickell_bias_review}.

    \item In the Static Model, treatment in other time periods is correlated with model error. Specifically, $\text{Temp}_{i,t-1}$ is correlated with $e_{i,t}$, which leads to biased treatment effects. I term this \textbf{dynamic bias}, which is discussed in detail in Section \ref{sec:dynamic_bias}. 
    
    \end{enumerate}

\paragraph{Variation on dynamic bias:} Before showing simulation results, I introduce a variation on the Static Model, which I call the Delta Model, often implemented in applied work that also leads to dynamic bias. 

      \begin{equation}
  \text{Delta Model:}\footnote{\text{Since Temp$_{i,t}$ only impacts GDP$_{it}$ and not GDP$_{i,t-1}$, the effect of  Temp$_{i,t}$ on the transformed outcome, $\Delta \text{GDP}_{i,t}$,} \text{ is the same the effect of  Temp$_{i,t}$ on the untransformed outcome $\text{GDP}_{i,t}$ }.} \quad  \underbrace{\Delta \text{GDP}_{i,t}}_{\text{GDP}_{i,t} - \text{GDP}_{i,t-1}} = a_i + \tau_0 \text{Temp}_{i,t}  +  \underbrace{\eta_{i,t}}_{(1 - \rho_{10}) \text{GDP}_{i,t -1}  +  \epsilon_{i,t}}.
\end{equation}

Some researchers suspect that highly persistent outcomes ($\rho_{10}$ close to 1) can cause problems with their analysis, so instead they study transformations of their outcomes, such as differences or growth. When taking the difference on the left-hand side, this imposes a $\rho_{10}$ coefficient of 1, since $1 \times \text{GDP}_{i,t-1}$ is subtracted from the model. The difference between 1 and true $\rho_{10}$  remains in the error of the model and therefore $\eta_{i,t} := (1 - \rho_{10}) \text{GDP}_{i,t -1}  +  \epsilon_{i,t} $.\footnote{Note here that only the outcome ($GDP_{i,t}$) is being transformed through differences - the variables on the right-hand side are not being differenced - therefore this transformation is not equivalent to the the first-differences transformation.} A type of dynamic bias occurs in this model because like in the Static Model, part of the outcome remains in the error term. It is the case that $\text{Temp}_{i,t-1}$ is correlated with $\eta_{i,t}$, discussed in detail in Appendix \ref{append:growth_vs_level}, also leading to bias. I call this type of dynamic bias \textbf{transformation bias}. 

\subsubsection{Simulation}

\begin{figure}[h!]
    \centering
    \includegraphics[scale=0.5]{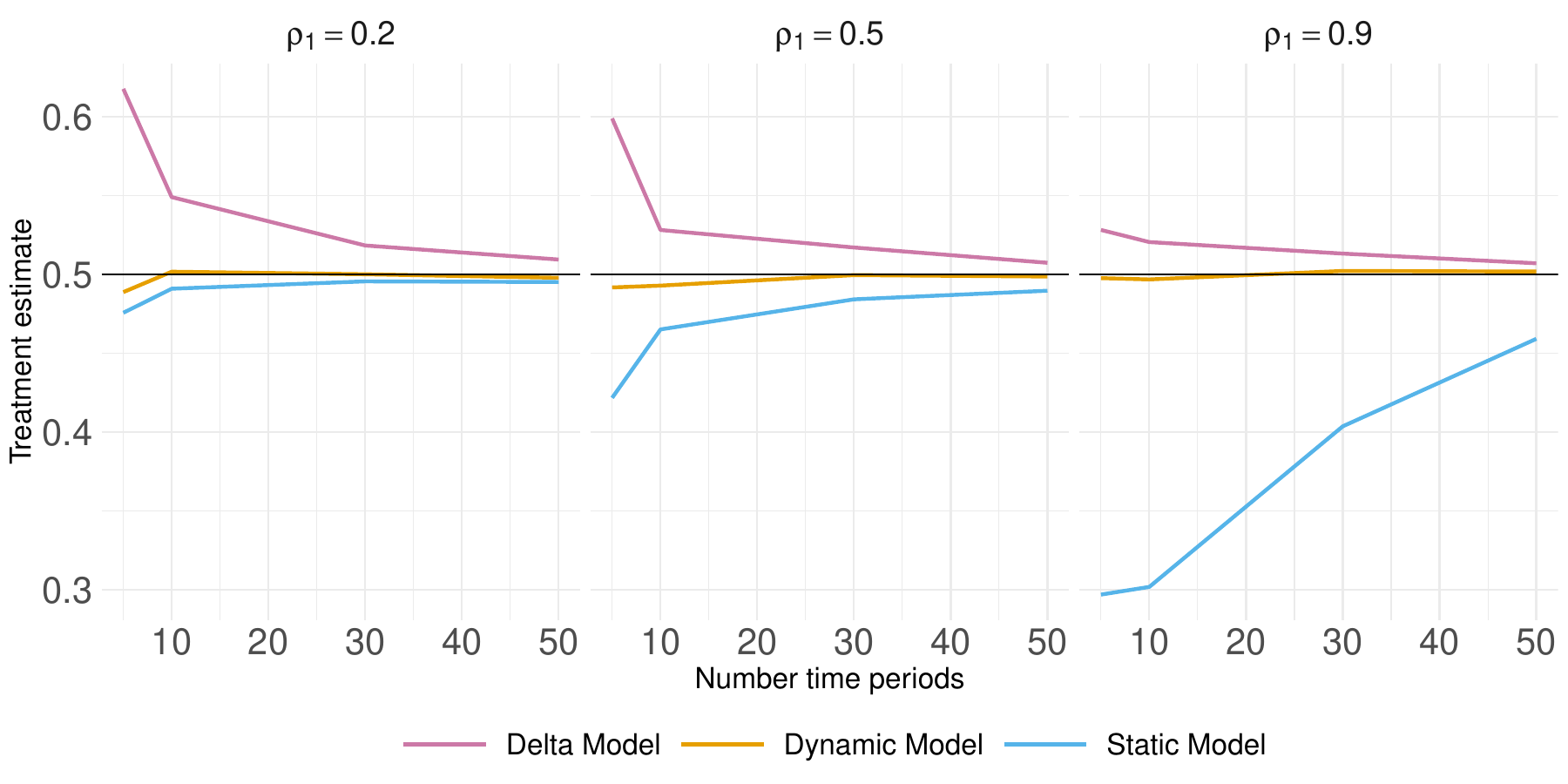}
    \caption{Bias of three different models.}
    \label{fig:growth_rate_into}
\end{figure}

Although all three models (Dynamic Model, Static Model, and Delta Model) lead to biased treatment effect estimates, the magnitude of the bias varies greatly. I run a simulation and present the results in Figure \eqref{fig:growth_rate_into} to highlight some key takeaways from these biases. I generate datasets based on the true model given in Equation \eqref{eq:true_model}  with random treatment. 
I set the true treatment effect $\tau_0 = .5$, and simulate a variety of datasets. Each of the three panels in the figure corresponds to a different value of the correlation between past outcomes and current outcomes $\rho_{10} \in (.2,.5,.9)$ used to generate the data. I simulate datasets with 1000 units and vary the number of time periods along the x-axis.  I estimate the treatment effects using OLS to estimate the three models above, and plot the value of the treatment effect estimates on the y-axis. The number of time periods in the dataset and the magnitude of $\rho_{10}$ significantly impact the magnitude of the bias.

The Dynamic Model (Equation \ref{eq:dynamic_model_intro}) leads to the smallest treatment effect estimate bias out of all three models regardless of the value of $\rho_{10}$. The intuition for this is that only the Dynamic Model explicitly controls for past outcome, and so only in this model does treatment remain strictly exogenous. 

As for whether the Static Model or Delta Model has the largest bias, that depends on the value of $\rho_{10}$. When $\rho_{10}$ is close to 1 (the right most panel), the bias of the Delta Model is smaller than the bias of the Static Model. This is because in the Delta Model, the past outcome with a coefficient of 1 is subtracted from the current outcome to create the transformed outcome $\Delta \text{GDP}_{i,t}$. Therefore if $\rho_{10}$ is close to 1, this subtracting is sort of controlling for the past outcome. The opposite is true when $\rho_{10}$ is small (close to 0, the left-most panel) - then subtracting out a value of 1 leads to more bias. 

One key lesson for applied work is that the fewer the number of time periods in the panel dataset, the worse the bias of all these methods. This bias is important to keep in mind when shortening panel datasets; panel datasets are often shortened in environmental research when studies differentiate between temperature and climate change by using ``long differences''.\footnote{Papers that use long differences include \cite{nordhaus2006geography}, \cite{deryugina2014does}, \citep{burke2015global}.} When researchers implement long differences, they normally reduce the length of the panel, which in turn increases the bias of the above models. On the other hand, the longer the difference, the smaller the correlation between the outcomes in the two time periods, which can reduce the bias as well. 

\subsection{Endogenous Treatment}

Treatment effect estimates are even more biased in settings where treatment is not random and is instead related to past outcomes, making it endogenous. Treatment can be related to past outcomes due to selection into treatment. \cite{marx2022parallel}, \cite{ghanem2022selection} show that many economic models lead to treatment selection based on past outcomes. 

For example, selection into environmental policy treatment is often a function of past environmental conditions. Policies that target air pollution in specific areas are implemented because of those areas' past pollution levels \citep{chay2005does}. In Brazil, regional deforestation protection is based on previous years deforestation \citep{harding2021commodity, assunccao2023optimal}. In public finance, we might expect that health care spending in the past impacts health care plan choice, which then impacts future health care spending.\footnote{A key finding in \cite{kowalski2023reconciling} was that emergency room utilization before the Oregon Health Insurance Experiment began affected enrollment into health insurance within the experiment, which in turn affected emergency room utilization.} In development economics, the assignment of drought relief in southern India has been found to be a function of the region's past outcomes \citep{tarquinio2022politics}. In industrial economics,  past firm productivity, which is the outcome, impacts output choices in current time periods \citep{olley1992dynamics}. 

When treatment is endogenous, the bias does not disappear as the number of time periods increases for the Static and Delta model.  Since the treatment is endogenous, omitted variable bias arises when past outcomes are not explicitly included in the model. This omitted variable bias does not decrease, regardless of how large the number of time periods or units is. Simulations in Appendix \ref{sec:applied_endo_sims} demonstrate that even with a small degree of endogeneity, the bias in treatment effect estimates can be substantial.  Omitted variable bias can be removed by controlling for past outcomes and correcting for the resulting Nickell bias, which is smaller than the omitted variable bias caused by failing to control for past outcomes.

\section{Theoretical Results} 
\label{sec:theoretical_results}

Dynamic bias arises when there are dynamics in outcomes but researchers do not control for past outcomes when estimating treatment effects. Dynamic bias is explained and characterized in Section \eqref{sec:dynamic_bias}. Correcting dynamic bias requires controlling for past outcomes. Estimators that include past outcomes as controls still suffer from Nickell bias, which is discussed in Section \eqref{sec:nickell_bias_review}. Section \eqref{sec:nickell_bias_correction} provides the dynamic bias-corrected (DBC) estimator which eliminates Nickell bias.

\subsection{Notation}

I use $Y_{i,t}$ to denote the outcome, $Y_{i,t-h}$ to denote the $h^{th}$ lag of the outcome, $D_{i,t}$ to denote treatment, and $W_{i,t}$ to denote other covariates. I use the superscript $t$ to denote the vector of variables up to time period $t$, for example $D_{i}^t := (D_{i,1}, D_{i,2}, \cdots, D_{i,t})$. The concatenation of all possible regressors is given by $X_{i,t} := (D_{i}^t, W_{i}^t, Y_{i}^{t-1})$. 

Note that the covariates $W_{i,t}$ can include time period dummies. That is for each time period $t$ I can include a dummy $Q_t$ with 1 in the $t^{th}$ position. This enables the model to include time effects.

 I use $\indep$ to denote independence, and $\perp$ for uncorrelated.  I use bar notation for averages over time, for example $\Bar{Y}_{i,-1} := \frac{1}{T} \sum_{t=1}^T Y_{i,t-1}$ and $\Bar{Y}_{i} := \frac{1}{T} \sum_{t=1}^T Y_{i,t}$. I use tilde to denote the within transformation, for example, 
 
 \begin{equation}
 \label{eq:def_within_transformation}
     \Tilde{Y}_{i, t-1} := Y_{i,t-1} - \bar{Y}_{i,-1}.
 \end{equation}


\subsection{Setup}
I begin by introducing the setting under the potential outcomes framework (\cite{neyman1923}, \cite{rubin1974estimating}). I observe outcomes $Y_{i,t}$ for a sample of units $i = 1, \dots, N$ for time periods $t = 1, \dots, T$. The number of time periods $T$ is fixed, but the cross-sectional dimension $N$ grows with more observations. Therefore, the formal asymptotic results are for the large $N$ setting where $N \rightarrow \infty$ and $T$ is fixed. This is also sometimes known as the ``short panel''.

For now, let us assume that the treatment $D_{i,t}$ is continuous. 
For each value $d$ of the treatment $D_{i,t}$, unit $i$ has a corresponding potential outcome $Y_{i,t}(D_{i,t} = d)$.  In the example in \cite{dell2012temperature}, $D_{i,t}$ is temperature and $Y_{i,t}$ is country GDP. The authors are interested in estimating the
``contemporary causal effect of temperature on the development
process''.\footnote{One may also be interested in the estimation of the long term effect of treatment, as studied in \cite{hausman1985taxes}, as opposed to the contemporary. I focus on the contemporary effect in this paper as it is the most common effect of interest in a set of surveyed applied papers.} Formally, the causal object of interest is average effect of marginally increasing temperature in the current time period ($D_{i,t}$) on GDP, holding all else fixed. This is also referred to the contemporary average partial derivative (APD) and is written as
\begin{equation}
\label{eq:APD}
    \tau_{0} = \E \bigg[ \frac{\partial Y_{i,t}(D_{i,t})}{\partial D_{i,t}} \bigg],
\end{equation}

where the expectation is over all units and time periods.

I impose structure on the potential outcomes to allow estimation. In the rest of this section, I showcase that dynamic bias is a problem even in a simple (though common) stylized model. I use the formula for the bias in this example to provide intuition. However, when I provide a bias correction in Section  \ref{subsection:estimation} I introduce a more general model (Section  \ref{sec:more_flexiable_model}), to which the DBC correction also applies.

\begin{assumption}(Stylized Example One: Random Treatment.)
\label{ass:simple_random_treatment}
Let the true vector of parameters be given by $\theta_0 := (\rho_{10}, \tau_{0})$. The errors $\epsilon_{i,t}$ are i.i.d mean zero with variance $\sigma^2_{\epsilon, i}$ and the errors $u_{i,t}$ are i.i.d mean zero with variance $\sigma^2_{u, i}$. The true underlying  has the following structure:
    \begin{equation}
\label{eq:stylized_potential_outcome}
    Y_{i,t}(D_{i,t}) = a_i + \tau_{0} D_{i,t} + \rho_{10} Y_{i,t-1} + \epsilon_{i,t},
\end{equation}
\begin{equation}
\label{eq:stylized_treatment_outcome}
    D_{i,t} = c_i + u_{i,t}.
\end{equation}
\end{assumption}

 This simple example is based off the setting of \cite{dell2012temperature}. I include $Y_{i,t-1}$ in the outcome model because economic theory tells us that past GDP ($Y_{i,t-1}$) impacts current GDP ($Y_{i,t}$) \citep{solow1956contribution}.  As is standard in panel data, I impose additive fixed effects $a_i$, often called individual fixed effects,  which can be correlated with covariates.  In the treatment model in Equation \eqref{eq:stylized_treatment_outcome},treatment $D_{i,t}$ is also a function of individual  fixed effects $c_i$.\footnote{It is important to note dynamic bias remains even if treatment is not a function of fixed effects and $D_{i,t} = u_{i,t}$.} This structure mimics \cite{dell2012temperature}, who write that treatment is correlated with country fixed effects $a_i$. The presence of fixed effects in the true structural model generates the need to control for fixed effects in treatment effect estimation.

However, in the data we don't observe the fixed effects. Further, they cannot be estimated consistently in short panels, which is known as the incidental parameter problem \citep{neyman1948consistent}. I bypass the estimation of fixed effects by using within-estimation, which leads to the same estimates of $\theta = (\rho_{10},\tau_0)$ as would explicitly estimating fixed effects by including fixed effect dummies for each unit.

\begin{assumption} (Weak Exogeneity.) 
\label{ass:exogeneity}

\begin{equation}
      \quad \epsilon_{i,t}, u_{i,t} \perp (X_{i}^t, a_i), \quad X_{i}^t := (X_{i,1}, \dots ,X_{i,t-1}, X_{i,t} ). 
     \end{equation} 
\end{assumption}

The weak exogeneity assumption says the that error has zero conditional expectation given current and past covariates. Because the structural outcome model  allows for lagged outcomes to impact current outcomes, I cannot make the more familiar strict exogeneity assumption, which would require that the regressors are uncorrelated with the error term across all time periods, meaning that current, past, and future values of the regressors do not influence the current error term.

\begin{assumption} (Independence across $i$)
\label{ass:independence}
The data vectors $Z_i = \{(Y_{i,t}, X_{i,t}')'\}_{t = 1}^T$ are i.i.d. across units $i$. 
\end{assumption}



\subsection{Dynamic Bias}
\label{sec:dynamic_bias}

Given the model laid out above,  I now introduce dynamic bias and show how it impacts treatment effect estimates. Dynamic bias arises when there is a relationship between past outcomes and current outcomes in the true model but past outcomes are not included in the estimation  model. A crucial feature of  dynamic bias is that it arises even when the treatment is random.

\subsubsection{Causal Object}

Given the simple stylized model presented in Assumption \eqref{ass:simple_random_treatment}, in particular the homogeneous treatment assumption in Equation \ref{eq:stylized_potential_outcome}, the treatment object of interest is simply $\tau_0$. This follows using our definition of the APD from Equation \eqref{eq:APD} along with potential outcome given in Equation \ref{eq:stylized_potential_outcome}.

\begin{equation}
     \E \bigg[ \frac{\partial Y_{i,t}(D_{i,t})}{\partial  D_{i,t}} \bigg] = \E \Bigg[ \frac{\partial \big(a_i + \tau_{0} D_{i,t} + \rho_{10} Y_{i,t-1} + \epsilon_{i,t} \big)}{\partial D_{i,t}} \Bigg] = \tau_0.
\end{equation}

In the context of  \cite{dell2012temperature} this would represent the contemporary causal effect of temperature on the development process.

\subsubsection{Estimation}

In settings where treatment assignment is random, applied researchers often rely on a static model that does not account for past outcomes to estimate $\tau_0$. I write the static model in Equation \eqref{eq:dell_simple}.

\paragraph{Model 1: Static Model}

\begin{equation}
\label{eq:dell_simple}
    Y_{i,t} = a_i + \tau^D D_{i,t} + e_{i,t},
\end{equation}

Given the true model presented in Assumption \ref{ass:simple_random_treatment}, the error in the static model is given by $e_{i,t} := \rho_{10} \text{Y}_{i,t -1} + \epsilon_{i,t}$. 
Since applied researchers do not observe fixed effects $a_i$, they can not use OLS to estimate equation \eqref{eq:dell_simple} directly. Instead they have to estimate the fixed effects with dummy variables, or by using the within or first difference transformation. I remove the fixed effects by using the within-transformation, which recall was defined in Equation \eqref{eq:def_within_transformation}. 

\paragraph{Model 2: Within-Transformed Static Model}
\begin{equation}
\label{eq:dell_simple_within}
    \Tilde{Y}_{i,t} = \tau^D \Tilde{D}_{i,t} + \Tilde{e}_{i,t}.
\end{equation}
\paragraph{OLS Estimator.} To estimate the treatment effect, OLS is used to estimate Equation \eqref{eq:dell_simple_within}. I call the coefficient estimated this way $\hat{\tau}^{D}$.  Unfortunately $\hat{\tau}^{D}$  is inconsistent—no matter how large the number of units grows, the estimator does not converge to the true parameter value.

\begin{theorem}
\label{theorem:omit_lag_y} (Dynamic Bias.)
Under Assumptions \eqref{ass:simple_random_treatment}, \eqref{ass:exogeneity}, \eqref{ass:independence}, and that $Var(D_{i,t}) > 0$, running OLS to estimate the static model in Equation \eqref{eq:dell_simple_within} leads to a biased treatment effect estimator $\hat{\tau}^D$.

\begin{equation}
\begin{aligned}
        \plim_{N \rightarrow \infty} \hat{\tau}^{D} &=  \tau_0 
        - \frac{\rho_{10}\tau_0 }{T(T-1)}\bigg[ \frac{T}{1- \rho_{10}} - \frac{1 - \rho_{10}^T}{(1 - \rho_{10}^2)}\bigg] .
\end{aligned}
\end{equation} 

\end{theorem}

The proof of Theorem \eqref{theorem:omit_lag_y} is given in Appendix \ref{appendix:dynamic_bias}. This bias is what I term dynamic bias. This bias is of order $1/T$, and it diminishes as the number of time period increases.  

\subsubsection{Dynamic Bias Intuition}

Dynamic bias arises due to the estimation of fixed effects, which are typically estimated by including dummies, applying within-transformation, or first-differencing the data. This estimation \textit{generates} confounding if past outcomes are not controlled for. In classic cross-sectional causal inference, a covariate is a confounder only if it is correlated with \textit{both} the treatment and the outcomes. This reasoning explains why many researchers opt for static panel regressions; in their contexts, treatment is random, so although past outcomes affect current outcomes, they are not correlated with the treatment and, therefore, are not classic confounders \citep{angrist2009mostly}. Due to fixed effects estimation, a past outcome is a generated confounder and generates bias if it is correlated with \textit{either} the treatment or the outcome.

 Estimation of models with fixed effects requires strict exogeneity for unbiased estimation. When treatment is random, and the past outcome is controlled for in the model, then treatment is a strictly exogenous regressor. However, when the past outcome is not controlled for, the past outcome is part of the model error, and treatment is no longer strictly exogenous. This leads to biased treatment effects. 

To see algebraic intuition for why strict exogeneity does not hold, consider the within-transformation approach for estimating fixed effects. After applying the within transformation, the newly transformed variables become functions of data from all time periods. Consequently, the errors $\Tilde{e}_{i,t}$ are functions of errors $e_{i,t}$ across all time periods, causing the within-transformed error to be correlated with the treatment, which in turn leads to endogeneity.

\begin{enumerate}
    \item $\widetilde{e}_{i,t}$ is a function of $e_{i,t + 1}$. 
    
    (Because of the definition of the within transformation: $\widetilde{e}_{i,t} = e_{i,t} - \frac{1}{T} \sum_{s = 1}^T e_{i,s}$.)

    \item $e_{i,t + 1}$ is a function of $\text{Y}_{i,t}$. 
    
    (Because $e_{i,t + 1} := \rho_{10} \text{Y}_{i,t} + \epsilon_{i,t + 1}$ by definition of the Static Model error.)
    \item $\text{Y}_{i,t}$ is a function of ${\text{D}}_{i,t}$
    
    (Because $ \text{Y}_{i,t} = a_i + \tau_0 \text{D}_{i,t} +  \rho_{10} \text{Y}_{i,t - 1} + \epsilon_{i,t}$ by definition of the true model in Equation \eqref{eq:stylized_potential_outcome}.)
\end{enumerate}

Therefore $\widetilde{e}_{i,t}$ is a function of ${D}_{i,t}$.


However, if past outcomes had been included in the model, the treatment would have been a strictly exogenous regressor. Dynamic bias occurs when the outcome lag is not considered, leading to potentially large biases in treatment effect estimates. This dynamic bias can be mitigated by including past outcomes in the regression model.

However, even when past outcomes are included, the well-known Nickell bias remains. In the next section, I will discuss Nickell bias and then demonstrate in Section \ref{sec:simulation_study} that, in simulations, Nickell bias is much smaller than dynamic bias. Regardless of the values of $\rho_{10}$ or $\tau_{0}$, I show that in simulation that dynamic bias is consistently larger than Nickell bias.

\section{Debiased Estimator (DBC)}
\label{subsection:estimation}
  The DBC estimator works both when treatment is random and when treatment is a function of past outcomes and is therefore endogenous.

To explain how the bias correction works, I assume the following simple data-generating processes for the remainder of the section. However, the results of the paper hold for a richer class of models.\footnote{The richer class of models are given in Section \ref{sec:more_flexiable_model} below.} Equation \eqref{eq:treatment_model} shows that the treatment is a function of the past outcome. Recall biases get worse if, in addition to past outcome not being controlled for, treatment is related to past outcomes in the true model (not random). This is because in addition to dynamic bias there is now omitted variable bias. Equation \eqref{eq:outcome_model} shows that outcomes are a function of the past outcome and treatment. This simple example highlights the causal parameter of interest, inferential goal, and key assumptions of this approach.

\begin{assumption}(Stylized Example Two: Endogenous Treatment.)
\label{ass:dgp_endo_treat}
The true underlying data generating  has the following structure:
    \begin{equation}
\label{eq:outcome_model}
    Y_{i,t}(D_{i,t}) = a_i + \tau_{0} D_{i,t} + \rho_{10} Y_{i,t-1} + \epsilon_{i,t},
\end{equation}
\begin{equation}
\label{eq:treatment_model}
    D_{i,t} = c_i + \rho_{20} Y_{i,t-1} + u_{i,t},
\end{equation}
\end{assumption}

where as before $\epsilon_{i,t}$ and $u_{i,t}$ are i.i.d mean zero true errors with variance $\sigma^2_{i,\epsilon}$ and $\sigma^2_{i,u}$. 

\subsection{Dynamic Model}
\label{sec:nickell_bias_review}

Nickell bias occurs when a regression includes both estimated fixed effects and past outcome variables as regressors. Since fixed effects are not observed, they have to be estimated with dummies, the within-transformation, or first differences. These transformations impact all parts of the model, including the error terms. Taking the within transform as an example, the transformed error term becomes a function of error terms in all time periods. Therefore, if a regressor is a function of past error terms (like lagged outcome variables), it becomes correlated with the new within-transformed errors. 

\paragraph{Model 3: Within-Transformed Dynamic Model}

\begin{equation}
\label{eq:ols_within}
    \Tilde{Y}_{i,t} =  {\tau}_0 \Tilde{D}_{i,t} + {\rho}_{10} \Tilde{Y}_{i,t-1} +  \Tilde{\epsilon}_{i,t}
\end{equation}
\begin{equation}
\label{eq:ols_within_treat}
    \Tilde{D}_{i,t} =  {\rho}_{20} \Tilde{Y}_{i,t-1}  + \Tilde{u}_{i,t}
\end{equation}

\paragraph{OLS Estimator.} One could estimate treatment effects with a dynamic model by using OLS to estimate Equation \eqref{eq:ols_within}. This OLS regression leads to estimates of $\hat{\tau}^{NB}$ and $\hat{\rho}_1^{NB}$ that have Nickell bias. Though less common in practice, researchers could also use OLS to estimate Equation \eqref{eq:ols_within_treat} to get the estimate $\hat{\rho}_2^{NB}$. Our vector of estimated OLS parameters is $\hat{\theta}^{NB} := (\hat{\rho}_1^{NB}, \hat{\tau}^{NB}, \hat{\rho}_2^{NB}) $. The estimated residuals for a given parameter vector $\hat{\theta}$ are given by $\Tilde{\epsilon}_{i,t}(\hat{\theta}) := \Tilde{Y}_{i,t}  - \hat{\tau} \Tilde{D}_{i,t} - \hat{\rho}_{1} \Tilde{Y}_{i,t-1} $ and $\Tilde{u}_{i,t}(\hat{\theta}) := \Tilde{D}_{i,t}  - \hat{\rho}_2 \Tilde{Y}_{i,t -1}$ . By definition we have that $\Tilde{\epsilon}_{i,t}(\theta_0) = \Tilde{\epsilon}_{i,t}$ and $\Tilde{u}_{i,t}(\theta_0) = \Tilde{u}_{i,t} $.

To see concretely how the Nickell bias arises, focus on the OLS estimation of Equation \eqref{eq:ols_within}. I use the familiar formula for our OLS coefficients \big($\hat{\theta} = (\mathbb{X}'\mathbb{X})^{-1}\mathbb{X}'\mathbb{Y}$\big) plugging in our stack of covariates, a $(N \cdot N_T) \times 2$ matrix $\mathbb{X}$ whose row $i,t$ is  $\mathbb{X}_{i,t} = [\Tilde{Y}_{i, t-1}  \quad \Tilde{D}_{i,t}]'$, and the outcomes given in vector $\mathbb{Y}$ where the $i,t$-th element is $\Tilde{Y}_{i, t}$. 

\begin{subequations}
\label{eq:within_hat}
\begin{align}
         \begin{bmatrix}
\hat{\rho}_1^{NB}  \\
\hat{\tau}^{NB} 
\end{bmatrix} &=  \big(    
\mathbb{X}' 
\mathbb{X} \big)^{-1}   \big(   
\mathbb{X}' 
\mathbb{Y}
 \big) \\
&=         \underbrace{ \begin{bmatrix}
\rho_{10}  \\
\tau_0 
\end{bmatrix}}_{\ref{eq:within_hat}.1} +   \underbrace{\bigg( \frac{1}{N}   
\mathbb{X}'\mathbb{X} \bigg)^{-1}}_{\ref{eq:within_hat}.2}  \underbrace{\bigg( \frac{1}{N} \sum_{i=1}^N \sum_{t=1}^T    
\begin{bmatrix}
{\Tilde{Y}_{i, t-1}}  \\
{\Tilde{D}_{i,t}} 
\end{bmatrix}   
\begin{bmatrix}
{\Tilde{\epsilon}_{i, t}} 
\end{bmatrix} \bigg) }_{\ref{eq:within_hat}.3}
\end{align}
\end{subequations}
If the errors were strictly exogenous the expectation of term \ref{eq:within_hat}.3 would be zero, and the OLS estimator would be unbiased. This would imply that $\E[{\Tilde{Y}_{i, t-1}} {\Tilde{\epsilon}_{i, t}} ] = 0$ and $\E[{\Tilde{D}_{i,t}}{\Tilde{\epsilon}_{i, t}}  ] = 0$. However, given the model in Assumption \eqref{ass:dgp_endo_treat} the expectation of term \ref{eq:within_hat}.3 is not zero. This is because $\E[{\Tilde{Y}_{i, t-1}} {\Tilde{\epsilon}_{i, t}} ] \neq 0$ and $\E[{\Tilde{D}_{i,t}} {\Tilde{\epsilon}_{i, t}}  ] \neq 0$ . As an example, let us focus on why $\E[{\Tilde{Y}_{i, t-1}} {\Tilde{\epsilon}_{i, t}} ] \neq 0$. It follows that ${\Tilde{Y}_{i, t-1}}$  and ${\Tilde{\epsilon}_{i, t}}$ are correlated by observing that they are both correlated with {$\epsilon_{i,t -1}$}. First, ${\Tilde{Y}_{i, t-1}}$ is a function of ${Y}_{i,t-1}$, which itself is a function of {$\epsilon_{i,t -1}$}. Second, ${\Tilde{\epsilon}_{i, t}} = {\epsilon}_{i,t} - \bar{\epsilon}_{i,t}$ and $\bar{\epsilon}_{i,t} = \frac{1}{T} (e_{i,0} + \cdots {\epsilon_{i,t -1}}$ $+ \cdots + \epsilon_{i,T} )$. Similar logic is used to understand why $\E[{\Tilde{D}_{i,t}} {\Tilde{\epsilon}_{i, t}}  ] \neq 0$.

\subsection{Bias Correction}
\label{sec:nickell_bias_correction}

\subsubsection{OLS Intuition}

 Though my de-biased DBC estimator uses a GMM framework, I first build an understanding for how the bias correction works using OLS. Focusing on the OLS estimates in Equation \eqref{eq:within_hat} 
, I calculate an expression for the asymptotic OLS estimator bias (term \ref{eq:within_hat}.2 + \ref{eq:within_hat}.3). The bias correction is created by calculating an estimate of this bias term, and the de-biased estimator by re-centering the GMM moment condition.   

Rewriting Equation \eqref{eq:within_hat}, I get an expression of the asymptotic bias of the OLS estimator.  

\begin{equation}
\begin{aligned}
\text{Bias} &=  \plim_{N \rightarrow \infty}  
  \begin{bmatrix}
\hat{\rho}_1^{NB}  \\
\hat{\tau}^{NB} 
\end{bmatrix} -  \begin{bmatrix}
{\rho}_{10}  \\
{\tau}_{0} 
\end{bmatrix} \\
&= \plim_{N \rightarrow \infty} \Bigg[  \underbrace{\bigg( \frac{1}{N}   
\mathbb{X}'\mathbb{X} \bigg)^{-1}}_{\ref{eq:within_hat}.2} \Bigg]  \plim_{N \rightarrow \infty}  \Bigg[  \underbrace{\bigg( \frac{1}{N} \sum_{i=1}^N \sum_{t=1}^T    
\begin{bmatrix}
{\Tilde{Y}_{i, t-1}}  \\
{\Tilde{D}_{i,t}} 
\end{bmatrix}   
\begin{bmatrix}
{\Tilde{\epsilon}_{i, t}} 
\end{bmatrix} \bigg) }_{\ref{eq:within_hat}.3} \Bigg]
 \end{aligned}
\end{equation}
The bias correction comes from solving for $[\ref{eq:within_hat}.3]$ and subtracting the estimated bias out. Therefore the bias-corrected estimate $\hat{\theta}^{DBC} = \hat{\theta}^{NB} - \hat{\text{Bias}}$.
Since I am interested in $\theta_0 = (\rho_{10}, \tau_0, \rho_{20})$, not just $\rho_{10}$ and $\tau_0$,  I need to estimate both Equations $\eqref{eq:ols_within}$ and $\eqref{eq:ols_within_treat}$. In order to estimate both equation simultaneously, I cast the problem as a GMM system. 

\subsubsection{GMM Bias Correction} The original moment conditions, which contain bias, for estimating the parameters in Equations $\eqref{eq:ols_within}$ and $\eqref{eq:ols_within_treat}$ are given in Equation \eqref{eq:biased_moments}.  I index the moment conditions by $iT$ to make it clear that they are functions of the data and that the number of time periods $T$ is fixed. 
\begin{equation}
\label{eq:biased_moments}
\mathbf{m}_{iT}(\theta) = 
   \begin{bmatrix}
            m_{\rho_1, iT}(\theta) \\
        m_{\tau, iT}(\theta) \\
        m_{\rho_2, iT}(\theta) 
    \end{bmatrix} =   \begin{bmatrix}
            \frac{1}{T} \sum_{t=1}^T \Tilde{Y}_{i,t-1}\Tilde{\epsilon}_{i,t}(\theta) \\
        \frac{1}{T} \sum_{t=1}^T \Tilde{D}_{i,t}\Tilde{\epsilon}_{i,t}(\theta) \\
        \frac{1}{T} \sum_{t=1}^T \Tilde{Y}_{i,t-1}\Tilde{u}_{i,t}(\theta)
    \end{bmatrix}  
\end{equation}


 Because of Nickell bias, the expectation of these moment equations are not zero at the true parameter $\theta_0$. For each moment in $\mathbf{m}_{iT}(\theta)$, however, I can create a new de-biased moment  by subtracting the mean of each original moment equation at $\theta_0$. By  construction, this  new moment that is mean zero at the true parameters. This term that I subtract out is labeled $\mathbf{b}_{iT}(\theta_0)$ and is written

\begin{equation}
\label{eq:bias_correction}
\mathbf{b}_{iT}(\theta_0) =  \begin{bmatrix}
            b_{\rho_1,iT}(\theta_0) \\
        b_{\tau, iT}(\theta_0) \\
        b_{\rho_2, iT}(\theta_0) 
    \end{bmatrix} =  \E \begin{bmatrix}
            \frac{1}{T} \sum_{t=1}^T \Tilde{Y}_{i,t-1}\Tilde{\epsilon}_{i,t}(\theta_0)  \\
        \frac{1}{T} \sum_{t=1}^T \Tilde{D}_{i,t}\Tilde{\epsilon}_{i,t}(\theta_0)  \\
        \frac{1}{T} \sum_{t=1}^T \Tilde{Y}_{i,t-1}\Tilde{u}_{i,t}(\theta_0)
    \end{bmatrix} .
\end{equation}

The analytic bias correction comes from solving for $\mathbf{b}_{iT}(\theta_0)$. 

\paragraph{Bias Correction Formula Intuition.} A detailed calculation $\mathbf{b}_{iT}(\theta_0)$ is given in Appendix \ref{sec:bias_corr_simple}. However, here, I give intuition for how the calculation works. As an example, consider $b_{\rho_1,iT}(\theta) = \E[\frac{1}{T} \sum_{t=1}^T \Tilde{Y}_{i,t-1}\Tilde{\epsilon}_{i,t}(\theta_0)]$. To calculate this expectation, we need to derive how $\Tilde{Y}_{i,t-1}$ relates to $\Tilde{\epsilon}_{i,t}$. To do this, it is helpful to unpack $\Tilde{Y}_{i,t-1}$ by looking just at ${Y}_{i,t-1}$ and how it depends on $\epsilon_{i,t}$ terms in all time periods. I use the outcome model in Assumption \ref{ass:dgp_endo_treat} to write $Y_{i,t-1}$ as a sum of past outcomes and errors by iteratively plugging in the definition of past outcomes and treatment.\footnote{This formula uses the fact that treatment is varying over time. If it is the case that treatment is ``absorbing''  (once a unit starts treatment they stay treated), a different formula applies, given in Appendix \ref{sec:bias_corr_formula_absorbing}.}:

\begin{equation}
\begin{aligned}
    Y_{i,t -1} &= a_i \sum_{j=0}^{t-2} (\rho_{10} + \tau_0 \rho_{20})^j + (\rho_{10} + \tau_0 \rho_{20})^{t-2} Y_{i,0} + \tau_0 \sum_{j=0}^{t-2} (\rho_1 + \tau \rho_2)^j (c_i + u_{i,t-j-1}) \\
    &+ \sum_{j=0}^{t-2} (\rho_{10} + \tau_0 \rho_{20})^j \epsilon_{i,t-j-1}
\end{aligned}
\label{eq:lag_y_spiral_sum}
\end{equation}

Given equation \eqref{eq:lag_y_spiral_sum}, it is clear how $Y_{i,t-1}$ relates to $\epsilon_{i,t}$ in all time periods. The dependence comes from the $\sum_{j=0}^{t-2} (\rho_1 + \tau \rho_2)^j \epsilon_{i,t-j-1}$ term. Using the properties of geometric sums and accounting for the fact that I am calculating the expectation of the within-transformed $\Tilde{Y}_{i,t-1}$ and $\Tilde{\epsilon}_{i,t}$, I calculate the following bias correction terms. 

\begin{lemma} (Bias Correction Terms.)
\label{theorem:bias_correction_terms}
Under Assumptions 
 \ref{ass:exogeneity}, \ref{ass:independence},  \ref{ass:dgp_endo_treat}, \ref{ass:errors} the expression for $b(\theta_0)$ is the following. 
\begin{equation}
    b_{\rho_1, iT}(\theta_0) = \frac{- \sigma_{\epsilon, i}^2}{T} \bigg( \frac{T - 1}{1 - \phi(\theta_0)} - \frac{\phi(\theta_0) - \phi(\theta_0)^T}{(1 - \phi(\theta_0))^2}\bigg), 
\end{equation}

\begin{equation}
\begin{aligned}
     b_{\tau, iT}(\theta_0) 
 &= \rho_{20} \times  b_{\rho_1}(\theta_0), 
\end{aligned}
\end{equation}

\begin{equation}
\begin{aligned}
 b_{\rho_2, iT}(\theta_0) &= \tau_0 \times \frac{- \sigma_{u,i}^2 }{T^2} \bigg( \frac{T - 1}{1 - \phi(\theta_0)} - \frac{\phi(\theta_0) - \phi(\theta_0)^T}{(1 - \phi(\theta_0))^2}\bigg). 
\end{aligned}
\end{equation}

\end{lemma}

Here $\phi(\theta)  = (\rho_{10} + \tau_{0} \rho_{20}) $. Proofs are given in Appendix \ref{sec:bias_corr_simple}. The true variance parameters $\sigma_{\epsilon,i}^2$ and $\sigma_{u,i}^2$ are not observed and have to be estimated. The estimated bias correction terms are the following.

\paragraph{Estimated Bias Correction Terms.}

       \begin{equation}
   \hat{\sigma}_{\epsilon, iT}^2(\theta) = \frac{1}{T - 1}   \sum_{t = 1}^T \Tilde{\epsilon}_{i,t}(\theta)^2  
\end{equation}

\begin{equation}
    \hat{\sigma}_{u, iT}^2(\theta) = \frac{1}{T - 1}    \sum_{t = 1}^T \Tilde{u}_{i,t}(\theta)^2  
\end{equation} 

\begin{equation}
    \hat{b}_{\rho_1, iT}(\theta) = \frac{- \hat{\sigma}_{\epsilon, iT}(\theta)^2}{T} \bigg( \frac{T - 1}{1 - \phi(\theta)} - \frac{\phi(\theta) - \phi(\theta)^T}{(1 - \phi(\theta))^2}\bigg) 
\end{equation}

\begin{equation}
\begin{aligned}
     \hat{b}_{\tau, iT}(\theta) 
 &= \rho_2 \times  \hat{b}_{\rho_1, iT}(\theta). 
\end{aligned}
\end{equation}

\begin{equation}
\begin{aligned}
 \hat{b}_{\rho_2, iT}(\theta) &= \tau \times \frac{- \hat{\sigma}_{u, iT}(\theta)^2 }{T^2} \bigg( \frac{T - 1}{1 - \phi(\theta)} - \frac{\phi(\theta) - \phi(\theta)^T}{(1 - \phi(\theta))^2}\bigg) 
\end{aligned}
\end{equation}

Here $\E[\hat{\sigma}_{\epsilon, iT}^2(\theta_0)] = {\sigma}_{\epsilon, iT}^2$ and $\E[\hat{\sigma}_{u, iT}^2(\theta_0)] = {\sigma}_{u, iT}^2$.

\subsection{Bias-Corrected GMM Estimator}

The bias-corrected moment equations are the original OLS moment equations minus the bias correction term:
\begin{equation}
\label{eq:bc_is_plim}
    \mathbf{m}^{DBC}_{iT}(\theta) := \mathbf{m}_{iT}({\theta}) - \hat{\mathbf{b}}_{iT}(\theta). 
\end{equation}
Given the definition of the model, only at $\theta_0$ do the residuals in the original moment equation equal the true noise variables in the bias correction term. Consequently, the moment conditions for the the de-biased moment equations are satisfied:

\begin{equation}
\label{eq:debiased_moments}
\E \bigg[\mathbf{m}^{DBC}_{iT}(\theta_0)\bigg] = 
   \E \begin{bmatrix}
            m^{DBC}_{\rho_1, iT}(\theta_0) \\
        m^{DBC}_{\tau, iT}(\theta_0) \\
        m^{DBC}_{\rho_2, iT}(\theta_0) 
    \end{bmatrix} =  \E  \begin{bmatrix}
            \frac{1}{T} \sum_{t=1}^T \Tilde{Y}_{i,t-1}\Tilde{\epsilon}_{i,t}(\theta_0)  - \hat{b}_{\rho_1, iT}(\theta_0) \\
        \frac{1}{T} \sum_{t=1}^T \Tilde{D}_{i,t}\Tilde{\epsilon}_{i,t}(\theta_0) - \hat{b}_{\tau, iT}(\theta_0) \\
        \frac{1}{T} \sum_{t=1}^T \Tilde{Y}_{i,t-1}\Tilde{u}_{i,t}(\theta_0) - \hat{b}_{\rho_2, iT}(\theta_0) 
       \end{bmatrix} = 0 
\end{equation}

\paragraph{Bias-Corrected Estimator.}

The DBC estimator $\hat{\theta}_{DBC}$ is obtained by solving the GMM objective function under sample moment conditions: 

\begin{equation}
\label{eq:gmm_estimator}
\hat{\theta}^{DBC} = \argmin_{\theta}
\bigg(\frac{1}{N} \sum_{i = 1}^N
\mathbf{m}^{DBC}_{iT} \bigg)' 
\bigg(\frac{1}{N} \sum_{i = 1}^N
\mathbf{m}^{DBC}_{iT} \bigg).
\end{equation}




I now derive asymptotic properties. Since the proposed estimator is based on GMM, asymptotic normality is established following standard results from the GMM framework.
\begin{assumption}
\label{eq:ass_stationarity}
(Stationarity.) For some small $\delta_s > 0$
\begin{equation}
|\rho_1 + \tau \rho_2| \leq 1 - \delta_s
\end{equation}
\end{assumption}

This assumption ensures that the process is stationary, meaning that the statistical properties (such as the mean and variance) of the outcome are stable over time.

\begin{assumption}
\label{ass:GMM}
The parameter space $\Theta$ is compact, $\theta_0$ is the unique solution to Equation \ref{eq:debiased_moments}, and $\theta_0$ satisfies $E[\nabla_{\theta} \mathbf{m}^{DBC}(\theta_0)]$ having full column rank.  Furthermore, $\theta_0$ is in the interior of $\Theta$. 
    
\end{assumption}

\begin{assumption}
    \label{ass:errors}
    The noise $\epsilon_{i,t}$ and $u_{i,t}$ are independent across $i$ and $t$ with $\E[\epsilon_{i,t}] = 0$ and $\E[u_{i,t}] = 0$, and  $\E[\epsilon_{i,t}^2] = \sigma_{\epsilon,i}^2 < C$ fand  $\E[u_{i,t}^2] = \sigma_{u,i}^2 < C$. Also

    \begin{equation}
    \label{ass:markov_slln}
\begin{aligned}
            \max_i \E\big[ \| \epsilon_{i,t}\|^{4+ \delta_{\epsilon}} \big] < \infty \quad  \forall t \text{ for some } \delta_{\epsilon} > 0 \\
                        \max_i \E\big[ \| u_{i,t}\|^{4+ \delta_u} \big] < \infty \quad  \forall t \text{ for some } \delta_u > 0. \\  
\end{aligned}
    \end{equation}

   Errors are uncorrelated: $E[\epsilon_{i,t} u_{i,t}] = 0$ for all $i$ and $t$ and $E[\epsilon_{i,t}\epsilon_{is}] = 0$ and $E[u_{i,t} u_{is}] = 0$ for $t \neq s$ and $E[\epsilon_{i,t} u_{is}] = 0$. Finally, the initial values $Y_{i0}$ and $D_{i0}$  satisfy $\E[Y_{i0}^2] < \infty$ and $\E[D_{i0}^2] < \infty$ for all $i$.   
\end{assumption}

\begin{theorem}
\label{theorem:normality}
  Under Assumptions \eqref{ass:exogeneity}  \eqref{ass:independence} \eqref{ass:dgp_endo_treat}, \eqref{eq:ass_stationarity}, \eqref{ass:GMM}, \eqref{ass:errors} the limiting distribution as $T$ remains fixed and as $N \rightarrow \infty$ of the estimator presented in Equation \eqref{eq:gmm_estimator} is given by: 

\begin{equation}
    \sqrt{N} (\hat{\theta}^{DBC} - \theta_0) \xrightarrow{d} \mathcal{N}(0, G^{-1} \Omega G'^{-1}),
\end{equation}
with
\begin{equation}
    \Omega = \plim_{N \rightarrow \infty} \big[\frac{1}{N} \sum_{i = 1}^N \mathbf{m}^{DBC}_{iT}(\theta_0)\mathbf{m}_{iT}'^{DBC}(\theta_0)\big],
\end{equation}
and
\begin{equation}
    G = \plim_{N \rightarrow \infty} [\frac{1}{N} \sum_{i = 1}^N \nabla_{\theta_0} \mathbf{m}_{iT}^{DBC}(\theta_0) ]
\end{equation}

\end{theorem}

The complete expressions for $\Omega$, $G$ and the proof are given in Appendix \eqref{appendix:gmm_normal}. 

\subsubsection{More Flexible Model}
\label{sec:more_flexiable_model}
The bias correction formula for the more following more general model is given in Appendix \ref{appen:general_model}. 

\begin{assumption}  
\label{ass:add_model}

\begin{equation}
Y_{i,t} = a_i + \sum_{c=1}^{N_c} \tau_c (D_{i,t} \cdot W_{i,t}^c) + \beta_1 X_{1,i,t} + \rho_1 Y_{i,t-1} + \epsilon_{i,t}
\end{equation}

\begin{equation}
D_{i,t} = c_i + \rho_2 Y_{i,t-1} + \beta_2 X_{2,i,t} + u_{i,t}
\end{equation}
\end{assumption}

Allowing the past outcomes $Y_{i,t-1}$ through $Y_{i, t-h}$ to impact current potential outcomes follow from \cite{breitung2022bias} and general VAR structure follows from \cite{juodis2015iterative}. In an in-progress extension to this paper, I have a machine learning estimator that still imposes additive fixed effects and errors, but allows for more flexible modeling of the regressors, such as the interaction terms in the model above. A discussion of this extension is given in Appendix \ref{appendix:ML}.  Note that a general nonparmetric form with a non-separable fixed effect is not possible, especially with fixed $T$.\footnote{The current methods that control for time varying unobserved heterogeneity rely on large $T$ asymptotics so they can estimate latent factor structures, which is not the setting of this paper \citep{moon2017dynamic}. }

\section{Simulation Study}
\label{sec:simulation_study}

 In this section, I present a simulation study to compare the dynamic and Nickell bias along several dimensions. 
 I provide additional details for the simulation that I introduced in Section \ref{section:applied_motivation} as well as additional results comparing my DBC estimator to other Nickell bias correction procedures. 

\subsection{Dynamic Bias Versus Nickell Bias}
\label{sec:simulation_dynamic_vs_Nickell}

\subsubsection{Simulation Design with Random Treatment}

I start with a Monte Carlo simulation using the structural model presented in Assumption \ref{ass:simple_random_treatment}, which I reproduce here:


\begin{equation}
\label{eq:dell_outcome}
    Y_{i,t} =  a_{i} +  \tau_0 D_{i,t} + \rho_{10} Y_{i, t - 1} +  \epsilon_{i,t}.
\end{equation}
\begin{equation}\label{eq:treatment-model}
    D_{i,t} =  a_i  + u_{i,t}.\footnote{Here I keep the fixed effect term the same across the treatment and outcome model for simplicity, but exactly matching  Assumption \ref{ass:simple_random_treatment} by modify treatment to $ D_{i,t} =  c_i  + u_{i,t}$ would produce similar results.}
\end{equation}
In Equation \eqref{eq:dell_outcome}, the outcome is a function of the treatment and the past outcome. For the treatment model (Equation \ref{eq:treatment-model}), I make treatment a function of the fixed effect, in line with the common practice in applied research of adding fixed effects to account for correlation between treatment and a time-invariant individual-level term.\footnote{It is important to note dynamic bias remains even if treatment is not a function of fixed effects and $D_{i,t} = u_{i,t}$.} This model is the simplest setting, where $D_{i,t}$ is not impacted by past $Y_{i,t-1}$. The errors $u_{i,t}$ and $\epsilon_{i,t}$ are i.i.d random noise.

The individual fixed effects are drawn from a normal distribution $a_i \sim N(0,5).$  The treatment is set to $\tau_0 = .5$. I vary the values of $\rho_{10}$. \footnote{Though I show simulations for this setting, Nickell bias is smaller than dynamic bias in all DGPs I have studied. Additional simulation results for a wide variety of parameters are given in Appendix\ref{append:additional_simulations_dynamic_vs_Nickell}}


\paragraph{Simulation estimators:}

I compare the OLS estimates of three different models. 

\begin{enumerate}
    \item Within-Transformed Static Model:
    From Equation \eqref{eq:dell_simple_within}. This model does not include past outcome as a control, and therefore has no lag. 
\begin{equation}
    \Tilde{Y}_{i,t} = \tau^D \Tilde{D}_{i,t} + \Tilde{e}_{i,t},
\end{equation}
    
    \item  Within-Transformed Dynamic Model:
    From Equation \eqref{eq:ols_within}. This model does include past outcome as a control, and therefore has a lag. 

\begin{equation}
    \Tilde{Y}_{i,t} =  {\tau}_0 \Tilde{D}_{i,t} + {\rho}_{10} \Tilde{Y}_{i,t-1} +  \Tilde{\epsilon}_{i,t},
\end{equation}

    \item  Within-Transformed Delta Model:

\begin{equation}
    \Delta \Tilde{Y}_{i,t} =  {\tau}_0 \Tilde{D}_{i,t} + {\rho}_{10} \Tilde{Y}_{i,t-1} +  \Tilde{\eta}_{i,t},
\end{equation}

\end{enumerate}

I repeat the same data generating process for a range of different panel lengths. For each number of time periods $N_T:$ 3 - 50, I generate datasets and plot the OLS estimate of the Static, Dynamic, and Delta Models. 

\begin{figure}[h!]
    \centering
    \includegraphics[scale=0.5]{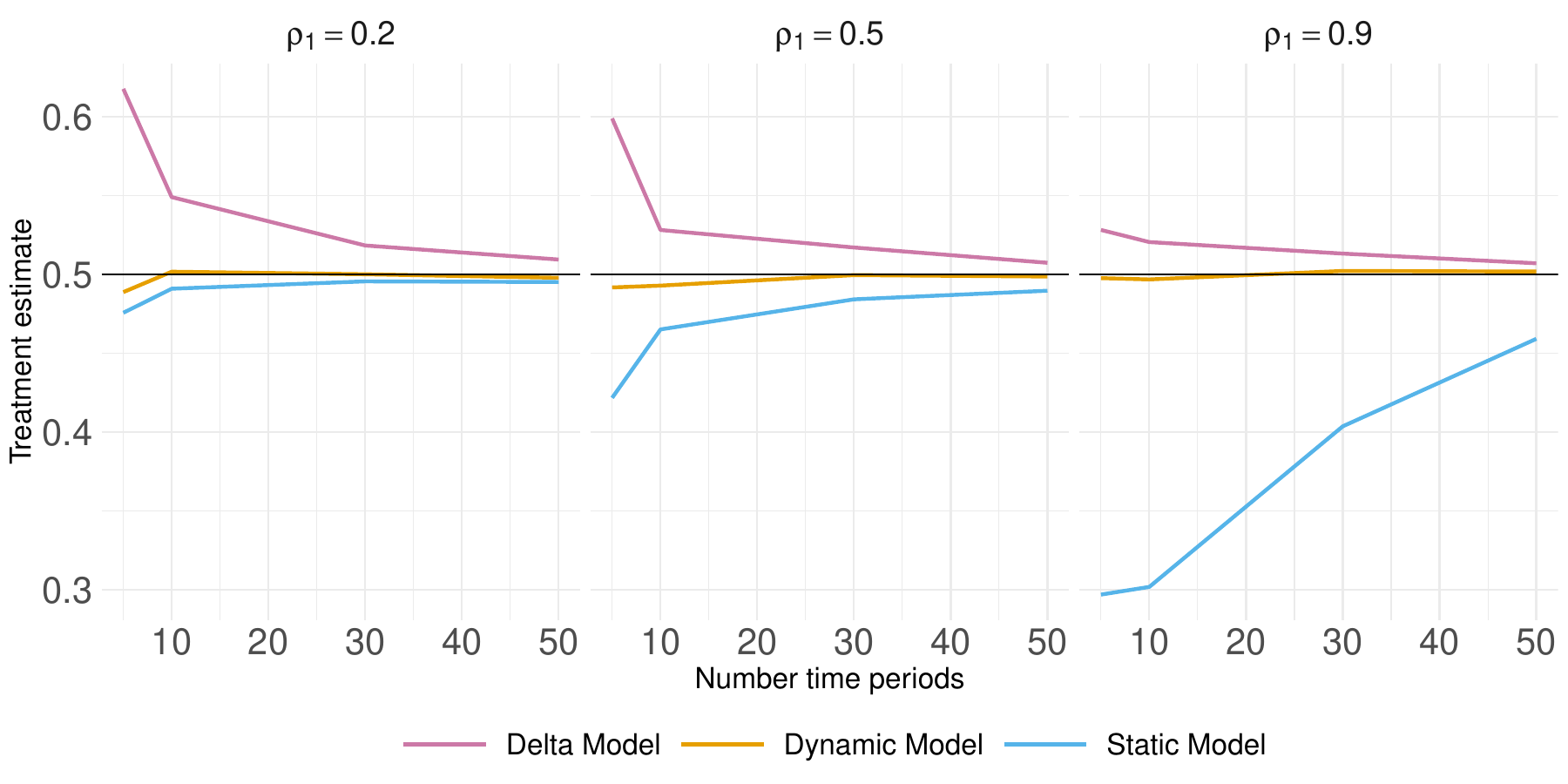}
    \caption{Bias of three different models.}
    \label{fig:growth_rate_sim}
\end{figure}

Figure \ref{fig:growth_rate_sim} (replicated from Section \ref{section:applied_motivation}) plots the results.
On the x-axis we have the number of time periods in the generated panel data set, and then the y-axis marks the treatment estimate. The true treatment effect value is marked with a black horizontal line.

The plots show that even when treatment is random, dynamic bias and transformation bias are larger than Nickell bias. The Nickell bias is small because treatment effect coefficient is biased only because the coefficient on the past outcome is estimated with bias, which then leads to bias in the other parameters in the model.

\subsubsection{Simulation Design with Endogenous Treatment}

Now let us consider the case when treatment is not random. I modify the DGP to follow the structural model outlined in Assumption \ref{ass:dgp_endo_treat}: 

\begin{equation}
\label{eq:sim_endo_outcome}
    Y_{i,t} =  a_{i} + \rho_{01} Y_{i, t - 1} +  \tau_0 D_{i,t} +  \epsilon_{i,t},
\end{equation}
\begin{equation}
\label{eq:sim_endo_treat}
    D_{i,t} = a_i +  \rho_{02} Y_{i, t - 1}  + u_{i,t}.
\end{equation}

I set $\rho_{20} = .1$, and generate results  analogously to those in the previous section, but now using the endogenous treatment simulation design. 

\begin{figure}[h!]
    \centering
    \includegraphics[scale=0.5]{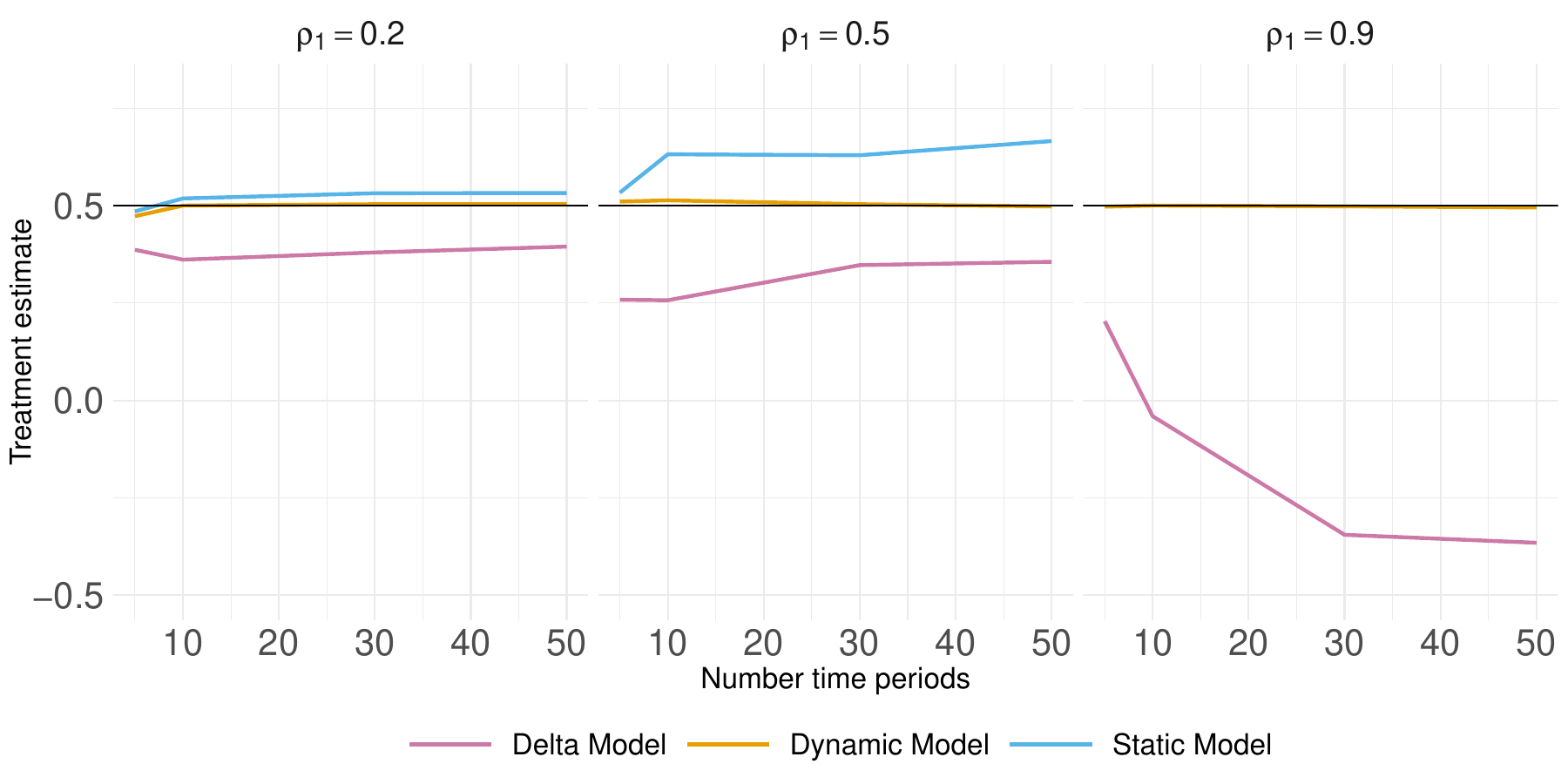}
    \caption{Bias of three different models.}
    \label{fig:growth_rate_into_endo_sim}
\end{figure}

Figure \ref{fig:growth_rate_into_endo_sim} present the results. The bias gets larger the larger the absolute value of $\rho_{20}$. Recall that when treatment is endogenous, the bias for the Static and Delta model does not disappear as the number of time periods increases omitted variable bias arises when past outcomes are not explicitly included in the model. Note in Figure \ref{fig:growth_rate_into_endo_sim}, for the case of $\rho_{10} = .9$, the bias under the Static Model is so large it is omitted from the plot.

\subsection{Bias Correction Simulation}
\label{sec:bias_corr_sim}

In these simulations I showcase the performance of my  bias-corrected estimator.  I also show how my estimator performs in comparison to the Arellano Bond estimator \citep{arellano1991some}.

I create a simulation data generating process with an endogenous treatment following Equations \eqref{eq:sim_endo_outcome} and \eqref{eq:sim_endo_treat}. Again, the individual fixed effects are drawn from a standard normal distribution $a_i \sim N(0,5).$ The initial values $D_{i0} \sim N(a_i, 1) $. The vector of the true parameters is given by $\theta_0 = (\rho_{10}, \tau_0, \rho_{20}) = (.2,.5,.3)$.

\paragraph{Simulation Estimators:}

\begin{enumerate}
        \item Within-Transformed Dynamic Model from Equation \eqref{eq:ols_within}. This model does include past outcome as a control, and therefore has a lag. 

\begin{equation}
    \Tilde{Y}_{i,t} =  {\tau}_0 \Tilde{D}_{i,t} + {\rho}_{10} \Tilde{Y}_{i,t-1} +  \Tilde{\epsilon}_{i,t},
\end{equation}    
    \item My bias-corrected GMM estimator, presented in the previous section in Equation \eqref{eq:gmm_estimator}

    \item Arellano Bond estimator. Implemented using the R package \texttt{plm}\footnote{Identifying equations $E(\Delta Y_{i,t} - \Delta X_{i,t} \beta_0) X_{i}^{t-2} = 0, t = 2, \dots, T.$. Our instruments are $Y_{i,t-2}, Y_{i,t-3}, Y_{i,t-4}, Y_{i,t-5}$, $D_{i,t-2}, D_{i,t-3}, D_{i,t-4}, D_{i,t-5}$ }. 
\end{enumerate}

I run 1000 Monte Carlo simulations with the number of units $N = 1000$ and the number of time periods $N_T = 5$.

\begin{table}[ht]
\centering
\begin{tabular}{rrrr}
  \hline
 & DBC $\hat{\tau}$  & OLS $\hat{\tau}$  & AB $\hat{\tau}$  \\ 
  \hline
Mean &  0.501 & 0.469 & 0.487 \\ 
  SD & 0.015 &  0.014 & 0.311  \\ 
  95\% cov &  0.950 & 0.440 & 0.960 \\ 
   \hline
\end{tabular}
\caption{Average estimates of the treatment effects $\tau_0$.}
\label{table:bias_corr_table_tau}
\end{table}

\begin{table}[ht]
\centering
\begin{tabular}{rrrr}
  \hline
 & DBC $\hat{\rho}_1$  & OLS $\hat{\rho}_1$  & AB $\hat{\rho}_1$  \\ 
  \hline
Mean &  0.198 & -0.026 & 0.187 \\ 
  SD & 0.020 &  0.014 & 0.099  \\ 
  95\% cov &  0.960 & 0.000 & 0.970 \\ 
   \hline
\end{tabular}
\caption{Average estimates of parameter on past outcome in outome equation $\rho_{10}$.}
\label{table:bias_corr_table_rho1}
\end{table}


Table \ref{table:bias_corr_table_tau} compares three different estimation strategies for treatment effects. The table compares my bias-corrected (DBC) estimator and ordinary least squares (OLS) estimates and Arellano Bond (AB) estimates. The DBC estimates are closest to the true treatment parameter $\tau_0 = .5$. The table presents the mean and standard deviation (sd) for each estimate, showing that DBC estimates yield proper 95\% coverage, while OLS does not. The Arellano Bond (AB) estimator uses several lagged variables as instruments. The data was generated with random i.i.d errors, and so by construction the instruments are valid. The results demonstrate that AB models also achieve proper 95\% coverage. I find that  proper coverage is achieved regardless of the instruments used. However, the instrument choice does have a substantial effect on the standard deviations of the estimates. Using fewer instruments leads to an even larger standard deviation for the AB estimates. The large standard deviations explain why in practice for a particular dataset (as opposed to our Monte Carlo setting here where we average over 1000 datasets) the choice of instruments can lead to very different treatment effect estimates. In the table \ref{table:bias_corr_table_tau}  present the AB results in which I selected the instruments that led to the smallest standard deviation for treatment ($\tau$), using all possible instruments. Yet, even in this case, AB estimation still leads to standard deviations that are larger in comparison to DBC.

Table \ref{table:bias_corr_table_rho1} compares the same three different estimation strategies for $\rho_{10} = .2$. While OLS led to biased estimates of treatment effects, the bias of the $\rho_{10}$ is much larger and even changes sign. Both my DBC method and AB are able to achieve proper coverage of the true $\rho_{10}$, but the standard deviation of method is again substantially smaller for my estimator.  

\section{Empirical Example} 
\label{sec:emperical_examples}

This paper is highly relevant for any applied research where past outcomes influence current outcomes. As discussed in the introduction, this includes studies focusing on variables such as agricultural yields, human capital, labor market outcomes, and migration flows.

To illustrate the application of my method, I focus on the relationship between temperature (as the treatment) and GDP (as the outcome). This relationship has been extensively explored in prior literature, and for this analysis, I utilize data from the seminal study by \cite{dell2012temperature}.

\subsection{GDP and Temperature}

\cite{dell2012temperature} explores the relationship between temperature ($D_{i,t}$) and GDP ($Y_{i,t}$), and examines how rising temperatures can impact economic growth. The authors' results suggest that warmer temperatures can lead to reduced agricultural yields, decreased labor productivity, and increased health risk, all of which can hinder economic outcomes. The main result of their paper focuses on how higher temperatures affect poorer countries.

I use this empirical example to highlight two main points. First, I use the example to show how dynamic bias, transformation bias, and Nickell bias compare in a real world setting. This is discussed in detail in Section \ref{sec:emperical_comparing_bias}. Second, I highlight how my bias correction performs in practice and compare it to Arellano Bond. This is discussed in detail in Section \ref{sec:emperical_bias_correction}.

\subsubsection{Comparing Bias}
\label{sec:emperical_comparing_bias}

In Table \ref{table:all_reg}, I present estimates from regressions of GDP growth and levels on temperature using data from \cite{dell2012temperature}.  The main results of \cite{dell2012temperature} focus on the impact of temperature on the GDP of poor countries, so my sample is restricted to poor countries. I do this for simplicity so that this analysis can focus on one treatment effect estimate. I exactly replicate the original analysis that uses all countries \cite{dell2012temperature} in Appendix \ref{append:additional_empirics}. The replication there is consistent with the results presented here. 

In Table \ref{table:all_reg}, I present the subset of results focused on poor countries. In the first two columns, I run baseline regressions where I regress GDP growth on temperature in the first column and then the second column shows how these estimates change once lagged GDP growth is included. In the third column and fourth columns I include all the original controls used in \cite{dell2012temperature}, which are 371 region and time controls . The third column of my Table \ref{table:all_reg} follows the same specification given in the second column of Table 2 in \cite{dell2012temperature} - the outcome is GDP growth, and the past outcome is not controlled for.  The treatment effect estimate without controlling for the lagged outcome is -1.421. Then in Column 4 I include lagged GDP growth, and the treatment effect estimate changes to -1.279. This is a 10\% difference in treatment effects that is significant at the $p < .1$ level.\footnote{To test whether the treatment estimate in Column 3 $(\hat{\tau}_3)$ is statistically significantly smaller than the treatment estimate in Column 4, $(\hat{\tau}_4)$  I conduct Wald test and the resulting p-value is .06. The Wald test requires accounting for the correlation between $\hat{\tau}_3$ and $\hat{\tau}_4$. These two estimates are highly correlated, which is why even though the estimates have a large variance, the null hypothesis is still rejected at the p < .1 level. }


\begin{table}[!htbp] \centering 
  \caption{} 
  \label{table:all_reg} 
\begin{tabular}{@{\extracolsep{5pt}}lcccccc} 
\\[-1.8ex]\hline 
\hline \\[-1.8ex] 
 &  \multicolumn{4}{c}{\textit{Outcome: GDP Growth }}  & \multicolumn{2}{c}{\textit{Outcome: GDP Level}} \\ 
\cline{2-7} 
\\[-1.8ex] & (1) & (2) & (3) & (4) & (5) & (6) \\ 
\hline \\[-1.8ex] 
Temperature & $-$1.139$^{***}$ & $-$1.052$^{***}$ & $-$1.421$^{***}$ & $-$1.279$^{***}$ & 174.202$^{***}$ & $-$19.330$^{**}$ \\ 
  & (0.244) & (0.247) & (0.397) & (0.401) & (50.315) & (8.018) \\ 
  & & & & & & \\ 
Outcome Lag &  & 0.136 &  & 0.109$^{***}$ &  & 1.025$^{***}$ \\ 
  &  & (0.083) &  & (0.021) &  & (0.003) \\ 
  & & & & & & \\ 
Controls & No & No & Yes & Yes & Yes & Yes \\ 
  & & & & & & \\ 
\hline \\[-1.8ex] 
Observations & 2,452 & 2,389 & 2,452 & 2,389 & 2,754 & 2,691 \\ 
R$^{2}$ & 0.106 & 0.126 & 0.205 & 0.218 & 0.788 & 0.995 \\ 
Adjusted R$^{2}$ & 0.082 & 0.102 & 0.114 & 0.127 & 0.761 & 0.994 \\ 
\hline 
\hline \\[-1.8ex] 
\textit{}  & \multicolumn{6}{r}{$^{*}$p$<$0.1; $^{**}$p$<$0.05; $^{***}$p$<$0.01} \\ 
\end{tabular} 
\end{table}



These changes in coefficient occur despite the fact that I include the large number of time trend controls used in the original analysis. This highlights the point discussed in the introduction: time trend controls do not control for dynamics. To see this analytically, I run additional specifications to see how  treatment effect estimates are impacted by the controls. In Table \ref{table:all_reg} Column 1 and 2 I run the same specifications in Column 3 and 4 respectively, just without controls.  Adding controls increases the estimated coefficient, whereas controlling for lags decreases the estimate -- demonstrating that controls do not correct for dynamic relationships in outcomes. Adding the lagged outcome changes the treatment effect estimate by the same percentage regardless of whether or not controls are included.

It is important to note that GDP growth is calculated by transforming GDP levels. Looking at GDP growth rather then GDP levels may somewhat reduce dynamic bias, but it introduces transformation bias, as discussed in Section \ref{section:applied_motivation} and Appendix \ref{append:growth_vs_level}. The magnitudes of both biases depend on $\rho_{10}$ and the number of time periods. In this particular setting, the true $\rho_{10}$ is very close to 1 and the number of time periods is 30, so the transformation bias is not as large as the dynamic bias. 

To understand the impact of temperature on the original untransformed variable, I use GDP levels as the outcome in Table \ref{table:all_reg} Column 5 and Column 6. This is important because as discussed in \cite{nath2024much}, the economic literature is divided over whether temperature affects GDP levels or growth rates. Therefore, it is also important to study GDP levels. Levels of outcomes are often studied in economics papers. \footnote{Many economics papers study the levels of outcome variables without controlling for past outcomes. \cite{somanathan2021impact} study the impact of temperature on economic productivity without controlling for levels, \cite{annan2015federal} study the temperature effects of crop yield levels.} The impact of including past outcomes in the regression models is even greater when looking at GDP levels. Table \ref{table:all_reg} Column 5 shows that without controlling for past outcome the treatment estimate is unexpectedly positive (174.202), and only becomes negative (-19.330) as expected when controlling for past GDP level in Column 6.

\subsubsection{Comparing Bias Correction to Arellano Bond}
\label{sec:emperical_bias_correction}

In this section I compare the performance of my bias correction method to Arellano Bond using real data from \cite{dell2012temperature}. In the case of \cite{dell2012temperature}, the number of time periods is 30, and so the expectation is that Nickell bias is relatively small. 

I replicate the analysis of Table \ref{table:all_reg} Column 2 and present the resulting estimates of $\tau_0$ and $\rho_{10}$ are presented in the ``OLS with Lag'' columns in Table \ref{table:dem_tau_bc} and Table \ref{table:dem_rho_bc}, respectively. I do the analysis on a balanced panel; therefore, the OLS estimates of $\tau_0$ and $\rho_{10}$ are slightly different from those in Table \ref{table:all_reg} where I used the unbalanced panel of \cite{dell2012temperature}.\footnote{The extension to unbalanced panels can be implemented but is left for future work.} 

I then implement my bias correction method for both parameters; the results are given in the bias-corrected columns labeled ``DBC''. The bias-corrected method does not change the estimate of treatment $\tau_0$ effect much, but does lead to higher estimate of ${\rho}_{10}$. This is expected as Nickell bias is small in the setting when the number of time periods is 30, and also Nickell bias impacts the estimation of ${\rho}_{10}$ more than the estimation of $\tau_0$. 

The tables also include the Arellano Bond estimates. \cite{arellano1991some} is one of the most cited papers in economics as it was one of the first papers to correct for Nickell bias. This approach requires picking which past outcomes are used as instruments, but point estimates of both $\tau_0$ and $\rho_{10}$ are quite sensitive to this choice. I run three different Arellano Bond specifications, changing which lags used as instruments - these results are reported in columns ``AB 1'', ``AB 2'' and ``AB 3''.\footnote{For specification AB 1, I use the lag 2-5 for both variables, for AB 2 I used lags from 2-15, for AB 3 I use lags 10-15.} Looking first at the treatment point estimates in Table \ref{table:dem_tau_bc}, depending on the choice of instrument the point estimates go from negative (-.460) to positive and become an order of magnitude smaller (.060). A similar pattern occurs with the point estimates of $\hat{\rho}_{10}$ given in Table \ref{table:dem_rho_bc}. Depending on the choice of instrument, the point estimates go from negative (-.016) to positive and change by an order of magnitude larger (.101). Avoiding the instability associated with the choice of instrument is a benefit of my analytical approach. 

The standard errors for all methods were calculated with the panel bootstrap clustered at the country level. The standard errors of the bias-corrected method are smaller than the standard errors of Arellano Bond  method, and are close to the original OLS standard errors, for both $\hat{\tau}$ and $\hat{\rho}_{1}$. Depending on the instruments, the standard errors of AB vary, and can be twice as large as the OLS and DBC standard errors.

\begin{table}[ht]
\centering
\begin{tabular}{rrrrrr}
  \hline
 & DBC $\hat{\tau}$  & OLS with Lag $\hat{\tau}$  & AB 1 $\hat{\tau}$ & AB 2 $\hat{\tau}$ & AB 3 $\hat{\tau}$\\ 
  \hline
Mean &  -1.145 & -1.143  & 0.060 & -0.283  & -0.460\\ 
  SE & 0.164 &  0.166 & 0.345 & 0.332 & 0.415 \\ 
   \hline
\end{tabular}
\caption{Estimates of $\tau_0$.}
\label{table:dem_tau_bc}
\end{table}

\begin{table}[ht]
\centering
\begin{tabular}{rrrrrr}
  \hline
 & DBC $\hat{\rho}_1$  & OLS with Lag $\hat{\rho}_1$  & AB 1 $\hat{\rho}_1$ & AB 2 $\hat{\rho}_1$ & AB 3 $\hat{\rho}_1$\\ 
  \hline
Mean &   0.200 & 0.163  & 0.094 &  0.101  & -0.016\\ 
  SE & 0.094 &  0.090 & 0.100 & 0.108 & 0.141 \\ 
   \hline
\end{tabular}
\caption{Estimates of $\rho_{10}$.}
\label{table:dem_rho_bc}
\end{table}

\section{Conclusion}
\label{sec:conclusion}

This paper identifies a source of bias in fixed effects panel models, which I term dynamic bias, which occurs when past outcomes are left out of the model. Economic research often uses static models even in settings in which economics theory tells us that past outcomes directly influence current outcomes\citep{griliches1963sources},\citep{cunha2007technology}, \citep{blanchard1988beyond},\citep{massey1993theories}. Empirical papers studying these outcomes often run fixed effects analysis without controlling for past outcomes.\footnote{Examples include  \cite{annan2015federal}, \cite{burke2015global}, \cite{cho2017effects}, \cite{jessoe2018climate}, \cite{drabo2015natural}, \cite{mahajan2020taken}, \cite{missirian2017asylum}, \cite{graff2018temperature} \cite{garg2020temperature}.}

Through simulations I show that dynamic bias is larger than the well-known Nickell bias. While Nickell bias comes from including past outcomes in the model, dynamic bias occurs when past outcomes are excluded. Through simulations and real-world data, I demonstrate that ignoring past outcomes can lead to significantly biased treatment effect estimates, even when treatments, such as temperature, are randomly assigned. This bias is especially problematic in contexts like environmental economics, where past environmental conditions affect current environmental outcomes.

To solve this challenge, I developed a new estimator (the DBC estimator) that corrects both dynamic bias and Nickell bias. The estimator works well even when the number of time periods is small (fixed-T panels). It provides more accurate estimates of treatment effects by properly accounting for the influence of past outcomes. The DBC estimator performs better than alternative methods, such as instrumental variable techniques, which can suffer from weak instruments and larger standard errors.

I applied the DBC estimator to study the impact of temperature shocks on GDP, a key question in environmental economics. The results show that properly accounting for dynamic biases significantly changes the estimated effects. For instance, correcting for dynamic biases reduces the estimated impact of temperature shocks on GDP growth by 10\% and on GDP levels by 120\%.

This research offers a practical solution for applied researchers dealing with dynamic settings. It highlights the importance of including past outcomes to avoid large biases in treatment effect estimates. Future work will focus on extending this estimator to more complex models by incorporating machine learning and adapting it for spatial data, where additional geographic dynamics play an important role.

\newpage

\bibliographystyle{plainnat} 
\bibliography{main.bib}

\appendix

\section{Dynamic Bias}
\label{appendix:dynamic_bias}

\subsection{Proof Theorem \ref{theorem:omit_lag_y}}
\label{appendix:dynamic_bias_proof}

Under Assumptions \eqref{ass:simple_random_treatment}, \eqref{ass:exogeneity}, \eqref{ass:independence} and that $Var(D_{i,t}) > 0$ running OLS to estimate the static model in Equation \eqref{eq:dell_simple_within} leads to a biased treatment effect estimator $\hat{\tau}^D$.  Given that Equation \eqref{eq:dell_simple_within} is a univariate panel regression, I have a simple formula for the OLS solution \citep{wooldridge2010econometric}.

\begin{equation}
    \begin{aligned}
        \hat{\tau}^D &= \frac{\sum_{i = 1}^N \sum_{t = 1}^T Cov(\Tilde{D}_{i,t}, {Y}_{i,t} )}{\sum_{i = 1}^N  \sum_{t = 1}^T  Var(\Tilde{D}_{i,t})}\\
        &= \frac{\sum_{i = 1}^N \sum_{t = 1}^T  Cov(\Tilde{D}_{i,t}, \rho_{10} {Y}_{i,t-1} + \tau_0 {D}_{i,t} + {\epsilon}_{i,t} )}{\sum_{i = 1}^N \sum_{t = 1}^T 
 Var(\Tilde{D}_{i,t})} \\
    \end{aligned}
\end{equation}

\subsubsection{Numerator}

\begin{equation}
    Cov(\Tilde{D}_{i,t}, \rho_{10} {Y}_{i,t-1} + \tau_0 {D}_{i,t} + {\epsilon}_{i,t} ) = \underbrace{Cov(\Tilde{D}_{i,t}, \rho_{10} {Y}_{i,t-1})}_{N.1} + \underbrace{Cov(\Tilde{D}_{i,t},  \tau_0 {D}_{i,t}  )}_{N.2} + \underbrace{Cov(\Tilde{D}_{i,t}, {\epsilon}_{i,t} )}_{N.3}
\end{equation}

\begin{equation}
   \begin{aligned}
     N.1 =   Cov(\Tilde{D}_{i,t}, \rho_{10} {Y}_{i,t-1}) &= \rho_{10} Cov({D}_{i,t} - \frac{1}{T}\sum_{s=1}^T {D}_{i,s},   {Y}_{i,t-1}  ) \\
        &= - \frac{\rho_{10}}{T} Cov( \sum_{s=1}^{t-1} {D}_{i,s},   {Y}_{i,t-1})\\
        &= - \frac{\rho_{10}}{T} Cov( \sum_{s=1}^{t-1} {D}_{i,s},    \tau_0 \sum_{j=0}^{t-2} (\rho_{10})^{j} (D_{i,t-j-1})) \\
     &= - \frac{\rho_{10}\tau_0}{T} \sum_{s=1}^{t-1}  Cov( {D}_{i,s},     \sum_{j=0}^{t-2} (\rho_{10})^{j } (D_{i,t-j-1})) \\
      &= - \frac{\rho_{10}\tau_0}{T} \sum_{s=1}^{t-1}  Cov( {D}_{i,s},     (\rho_{10})^{t-s-1 } (D_{i,s})) \\
      &= - \frac{\rho_{10}\tau_0 Var(D_{i,t})}{T} \sum_{s=1}^{t-1}  (\rho_{10})^{t-s-1 }   
   \end{aligned}
\end{equation}

This follows by plugging in for $Y_{i,t-1}$, which follows from \ref{eq:y_function_past}. 
Therefore:  

\begin{equation}
    \begin{aligned}
       \sum_{t=1}^T Cov(\Tilde{D}_{i,t}, \rho_{10} {Y}_{i,t-1})  = - \frac{\rho_{10}\tau_0 Var(D_{i,t})}{T}\bigg[ \frac{T}{1- \rho_{10}} - \frac{1 - \rho_{10}^T}{(1 - \rho_{10}^2)}\bigg]
    \end{aligned}
\end{equation}

The next term in the numerator is:

\begin{equation}
    \begin{aligned}
    N.2 =    Cov(\Tilde{D}_{i,t},  \tau_0 {D}_{i,t}  ) &= \tau_0 Cov({D}_{i,t} - \frac{1}{T}\sum_{t=1}^T {D}_{i,t},   {D}_{i,t}  ) \\
        &= \tau_0 Cov({D}_{i,t}, {D}_{i,t}  ) - \tau_0 Cov( \frac{1}{T}\sum_{t=1}^T {D}_{i,t}, {D}_{i,t}  )  \\
        &= \tau_0 \bigg[1 - \frac{1}{T}\bigg] Var(D_{i,t})
        \end{aligned}
\end{equation}

The final term in the numerator is: 
\begin{equation}
 N.3 =   Cov(\Tilde{D}_{i,t}, {\epsilon}_{i,t} ) = 0 
\end{equation}

Putting these parts together, and including the outside sums, our full numerator is.  

\begin{equation}
    \begin{aligned}
  \sum_{i = 1}^N \sum_{t = 1}^T Cov(\Tilde{D}_{i,t}, {Y}_{i,t} )  &=  \sum_{i = 1}^N \bigg( T \tau_0 \bigg[1 - \frac{1}{T}\bigg] Var(D_{i,t})    - \frac{\rho_{10}\tau_0 Var(D_{i,t})}{T}\bigg[ \frac{T}{1- \rho_{10}} - \frac{1 - \rho_{10}^T}{(1 - \rho_{10}^2)}\bigg]  \bigg)  \\
  &=  \sum_{i = 1}^N \bigg( T \tau_0 \bigg[1 - \frac{1}{T}\bigg] \sigma_i^2    - \frac{\rho_{10}\tau_0 \sigma_i^2}{T}\bigg[ \frac{T}{1- \rho_{10}} - \frac{1 - \rho_{10}^T}{(1 - \rho_{10}^2)}\bigg]  \bigg)  \\
    \end{aligned}
\end{equation}

\subsubsection{Denominator}
\begin{equation}
    \text{Var}\left(D_{it} - \bar{D}_i\right) = \text{Var}(D_{it}) + \text{Var}(\bar{D}_i) - 2 \cdot \text{Cov}(D_{it}, \bar{D}_i)
\end{equation}
where:

\begin{equation}
    \begin{aligned}
        \text{Var}(D_{it}) &= \sigma^2_i \\
        \text{Var}(\bar{D}_i) &= \frac{\sigma^2_i}{T} \\
        \text{Cov}(D_{it}, \bar{D}_i) &= \frac{\sigma^2_i}{T}
    \end{aligned}
\end{equation}

Substituting these into the variance formula:

\begin{equation}
    \text{Var}\left(D_{it} - \bar{D}_i\right) = \sigma^2_i + \frac{\sigma^2_i}{T} - 2 \cdot \frac{\sigma^2_i}{T} = \sigma^2_i \left(1 - \frac{1}{T}\right)
\end{equation}

Hence the denominator is 

\begin{equation}
    \sum_{i = 1}^N \sum_{t = 1}^T 
 Var(\Tilde{D}_{i,t}) =  \sum_{i = 1}^N \sum_{t = 1}^T \sigma^2_i \left(1 - \frac{1}{T}\right) 
\end{equation}

\subsubsection{Combining Numerator and Denominator}

\begin{equation}
    \begin{aligned}
\hat{\tau}^D &=  \frac{  \sum_{i = 1}^N \bigg( T \tau_0 \big[1 - \frac{1}{T}\big] \sigma^2_i    - \frac{\rho_{10}\tau_0 \sigma^2_i}{T}\bigg[ \frac{T}{1- \rho_{10}} - \frac{1 - \rho_{10}^T}{(1 - \rho_{10}^2)}\bigg]  \bigg)}{\sum_{i = 1}^N \sum_{t = 1}^T \sigma^2_i \left(1 - \frac{1}{T}\right)}  \\
&= \tau_0  - \frac{  \sum_{i = 1}^N   \bigg( \frac{\rho_{10}\tau_0 \sigma^2_i}{T}\bigg[ \frac{T}{1- \rho_{10}} - \frac{1 - \rho_{10}^T}{(1 - \rho_{10}^2)}\bigg]  \bigg)}{\sum_{i = 1}^N \sum_{t = 1}^T \sigma^2_i \left(1 - \frac{1}{T}\right)}  \\
&= \tau_0  - \frac{ \frac{\rho_{10}\tau_0 }{T}\bigg[ \frac{T}{1- \rho_{10}} - \frac{1 - \rho_{10}^T}{(1 - \rho_{10}^2)}\bigg]  }{\sum_{t = 1}^T  \left(1 - \frac{1}{T}\right)}  \\
&= \tau_0  - \frac{ \frac{\rho_{10}\tau_0 }{T}\bigg[ \frac{T}{1- \rho_{10}} - \frac{1 - \rho_{10}^T}{(1 - \rho_{10}^2)}\bigg]  }{(T - 1)}  \\
&= \tau_0  - \frac{\rho_{10}\tau_0 }{T(T-1)}\bigg[ \frac{T}{1- \rho_{10}} - \frac{1 - \rho_{10}^T}{(1 - \rho_{10}^2)}\bigg]  \\
    \end{aligned}
\end{equation}

\section{Bias Correction Formulas}
\label{sec:bias_corr_formula}

\subsection{Proof for Lemma \ref{theorem:bias_correction_terms}}
\label{sec:bias_corr_simple}

Under Assumptions \ref{ass:exogeneity}, \ref{ass:independence}, \ref{ass:dgp_endo_treat} the following bias correction term is calculated.

\begin{equation}
\mathbf{b}(\theta_0) =  \begin{bmatrix}
       \begin{pmatrix}
            b_{\rho_1}(\theta_0) \\
        b_{\tau}(\theta_0) \\
        b_{\rho_2}(\theta_0) 
       \end{pmatrix}
    \end{bmatrix} =  \E \begin{bmatrix}
       \begin{pmatrix}
            \frac{1}{T} \sum_{t=1}^T \Tilde{Y}_{i,t-1}\Tilde{\epsilon}_{i,t}(\theta_0)  \\
        \frac{1}{T} \sum_{t=1}^T \Tilde{D}_{i,t}\Tilde{\epsilon}_{i,t}(\theta_0)  \\
        \frac{1}{T} \sum_{t=1}^T \Tilde{Y}_{i,t-1}\Tilde{u}_{i,t}(\theta_0)
       \end{pmatrix}
    \end{bmatrix} 
\end{equation}

There are three expectations that have to be solved in order to calculate $\mathbf{b}(\theta_0)$. I have to solve for 1) $b_{\rho_1}(\theta_0)$  2) $b_{\tau}(\theta_0)$  3) $b_{\rho_2}(\theta_0)$.   
I first solve for $b_{\rho_1}(\theta_0) = \E_{\theta_0} [ \frac{1}{T} \sum_{t=1}^T \Tilde{Y}_{i,t-1}\Tilde{\epsilon}_{i,t}(\theta_0)]$. 

\subsubsection{Bias Term 1}
\label{sec:bias_term_1}

Equations \eqref{eq:y_tilde_start} to \eqref{eq:homo_geo_sum} below I outline the general argument for how to derive $b_{\rho_1}(\theta_0)$. Then after \eqref{eq:homo_geo_sum} I provide additional detail for each step. 

\begin{align}
b_{\rho_1}(\theta_0) &= \E_{\theta_0} \bigg[ \frac{1}{T} \sum_{t=1}^T \Tilde{Y}_{i,t-1}\Tilde{\epsilon}_{i,t}(\theta_0)  \bigg] \label{eq:y_tilde_start}
\\ 
        &= - \E_{\theta_0}  \bigg[  \Bar{Y}_{i,-1} \Bar{\epsilon}_{i}(\theta_0) \bigg] \label{eq:y_tilde_to_bar} \\
   &= - \E_{\theta_0}  \bigg[ \frac{1}{T}  \left\{ \sum_{l = 0}^{T - 2} \bigg(\sum_{j = 0}^{l} \phi_0^j \bigg)   \epsilon_{i, T - 1 - l}(\theta_0) \right\} \bar{\epsilon}_{i}(\theta_0) \bigg]  \label{eq:y_bar_plug} \\
      &= -   \frac{1}{T}  \sum_{l = 0}^{T - 2} \bigg(\sum_{j = 0}^{l} \phi_0^j \bigg) \E_{\theta_0} [   \epsilon_{i, T - 1 - l}(\theta_0) \bar{\epsilon}_{i}(\theta_0) ]  \\
  &= - \frac{1}{T^2} \sum_{l = 0}^{T - 2} \bigg( \sum_{j = 0}^l\phi^j_0 \bigg) \sigma^2_{\epsilon,iT}(\theta_0) \label{eq:variance_plug} \\
  &= - \frac{1}{T^2} \sum_{l = 0}^{T - 2} \bigg(\frac{1 - \phi^{l +1}}{1 - \phi} \bigg) \sigma^2_{\epsilon, iT}(\theta_0) \label{eq:geo_sum_1} \\
  &= - \frac{\sigma_{\epsilon, iT}^2}{T} \bigg( \frac{T - 1}{1 - \phi} - \frac{\phi - \phi^T}{(1 - \phi)^2}\bigg)  \label{eq:homo_geo_sum}
\end{align}

Now I explain the steps in greater detail. 

\paragraph{Steps for Line \eqref{eq:y_tilde_to_bar}}

\begin{align}
      \E_{\theta_0} \bigg[ \frac{1}{T} \sum_{t=1}^T \Tilde{Y}_{i,t-1}\Tilde{\epsilon}_{i,t}(\theta_0)  \bigg]  &= \E_{\theta_0} \bigg[\frac{1}{T} \sum_{t=1}^T ({Y}_{i,t-1} - \Bar{Y}_{i,-1})({\epsilon}_{i,t}(\theta_0) - \Bar{\epsilon}_{i}(\theta_0)) \bigg]  \\
        &=   \E_{\theta_0} \bigg[  \frac{1}{T} \sum_{t=1}^T ({Y}_{i,t-1} - \Bar{Y}_{i,-1})({\epsilon}_{i,t}(\theta_0)) \bigg] \\
        &=  - \E_{\theta_0}  \bigg[  \frac{1}{T} \sum_{t=1}^T  \Bar{Y}_{i,-1}{\epsilon}_{i,t}(\theta_0) \bigg] \\
       &= - \E_{\theta_0}  \bigg[\Bar{Y}_{i,-1} \frac{1}{T}\sum_{t=1}^T  {\epsilon}_{i,t}(\theta_0) \bigg] \\
        &= - \E_{\theta_0}  \bigg[  \Bar{Y}_{i,-1} \Bar{\epsilon}_{i}(\theta_0) \bigg]
\end{align}

Because of weak indepdence I have that the covariance of $Y_{i,t -1}$ and $\epsilon_{i,t}$ is zero, and the mean of $\epsilon_{i,t}$ is zero. Therefore $ \E_{\theta_0} [  \frac{1}{T} \sum_{t=1}^T {Y}_{i,t-1} {\epsilon}_{i,t} ] = 0$.

\paragraph{Steps for Line \eqref{eq:y_bar_plug}}

Here I derive an expression for $\Bar{Y}_{i,-1}$ as a sum of initial and past values. 


I start by first deriving an expression for $Y_{i,t}$ as a sum of initial and past values. 

\begin{equation}
    D_{i,1} = c_i + \rho_2 Y_{i,0} + u_{i,1}
\end{equation}
\begin{equation}
\begin{aligned}
        Y_{i,1} &=  a_i + \rho_1 Y_{i,0} + \tau D_{i,1} + \epsilon_{i1} \\
        &= a_i + \rho_1 Y_{i,0} + \tau [c_i  + \rho_2 Y_{i, 0} + u_{i1}] + \epsilon_{i1} \\
        &= a_i + (\rho_1 + \tau \rho_2) Y_{i,0}  + \tau (c_i + u_{i1}) + \epsilon_{i1}
\end{aligned}
\end{equation}

\begin{equation}
\begin{aligned}
        Y_{i,2} &= a_i + (\rho_1 + \tau \rho_2) Y_{i,1}  + \tau (c_i + u_{i1}) + \epsilon_{i2}\\
        &= a_i + (\rho_1 + \tau \rho_2) \bigg[ a_i + (\rho_1 + \tau \rho_2) Y_{i,0}  + \tau (c_i + u_{i0}) + \epsilon_{i1} \bigg]  + \tau (c_i + u_{i1}) + \epsilon_{i2}
\end{aligned}
\end{equation}

Notice the recursive nature in the terms, use this pattern to generalize \( Y_{i,t} \):

\begin{equation}
Y_{i,t} = a_i + (\rho_1 + \tau \rho_2) Y_{i,t-1} + \tau (c_i + u_{i,t}) + \epsilon_{i,t}
\label{eq:Y_it_recursive}
\end{equation}

Generalizing this pattern for any \( t \):

\begin{equation}
\label{eq:y_function_past}
\begin{aligned}
    Y_{i,t} &= a_i \sum_{j=0}^{t-1} (\rho_1 + \tau \rho_2)^j + (\rho_1 + \tau \rho_2)^t Y_{i,0} + \tau \sum_{j=0}^{t-1} (\rho_1 + \tau \rho_2)^j (c_i + u_{i,t-j}) \\
    &+ \sum_{j=0}^{t-1} (\rho_1 + \tau \rho_2)^j \epsilon_{i,t-j}
\end{aligned}
\end{equation}

Recall $\phi_0 = (\rho_{10} + \tau_0 \rho_{20})$. The expression for $\Bar{Y}_{i,-1}$ follows:
\begin{align}
       \Bar{Y}_{i,-1} &= \frac{1}{T}\bigg[ \bigg(\sum_{l = 0}^{T - 2} (T - l - 1) \phi_0^l \bigg) a_i  + \bigg(\sum_{l = 0}^{T - 1} \phi_0^l \bigg) Y_{i,0} + \sum_{l = 0}^{T - 2}\bigg(\sum_{j = 0}^{l} \phi_0^j \bigg) \tau_0 (c_i + u_{i, T - 1 - l}) \\
     &+ \sum_{l = 0}^{T - 2}\bigg(\sum_{j = 0}^{l} \phi_0^j \bigg) \epsilon_{i, T - 1 - l} \bigg] \label{eq:not_exogen}
\end{align}


\paragraph{Steps for Line \eqref{eq:variance_plug}}

Follows by the definition of variance. I use that the noise is homoscedastic over time, but heteroscedastic across individuals.  

\paragraph{Steps for Line \eqref{eq:geo_sum_1}}

Follows by rules of genometrics sums.

\subsubsection{Bias Term 2}

The second bias term follows almost an identical argument to what was outlined in Section \ref{sec:bias_term_1} for Term 1.

\begin{align}
b_{\tau}(\theta_0) &= \E_{\theta_0} \bigg[ \frac{1}{T} \sum_{t=1}^T \Tilde{D}_{i,t}\Tilde{\epsilon}_{i,t}(\theta_0)  \bigg] \label{eq:d_tilde_start} \\
&=  - \rho_2 \frac{\sigma_{\epsilon, iT}^2}{T} \bigg( \frac{T - 1}{1 - \phi} - \frac{\phi - \phi^T}{(1 - \phi)^2}\bigg)  \label{eq:final_treat}
\end{align}

It follows from the fact that. 
\begin{equation}
    \Bar{D}_{i} = \rho_2 \Bar{Y}_{i, -} + \Bar{e}_{i}
\end{equation}

Since $\Bar{Y}_{i, -}$ is solved above, it can be plugged in. The same exogeneity conditions hold.


\subsubsection{Bias Term 3}

The third bias term follows almost an identical argument to what was outlined in Section \ref{sec:bias_term_1} for Term 1. Again, since $\Bar{Y}_{i, -}$ is solved above, it can be plugged in. The same exogeneity conditions hold.

\begin{align}
b_{\rho_2}(\theta_0) &= \E_{\theta_0} \bigg[ \frac{1}{T} \sum_{t=1}^T \Tilde{Y}_{i,t-1}\Tilde{u}_{i,t}(\theta_0)  \bigg]  \\
&=  - \tau_0 \frac{\sigma_{u, iT}^2}{T} \bigg( \frac{T - 1}{1 - \phi} - \frac{\phi - \phi^T}{(1 - \phi)^2}\bigg) 
\end{align}

\subsection{Interaction Model}
\label{appen:general_model}

Thie bias correction for the more general model is proven here. 

\begin{equation}
\mathbf{b}(\theta_0) =  \begin{bmatrix}
            b_{\rho_1}(\theta_0) \\
        b_{\tau_c}(\theta_0) \\
        b_{\rho_2}(\theta_0) \\
        b_{\beta_1}(\theta_0) \\
        b_{\beta_2}(\theta_0) \\
    \end{bmatrix} =  \E \begin{bmatrix}
            \frac{1}{T} \sum_{t=1}^T \Tilde{Y}_{i,t-1}\Tilde{\epsilon}_{i,t}(\theta_0)  \\
        \frac{1}{T} \sum_{t=1}^T (\Tilde{D}_{i,t} W^c_{i,t}) \Tilde{\epsilon}_{i,t}(\theta_0)  \\
        \frac{1}{T} \sum_{t=1}^T \Tilde{Y}_{i,t-1}\Tilde{u}_{i,t}(\theta_0) \\
        \frac{1}{T} \sum_{t=1}^T \Tilde{X}_{1, i,t}\Tilde{\epsilon}_{i,t}(\theta_0) \\
        \frac{1}{T} \sum_{t=1}^T \Tilde{X}_{2,i,t}\Tilde{u}_{i,t}(\theta_0) 
    \end{bmatrix} 
\end{equation}

The outcome model has interactions with strictly exogenous covariates. Let there be $N_c$ different covariates $W_{i,t}^1, W_{i,t}^2, \dots, W_{i,t}^{N_c} $ that interact with treatment. The strictly exogenous regressors that do not interact with treatment are denoted by $X_{i,t}$.

The more general model that allows for interactions is given by:
\begin{equation}
Y_{i,t} = a_i + \sum_{c=1}^{N_c} \tau_c (D_{i,t} \cdot W_{i,t}^c) + \beta_1 X_{1,i,t} + \rho_1 Y_{i,t-1} + \epsilon_{i,t}
\end{equation}

\begin{equation}
D_{i,t} = c_i + \rho_2 Y_{i,t-1} + \beta_2 X_{2,i,t} + u_{i,t}
\end{equation}

Substitute \( D_{i,t} \) into the outcome equation:
\begin{align}
Y_{i,t} &= a_i + \sum_{c=1}^{N_c} \tau_c \left((c_i + \rho_2 Y_{i,t-1} + \beta_2 X_{2,i,t} + u_{i,t}) \cdot W_{i,t}^c\right) \nonumber \\
&\quad + \beta_1 X_{1,i,t} + \rho_1 Y_{i,t-1} + \epsilon_{i,t}
\end{align}

Distribute \( \tau_c \) and group terms by \( Y_{i,t-1} \): 
\begin{align}
Y_{i,t} &= a_i + \sum_{c=1}^{N_c} \tau_c c_i W_{i,t}^c + \sum_{c=1}^{N_c} \tau_c \rho_2 Y_{i,t-1} W_{i,t}^c \nonumber \\
&\quad + \sum_{c=1}^{N_c} \tau_c \beta_2 X_{2,i,t} W_{i,t}^c + \sum_{c=1}^{N_c} \tau_c u_{i,t} W_{i,t}^c \nonumber \\
&\quad + \beta_1 X_{1,i,t} + \rho_1 Y_{i,t-1} + \epsilon_{i,t}
\end{align}

Grouping by \( Y_{i,t-1} \): 
Let:
\begin{equation}
\gamma_t = \rho_1 + \sum_{c=1}^{N_c} \tau_c \rho_2 W_{i,t}^c
\end{equation}

\begin{equation}
\alpha_t = \sum_{c=1}^{N_c} \tau_c u_{i,t} W_{i,t}^c
\end{equation}

Then:
\begin{align}
Y_{i,t} &= a_i + \sum_{c=1}^{N_c} \tau_c c_i W_{i,t}^c + \gamma_t Y_{i,t-1} \nonumber \\
&\quad + \sum_{c=1}^{N_c} \tau_c \beta_2 X_{2,i,t} W_{i,t}^c + \alpha_t + \beta_1 X_{1,i,t} + \epsilon_{i,t}
\end{align}

Iterate the equation backwards: 
Substitute \( Y_{i,t-1} \) in terms of \( Y_{i,t-2} \):
\begin{align}
Y_{i,t-1} &= a_i + \sum_{c=1}^{N_c} \tau_c c_i W_{i,t-1}^c + \gamma_{t-1} Y_{i,t-2} \nonumber \\
&\quad + \sum_{c=1}^{N_c} \tau_c \beta_2 X_{2,i,t-1} W_{i,t-1}^c + \alpha_{t-1} + \beta_1 X_{1,i,t-1} + \epsilon_{i,t-1}
\end{align}

So:
\begin{align}
Y_{i,t} &= a_i + \sum_{c=1}^{N_c} \tau_c c_i W_{i,t}^c + \gamma_t \bigg( a_i + \sum_{c=1}^{N_c} \tau_c c_i W_{i,t-1}^c + \gamma_{t-1} Y_{i,t-2} \nonumber \\
&\quad + \sum_{c=1}^{N_c} \tau_c \beta_2 X_{2,i,t-1} W_{i,t-1}^c + \alpha_{t-1} + \beta_1 X_{1,i,t-1} + \epsilon_{i,t-1} \bigg) \nonumber \\
&\quad + \sum_{c=1}^{N_c} \tau_c \beta_2 X_{2,i,t} W_{i,t}^c + \alpha_t + \beta_1 X_{1,i,t} + \epsilon_{i,t}
\end{align}

Continue substituting backward iteratively until reaching \( Y_{i,0} \):
\begin{align}
Y_{i,t} &= \left( \prod_{j=0}^{t-1} \gamma_{t-j} \right) Y_{i0} \nonumber \\
&\quad + \sum_{k=0}^{t-1} \left( \prod_{j=0}^{k-1} \gamma_{t-j} \right) \left( a_i + \sum_{c=1}^{N_c} \tau_c c_i W_{i,t-k}^c \right. \nonumber \\
&\quad \left. + \sum_{c=1}^{N_c} \tau_c \beta_2 X_{2,i,t-k} W_{i,t-k}^c + \alpha_{t-k} + \beta_1 X_{1,i,t-k} + \epsilon_{i,t-k} \right)
\end{align}

Summarize the final expression for \( Y_{i,t} \):
\begin{align}
Y_{i,t} &= \left( \prod_{j=0}^{t-1} \gamma_{t-j} \right) Y_{i0} \nonumber \\
&\quad + \sum_{k=0}^{t-1} \left( \prod_{j=0}^{k-1} \gamma_{t-j} \right) \left( a_i + \sum_{c=1}^{N_c} \tau_c c_i W_{i,t-k}^c \right. \nonumber \\
&\quad \left. + \sum_{c=1}^{N_c} \tau_c \beta_2 X_{2,i,t-k} W_{i,t-k}^c + \sum_{c=1}^{N_c} \tau_c u_{i,t-k} W_{i,t-k}^c + \beta_1 X_{1,i,t-k} + \epsilon_{i,t-k} \right)
\end{align}

Expression plugging in for variables:
\begin{align}
\frac{1}{T} \sum_{t=1}^T Y_{i,t} &= \frac{1}{T} \sum_{t=1}^T \left( \left( \prod_{j=0}^{t-1} \gamma_{t-j} \right) Y_{i0} \right. \nonumber \\
&\quad \left. + \sum_{k=0}^{t-1} \left( \prod_{j=0}^{k-1} \gamma_{t-j} \right) \left( a_i + \sum_{c=1}^{N_c} \tau_c c_i W_{i,t-k}^c + \sum_{c=1}^{N_c} \tau_c \beta_2 X_{2,i,t-k} W_{i,t-k}^c  \right. \right. \nonumber \\
&\quad \left. \left. + \sum_{c=1}^{N_c} \tau_c u_{i,t-k} W_{i,t-k}^c + \beta_1 X_{1,i,t-k} + \epsilon_{i,t-k} \right) \right)
\end{align}

\paragraph{Expression for \( D \times W \) Average:}
Similarly, for \( D_{i,t} W_{i,t}^c \):
\begin{align}
D_{i,t} W_{i,t}^c &= \left( c_i + \rho_2 Y_{i,t-1} + \beta_2 X_{2,i,t} + u_{i,t} \right) W_{i,t}^c
\end{align}

Substitute \( Y_{i,t-1} \):
\begin{align}
D_{i,t} W_{i,t}^c &= \left( c_i + \rho_2 \left( a_i + \sum_{c=1}^{N_c} \tau_c c_i W_{i,t-1}^c + \gamma_{t-1} Y_{i,t-2} \right. \right. \nonumber \\
&\quad \left. \left. + \sum_{c=1}^{N_c} \tau_c \beta_2 X_{2,i,t-1} W_{i,t-1}^c + \alpha_{t-1} + \beta_1 X_{1,i,t-1} + \epsilon_{i,t-1} \right) \right. \nonumber \\
&\quad \left. + \beta_2 X_{2,i,t} + u_{i,t} \right) W_{i,t}^c
\end{align}

Iterating backward:
\begin{align}
D_{i,t} W_{i,t}^c &= \sum_{k=0}^{t-1} \left( \prod_{j=0}^{k-1} \gamma_{t-j-1} \right) \left( c_i W_{i,t-k}^c + \rho_2 \left( a_i + \sum_{c=1}^{N_c} \tau_c c_i W_{i,t-k}^c \right. \right. \nonumber \\
&\quad \left. \left. + \sum_{c=1}^{N_c} \tau_c \beta_2 X_{2,i,t-k} W_{i,t-k}^c + \alpha_{t-k} + \beta_1 X_{1,i,t-k} + \epsilon_{i,t-k} \right) \right)
\end{align}

Summarizing the result:
\begin{align}
D_{i,t} W_{i,t}^c &= \sum_{k=0}^{t-1} \left( \prod_{j=0}^{k-1} \gamma_{t-j-1} \right) \left( c_i W_{i,t-k}^c + \rho_2 \left( a_i + \sum_{c=1}^{N_c} \tau_c c_i W_{i,t-k}^c \right. \right. \nonumber \\
&\quad \left. \left. + \sum_{c=1}^{N_c} \tau_c \beta_2 X_{2,i,t-k} W_{i,t-k}^c + \alpha_{t-k} + \beta_1 X_{1,i,t-k} + \epsilon_{i,t-k} \right) \right)
\end{align}

Therefore the average can be written as:
\begin{align}
\frac{1}{T} \sum_{t=1}^T D_{i,t} W_{i,t}^c &= \frac{1}{T} \sum_{t=1}^T \sum_{k=0}^{t-1} \left( \prod_{j=0}^{k-1} \gamma_{t-j-1} \right) \left( c_i W_{i,t-k}^c + \rho_2 \left( a_i + \sum_{c=1}^{N_c} \tau_c c_i W_{i,t-k}^c \right. \right. \nonumber \\
&\quad \left. \left. + \sum_{c=1}^{N_c} \tau_c \beta_2 X_{2,i,t-k} W_{i,t-k}^c + \alpha_{t-k} + \beta_1 X_{1,i,t-k} + \epsilon_{i,t-k} \right) \right)
\end{align}

Therefore it follow by the same argument outlined in Appendix \ref{sec:bias_corr_simple}. Here $\phi = \rho_1 + \sum_{c=1}^{N_c} \tau_c \rho_2 \mu_c$. Here $\mu_{c}$ is the mean of covariate $W^c$
\begin{equation}
\mathbf{b}(\theta_0) =  \begin{bmatrix}
            b_{\rho_1}(\theta_0) \\
        b_{\tau_c}(\theta_0) \\
        b_{\rho_2}(\theta_0) \\
        b_{\beta_1}(\theta_0) \\
        b_{\beta_2}(\theta_0) \\
    \end{bmatrix} =   \begin{bmatrix}
            - \frac{\sigma_{\epsilon, iT}^2}{T} \bigg( \frac{T - 1}{1 - \phi} - \frac{\phi - \phi^T}{(1 - \phi)^2}\bigg) \\
                - \rho_2 \mu_{c} \frac{\sigma_{\epsilon, iT}^2}{T} \bigg( \frac{T - 1}{1 - \phi} - \frac{\phi - \phi^T}{(1 - \phi)^2}\bigg) \\
         - (\sum_{c=1}^{N_c} \tau_c \mu_{c}) \frac{\sigma_{u, iT}^2}{T} \bigg( \frac{T - 1}{1 - \phi} - \frac{\phi - \phi^T}{(1 - \phi)^2}\bigg) \\
    0 \\ 
    0 \\
    \end{bmatrix} 
\end{equation}

\subsection{Algebra interaction term}

Note:

\begin{equation}
\begin{aligned}
       \bar{Y}_{i, -1} = \frac{1}{T} \sum_{t=1}^T Y_{i,t} &= \frac{1}{T} \sum_{t=1}^T \left( \prod_{j=0}^{t-1} \gamma_{t-j} \right) Y_{i0} \\
       &+ \frac{1}{T}\sum_{t=1}^T \sum_{k=0}^{t-1} \left( \prod_{j=0}^{k-1} \gamma_{t-j} \right) \left( a_i + \sum_{c=1}^{N_c} \tau_c c_i W_{i,t-k}^c + \sum_{c=1}^{N_c} \tau_c u_{i,t-k} W_{i,t-k}^c + \beta X_{i,t-k} + \epsilon_{i,t-k} \right)
\end{aligned}
\end{equation}

We want to find $\E_{\theta_0}  \bigg[  \Bar{Y}_{i,-1} \Bar{\epsilon}_{i}(\theta_0) \bigg] \label{eq:y_tilde_to_bar}$. 

The part of $\Bar{Y}_{i,-1}$ we care about is $\frac{1}{T}\sum_{t=1}^T \sum_{k=0}^{t-1} \left( \prod_{j=0}^{k-1} \gamma_{t-j} \right) \epsilon_{i,t-k}$

\begin{equation}
    \begin{aligned}
       & \E_{\theta_0}  \bigg[ \left\{ \Bar{Y}_{i,-1} \right\} \Bar{\epsilon}_{i}(\theta_0) \bigg] \label{eq:y_tilde_to_bar} \\
       &= \E_{\theta_0}  \bigg[  \left\{ \frac{1}{T}\sum_{t=1}^T \sum_{k=0}^{t-1} \left( \prod_{j=0}^{k-1} \gamma_{t-j} \right) \epsilon_{i,t-k}   \right\} \Bar{\epsilon}_{i}(\theta_0) \bigg]
    \end{aligned}
\end{equation}

Let us consider a simple example of the expression above.  

\begin{equation}
    \begin{aligned}
        \E\bigg[ \sum_{t=1}^3\left\{ A_t B_t  \right\} C  \bigg] &=    \E\bigg[ \left\{ A_1 B_1 + A_2 B_2 +  A_3 B_3   \right\} C  \bigg] \\
        &= \E\bigg[ \left\{ A_1 B_1 C + A_2 B_2 C +  A_3 B_3 C  \right\}  \bigg] \\
         &=  \E\bigg[ A_1 B_1 C \bigg] + \E\bigg[ A_2 B_3 C \bigg] + \E\bigg[ A_3 B_3 C \bigg]  \\   
    \end{aligned}
\end{equation}

Consider $\E\bigg[ A_1 B_1 C \bigg]$. In the first line below I use the fact that $A$ is orthogonal to $B$ and $C$. 

\begin{equation}
    \begin{aligned}
        \E\bigg[ A_1 B_1 C \bigg] &= \E\bigg[ A_1 \bigg] \times \E\bigg[ B_1 C \bigg]\\
    \end{aligned}
\end{equation}


In our case 

\begin{equation}
    \begin{aligned}
    \E[A_t] &\sim  \E[ \left( \prod_{j=0}^{k-1} \gamma_{t-j} \right) ] \\
    \end{aligned}
\end{equation}

Below the first line uses the definition of $\gamma_{t-j}$. The second line uses the fact that the covariates are strictly exogenous variables that are iid across time and unit, so they are independent. The third line uses the fact that the expectation of a covaraiate is the same across time. 

\begin{equation}
    \begin{aligned}
       \E \bigg[  \prod_{j=0}^{k-1} \gamma_{t - j} \bigg] &= \E \bigg[ \prod_{j=0}^{k-1} \left\{\rho_1 + \sum_{c=1}^{N_c} \tau_c \rho_2 W^c_{i,t - j} \right\}  \bigg]   \\
       &=  \prod_{j=0}^{k-1} \left\{ \E \bigg[ \rho_1 + \sum_{c=1}^{N_c} \tau_c \rho_2 W^c_{i,t - j}   \bigg] \right\} \\
      &=  \prod_{j=0}^{k-1} \left\{ \rho_1 + \sum_{c=1}^{N_c} \tau_c \rho_2 E[W^c_{i,t - j}] \right\} \\
    &=  \prod_{j=0}^{k-1} \left\{ \rho_1 + \sum_{c=1}^{N_c} \tau_c \rho_2 \mu_c \right\} \\
    &= \left\{ \rho_1 + \sum_{c=1}^{N_c} \tau_c \rho_2 \mu_c \right\}^k \\
    &= \phi^k
    \end{aligned}
\end{equation}

Therefore we have 
\begin{equation}
    \begin{aligned}
       & \E_{\theta_0}  \bigg[ \left\{ \Bar{Y}_{i,-1} \right\} \Bar{\epsilon}_{i}(\theta_0) \bigg] \label{eq:y_tilde_to_bar} \\
       &=  \frac{1}{T}\sum_{t=1}^T \sum_{k=0}^{t-1} \left( \phi^k \right) \E_{\theta_0} [ \epsilon_{i,t-k}  \Bar{\epsilon}_{i}(\theta_0) ]
    \end{aligned}
\end{equation}

\begin{equation}
    \begin{aligned}
       & \E_{\theta_0}  \bigg[ \left\{ \Bar{Y}_{i,-1} \right\} \Bar{\epsilon}_{i}(\theta_0) \bigg] \label{eq:y_tilde_to_bar} \\
       &=  \frac{1}{T}\sum_{l=0}^{T-2} \sum_{j=0}^{l} \left( \phi^j \right) \E_{\theta_0} [ \epsilon_{i,T-1 - l}  \Bar{\epsilon}_{i}(\theta_0) ]
    \end{aligned}
\end{equation}

\section{Normality of bias-corrected Estimator}
\label{appendix:normality_bc_gmm}
\label{appendix:gmm_normal}

Proof for Theorem \ref{theorem:normality}. 

\subsection{Gradient}
\label{appendix:gradient}

The gradient of the bias-corrected estimator is the following. 

\begin{equation}
    \nabla_{\theta} \mathbf{m}^{DBC}_{NT}(\theta) = \frac{1}{N} \sum_{i = 1}^N \nabla_{\theta} \mathbf{m}^{DBC}_{iT}(\theta)
\end{equation}

Where the individual moment gradient is given by:

\begin{equation}
    \nabla_{\theta} \mathbf{m}^{DBC}_{iT}(\theta) =  \left( \begin{array}{ccc}
\nabla_{\rho_1} m^{DBC}_{\rho_1, iT}(\theta) &
\nabla_{\tau} m^{DBC}_{\rho_1, iT}(\theta)  &
\nabla_{\rho_2}m^{DBC}_{\rho_1, iT}(\theta)  \\
\nabla_{\rho_1} m^{DBC}_{\tau, iT}(\theta) &
\nabla_{\tau} m^{DBC}_{\tau, iT}(\theta)  &
\nabla_{\rho_2}m^{DBC}_{\tau, iT}(\theta)  \\ 
\nabla_{\rho_1} m^{DBC}_{\rho_2, iT}(\theta) &
\nabla_{\tau} m^{DBC}_{\rho_2, iT}(\theta)  &
\nabla_{\rho_2}m^{DBC}_{\rho_2, iT}(\theta)  \\
\end{array} \right)
\end{equation}
has the elements

\begin{equation}
\begin{aligned}
        \nabla_{\rho_1} m^{DBC}_{\rho_1, iT}(\theta) &= -\frac{1}{T} \sum_{t=1}^{T} (\Tilde{Y}_{i,t-1}) \Tilde{Y}_{i,t-1}  -  \nabla \Phi(\theta) \hat{\sigma}_{\epsilon, iT}^2(\theta) - 
        \Phi(\theta) \nabla_{\rho_1} \hat{\sigma}_{\epsilon, iT}^2(\theta),
\end{aligned}
\end{equation}

\begin{equation}
    \nabla_{\tau} m^{DBC}_{\rho_1, iT}(\theta) = -\frac{1}{T} \sum_{t=1}^{T} (\Tilde{Y}_{i,t-1})D_{i,t} -
    \rho_2 \nabla \Phi(\theta) \hat{\sigma}_{\epsilon, iT}^2(\theta)
    - \Phi(\theta) \nabla_{\tau} \hat{\sigma}_{\epsilon, iT}^2(\theta),
\end{equation}

\begin{equation}
\begin{aligned}
        \nabla_{\rho_2}m^{DBC}_{\rho_1, iT}(\theta) &= - \tau \nabla \Phi(\theta) \hat{\sigma}_{\epsilon, iT}^2(\theta) - \Phi(\theta) \nabla_{\rho_2} \hat{\sigma}_{\epsilon, iT}^2(\theta) \\
        &= - \tau \nabla \Phi(\theta) \hat{\sigma}_{\epsilon, iT}^2(\theta),
\end{aligned}
\end{equation}

\begin{equation}
    \nabla_{\rho_1} m^{DBC}_{\tau, iT}(\theta) = -\frac{1}{T} \sum_{t=1}^{T} (\Tilde{D}_{i,t-1})Y_{i,t-1} -
   \rho_2 \nabla \Phi(\theta) \hat{\sigma}_{\epsilon, iT}^2(\theta) - 
       \rho_2 \Phi(\theta) \nabla_{\rho_1} \hat{\sigma}_{\epsilon, iT}^2(\theta),
\end{equation}

\begin{equation}
    \nabla_{\tau}m^{DBC}_{\tau, iT} (\theta) =   -\frac{1}{T} \sum_{t=1}^{T} (\Tilde{D}_{i,t-1})D_{i,t} -
    \rho_2^2 \nabla \Phi(\theta) \hat{\sigma}_{\epsilon, iT}^2(\theta) - 
       \rho_2 \Phi(\theta) \nabla_{\tau} \hat{\sigma}_{\epsilon, iT}^2(\theta)
\end{equation}

\begin{equation}
\begin{aligned}
        \nabla_{\rho_2}m^{DBC}_{\tau, iT}(\theta) &= - \Phi(\theta) \hat{\sigma}_{\epsilon, iT}^2(\theta)  - \rho_2 \tau \nabla \Phi(\theta) \hat{\sigma}_{\epsilon, iT}^2(\theta) - \rho_2 \Phi(\theta) \nabla_{\rho_2} \hat{\sigma}_{\epsilon, iT}^2(\theta) \\
        &= - \Phi(\theta) \hat{\sigma}_{\epsilon, iT}^2(\theta) - \tau \nabla \Phi(\theta) \hat{\sigma}_{\epsilon, iT}^2(\theta),
\end{aligned}
\end{equation}

\begin{equation}
\begin{aligned}
        \nabla_{\rho_1}m^{DBC}_{\rho_2, iT}(\theta) &=  -  \tau \nabla \Phi(\theta) \hat{\sigma}_{u, iT}^2(\theta) - \tau \Phi(\theta) \nabla_{\rho_1} \hat{\sigma}_{u, iT}^2(\theta) \\
        &= - \tau \nabla \Phi(\theta) \hat{\sigma}_{u, iT}^2(\theta)
\end{aligned}
\end{equation}

\begin{equation}
\begin{aligned}
       \nabla_{\tau}m^{DBC}_{\rho_2, iT}(\theta) =  \nabla_{\rho_2}m^{DBC}_{\tau, iT}(\theta) &= - \Phi(\theta) \hat{\sigma}_{u, iT}^2(\theta)  - \rho_2 \tau \nabla \Phi(\theta) \hat{\sigma}_{u, iT}^2(\theta) - \tau \Phi(\theta) \nabla_{\tau} \hat{\sigma}_{u, iT}^2(\theta) \\
        &= - \Phi(\theta) \hat{\sigma}_{\epsilon, iT}^2(\theta) - \tau \nabla \Phi(\theta) \hat{\sigma}_{u, iT}^2(\theta),
\end{aligned} 
\end{equation}

\begin{equation}
    \nabla_{\rho_2}m_{\rho_2, iT}(\theta) = -\frac{1}{T} \sum_{t=1}^{T} (\Tilde{Y}_{i,t-1})Y_{i,t-1} -  \tau^2 \nabla \Phi(\theta) \hat{\sigma}_{u, iT}^2(\theta) - \tau \Phi(\theta) \nabla_{\rho_2} \hat{\sigma}_{u, iT}^2(\theta)(\theta),
\end{equation}

where

\begin{equation}
   \nabla \Phi(\theta) = - T^{-2} \sum_{l=1}^{T-2} \sum_{j=1}^{l} l (\rho_1 + \tau \rho_2)^{l-1}
\end{equation}

\begin{equation}
    \Phi(\theta) = - T^{-2} \sum_{l=1}^{T-2} \sum_{j=1}^{l} (\rho_1 + \tau \rho_2)^{l}
\end{equation}

\begin{equation}
\begin{aligned}
        \nabla_{\rho_1}\hat{\sigma}_{\epsilon, iT}(\theta) &= \frac{2}{T - 1} \sum_{t=1}^{T} (\Tilde{Y}_{i,t-1})\epsilon_{i,t}(\theta)\\
        \E[ \nabla_{\rho_1}\hat{\sigma}_{\epsilon, iT}(\theta)] &= 2 \Phi(\theta) {\sigma}_{\epsilon, iT}^2 T/(T - 1) 
\end{aligned}
\end{equation}

\begin{equation}
\begin{aligned}
        \nabla_{\tau}\hat{\sigma}_{\epsilon, iT}(\theta) &= \frac{2}{T - 1} \sum_{t=1}^{T} (\Tilde{D}_{i,t-1})\epsilon_{i,t}(\theta)\\
        \E[ \nabla_{\tau}\hat{\sigma}_{\epsilon, iT}(\theta)] &= 2 \rho_2 \Phi(\theta) {\sigma}_{\epsilon, iT}^2 T/(T - 1) 
\end{aligned}
\end{equation}

\begin{equation}
    \nabla_{\rho_2}\hat{\sigma}_{\epsilon, iT}(\theta) = 0,
\end{equation}

\begin{equation}
    \nabla_{\rho_1}\hat{\sigma}_{u, iT}(\theta) = 0,
\end{equation}

\begin{equation}
    \nabla_{\tau}\hat{\sigma}_{u, iT}(\theta) = 0,
\end{equation}

\begin{equation}
\begin{aligned}
        \nabla_{\rho_2}\hat{\sigma}_{u, iT}(\theta) &= \frac{2}{T - 1} \sum_{t=1}^{T} (\Tilde{Y}_{i,t-1})u_{i,t}(\theta)&= 2 \tau \Phi(\theta) {\sigma}_{u, iT}^2 T/(T - 1) 
\end{aligned}
\end{equation}

\subsection{GMM Normality}

\subsubsection{Consistency}

Given the assumptions of Theorem \ref{theorem:normality}, it follows by Theorem 14.1 from \cite{wooldridge2010econometric} that the GMM estimator is consistent. 

Theorem 14.1 from \cite{wooldridge2010econometric} requires that a) $\Theta \subset \mathbb{R}^3$ is compact, which is satisfied by assumption b) For each $\theta \in \Theta$, $\mathbf{m}^{DBC}(\cdot, \theta)$ is Borel measurable on $\mathcal{Z}$, which is satisfied by the moments being a polynomial in data and having compact parameters c) for each $\mathbf{z} \in \mathcal{Z}$, $\mathbf{m}^{DBC}(\mathbf{z}, \cdot)$ is continuous on $\Theta$. This follows from the fact moments are a polynomial in the random variables and $1/(1 - \phi) < 1/\delta_s$ d) $|\mathbf{m}_j^{DBC}(\mathbf{z}, \theta) | \leq b(\mathbf{z})$ for all $\theta \in \Theta$ and $j = 1,2,3$ where $b(\cdot)$ is a nonnegative function on $\mathcal{Z}$ such that $E[b(\mathbf{z})] < \infty$, this follows from the assumption on the bounded moments of the data and that the moments on only quadratic in the random variables e) The GMM weighting matrix $\hat{\Xi} \xrightarrow{p} {\Xi}_0$ a positive definite weighting matrix, in my case the weighting matrix is just the identity matrix so this is satisfied, f) $\theta_0$ is the unique solution to the problem, which the global identification condition that is satisfied by assumption. 

Therefore a random vector $\hat{\theta}$ exists that solves Equation \eqref{eq:gmm_estimator} and  $\hat{\theta} \xrightarrow{p} \theta_0$.  

\subsubsection{Normality}

Given the assumptions of Theorem \ref{theorem:normality}, it follows by Theorem 14.2 from \cite{wooldridge2010econometric} that the GMM estimator is asymptotically normal. 

Theorem 14.2 from \cite{wooldridge2010econometric} requires a) $\theta_0$ is in the interior of $\Theta$, which is satisfied by assumption b) $\mathbf{m}^{DBC}(\mathbf{z}, \cdot)$ is continuously differentiable on the interior of $\Theta$ for all $\mathbf{z}\in \mathcal{Z}$, this holds because the moments are ratios of polynomials with denominators bounded away from zero by the stationarity assumption c) each element of $\mathbf{m}^{DBC}(\mathbf{z}, \theta_0)$ has finite second moment, which follows from assumption bounding the moments of the data d) each element of $\nabla \mathbf{m}^{DBC}(\mathbf{z}, \theta)$ is bounded in absolute value by a function $b(\mathbf{z})$, where $E[b(\mathbf{z})] < \infty$ because the derivative of the polynomial is also bounded  e) $E[\nabla \mathbf{m}^{DBC}(\mathbf{z}, \theta)]$ is full rank follows by assumption. 

Therefore, it follows that the limiting distribution of $\hat{\theta}$ follows the equation given in Theorem \ref{theorem:normality}.

\begin{equation}
    \sqrt{N}( \hat{\theta}^{DBC} - \theta_0) \xrightarrow{d} N(0, [\Sigma_{T} + B_T(\theta_0) ]^{-1} S_{T}(\theta_0) [\Sigma_{T} + B_T(\theta_0) ]^{-1}
\end{equation}

With $S_T(\theta_0) = \plim_{N \rightarrow \infty} \frac{1}{N}\sum_{i=1}^N \mathbf{m}^{DBC}_{iT} \mathbf{m}^{DBC}_{iT}, $

\begin{equation}
   \Sigma_T = \plim_{N \rightarrow \infty} \frac{1}{NT} \sum_{i = 1}^N  \sum_{t = 1}^T  \begin{bmatrix}
 \tilde{Y}_{i,t-1}^2 &  \tilde{Y}_{i,t-1} \tilde{D}_{i,t} & 0 \\
  \tilde{D}_{i,t} \tilde{Y}_{i,t-1} &   \tilde{D}_{i,t}^2 & 0 \\
0 & 0 &  \tilde{Y}_{i,t-1}^2
\end{bmatrix}
\end{equation}

\begin{equation}
   B_T(\theta) =   \begin{bmatrix}
 \nabla \Phi(\theta) \sigma_{\epsilon,i}^2 - \frac{2 \Phi(\theta)^2 \sigma_{\epsilon,i}^2 T }{T -1}  & \rho_2(\nabla \Phi(\theta) \sigma_{\epsilon,i}^2 - \frac{2 \Phi(\theta)^2 \sigma_{\epsilon,i}^2 T }{T -1})  & - \tau \nabla \Phi(\theta) \sigma_{\epsilon,i}^2 \\
  \rho_2(\nabla \Phi(\theta) \sigma_{\epsilon,i}^2 - \frac{2 \Phi(\theta)^2 \sigma_{\epsilon,i}^2 T }{T -1}) &   \rho_2^2(\nabla \Phi(\theta) \sigma_{\epsilon,i}^2 - \frac{2 \Phi(\theta)^2 \sigma_{\epsilon,i}^2 T }{T -1}) & - \Phi(\theta) {\sigma}_{\epsilon, iT}^2 - \tau \nabla \Phi(\theta) {\sigma}_{\epsilon, iT}^2 \\
 - \tau \nabla \Phi(\theta) {\sigma}_{u, iT}^2 & - \Phi(\theta) {\sigma}_{u, iT}^2 - \tau \nabla \Phi(\theta) {\sigma}_{u, iT}^2 &  \tau^2(\nabla \Phi(\theta) \sigma_{u,i}^2 - \frac{2 \Phi(\theta)^2 \sigma_{u,i}^2 T }{T -1})
\end{bmatrix}
\end{equation}

\subsection{Local Identification}
\label{appendix:local_identification}

Let $\theta_0 \in \Theta \subset \mathbb{R}^p$ be the true parameter value. Suppose we have a set of moment conditions $\mathbb{E}[m^{DBC}(\theta)] = 0$, where $m: \mathbb{R}^d \times \Theta \rightarrow \mathbb{R}^q$ is a vector-valued function. The parameter $\theta_0$ is locally identified if the following conditions hold:

\begin{enumerate}
    \item \textbf{Continuity and Differentiability}:
    \begin{itemize}
        \item The moment condition function $m^{DBC}(\theta)$ is continuously differentiable with respect to $\theta$ in a neighborhood of $\theta_0$.
    \end{itemize}
    \item \textbf{Rank Condition}:
    \begin{itemize}
        \item The Jacobian matrix $J_{m}^{DBC}(\theta)$ of the expected moment condition function with respect to $\theta$, $J_{m}^{DBC}(\theta) = \frac{\partial}{\partial \theta} \mathbb{E}[m^{DBC}(\theta)] \big|_{\theta = \theta_0}$, has full column rank $p$ (where $p$ is the number of parameters).
    \end{itemize}
\end{enumerate}

Recall that the bias-corrected moments are the orginal OLS moment conditions minus the bias correction. 
\begin{equation}
\begin{aligned}
        \mathbf{m}^{DBC}(\theta) &:= \mathbf{m}({\theta}) - \mathbf{b}(\theta) = \begin{bmatrix}
          \frac{1}{T} \sum_{t=1}^T \Tilde{Y}_{i,t-1}\Tilde{\epsilon}_{i,t} - b_{\rho_1, iT}(\theta) \\
        \frac{1}{T} \sum_{t=1}^T \tilde{D}_{i,t}\Tilde{\epsilon}_{i,t} -b_{\tau, iT}(\theta) \\
        \frac{1}{T} \sum_{t=1}^T \Tilde{Y}_{i,t-1}\Tilde{u}_{i,t} - b_{\rho_2, iT}(\theta) 
    \end{bmatrix} \\
    &= \begin{bmatrix}
          \frac{1}{T} \sum_{t=1}^T \Tilde{Y}_{i,t-1}\Tilde{\epsilon}_{i,t}(\theta) - \frac{- \hat{\sigma}_{\epsilon, iT}(\theta)^2}{T} \bigg( \frac{T - 1}{1 - \phi(\theta)} - \frac{\phi(\theta) - \phi(\theta)^T}{(1 - \phi(\theta))^2}\bigg) \\
        \frac{1}{T} \sum_{t=1}^T \tilde{D}_{i,t}\Tilde{\epsilon}_{i,t}(\theta) - \rho_2 \frac{- \hat{\sigma}_{\epsilon, iT}(\theta)^2}{T} \bigg( \frac{T - 1}{1 - \phi(\theta)} - \frac{\phi(\theta) - \phi(\theta)^T}{(1 - \phi(\theta))^2}\bigg) \\
        \frac{1}{T} \sum_{t=1}^T \Tilde{Y}_{i,t-1}\Tilde{u}_{i,t}(\theta) - \tau  \frac{- \hat{\sigma}_{u, iT}(\theta)^2 }{T^2} \bigg( \frac{T - 1}{1 - \phi(\theta)} - \frac{\phi(\theta) - \phi(\theta)^T}{(1 - \phi(\theta))^2}\bigg)  
    \end{bmatrix}
\end{aligned}
\end{equation}

 In this section I provide conditions under which the Rank Condition holds. For the original OLS moment $\mathbf{m}(\theta)$  the corresponding Jacobian is $J_{m}(\theta) := \nabla_{\theta} \mathbf{m}(\theta)$. These are the original OLS moments, and therefore $\E[J_{m}(\theta)]$ is full rank when the columns of regressors are linearly independent.   For the bias correction $J_{b}(\theta) := \nabla_{\theta} \mathbf{b}(\theta)$. Due to linarity I have  $\E[J_{m}^{DBC}(\theta)] = \E[J_{m}(\theta)] - \E[J_{b}(\theta)]$. Therefore as long as columns of the 1) regressors are linearly independent, and 2) subtracting out $\E[J_{b}(\theta)]$ does not ruin the full rank of $\E[J_{m}(\theta)]$, then $\E[J_m^{DBC}(\theta)]$ has full column rank.  $\E[J_{b}(\theta)]$ will not break the full rank condition as long as it is ''small enough'', formalized below.  
\begin{proof}

In order to argue that $A- B$ is full rank, we need to argue that there does not exist an $x$ such that $(A - B)x = 0 $. 

This is true if 

\begin{equation}
    x'(A - B) x > c \|x \|^2
\end{equation}

which is the same as saying 

\begin{equation}
    x'(A' - B') x > c \|x \|^2
\end{equation}

which is the same as saying 

\begin{equation}
    x'(A - \frac{B + B'}{2})x \geq c \| x\|^2
\end{equation}

this will hold if 

\begin{equation}
    \lambda_{\max} (\frac{B + B'}{2}) \leq \lambda_{\min}(A)
\end{equation}

\end{proof}

Therefore we have full rank if

\begin{equation}
    \lambda_{\max} (\frac{\E[J_{b}(\theta)] + \E[J_{b}(\theta)]'}{2}) \leq \lambda_{\min}(\E[J_m^{DBC}(\theta)])
\end{equation}

\begin{equation}
 J_{m}(\theta) =    \begin{bmatrix}
-\frac{1}{T} \sum_{t=1}^{T} \tilde{Y}_{i,t-1}^2 & -\frac{1}{T} \sum_{t=1}^{T} \tilde{Y}_{i,t-1} \tilde{D}_{i,t} & 0 \\
-\frac{1}{T} \sum_{t=1}^{T} \tilde{D}_{i,t} \tilde{Y}_{i,t-1} & -\frac{1}{T} \sum_{t=1}^{T} \tilde{D}_{i,t}^2 & 0 \\
0 & 0 & -\frac{1}{T} \sum_{t=1}^{T} \tilde{Y}_{i,t-1}^2
\end{bmatrix}
 \end{equation}

\begin{equation}
   \E[J_b(\theta)]  =   \begin{bmatrix}
 \nabla \Phi(\theta) \sigma_{\epsilon,i}^2 - \frac{2 \Phi(\theta)^2 \sigma_{\epsilon,i}^2 T }{T -1}  & \rho_2(\nabla \Phi(\theta) \sigma_{\epsilon,i}^2 - \frac{2 \Phi(\theta)^2 \sigma_{\epsilon,i}^2 T }{T -1})  & - \tau \nabla \Phi(\theta) \sigma_{\epsilon,i}^2 \\
  \rho_2(\nabla \Phi(\theta) \sigma_{\epsilon,i}^2 - \frac{2 \Phi(\theta)^2 \sigma_{\epsilon,i}^2 T }{T -1}) &   \rho_2^2(\nabla \Phi(\theta) \sigma_{\epsilon,i}^2 - \frac{2 \Phi(\theta)^2 \sigma_{\epsilon,i}^2 T }{T -1}) & - \Phi(\theta) {\sigma}_{\epsilon, iT}^2 - \tau \nabla \Phi(\theta) {\sigma}_{\epsilon, iT}^2 \\
 - \tau \nabla \Phi(\theta) {\sigma}_{u, iT}^2 & - \Phi(\theta) {\sigma}_{u, iT}^2 - \tau \nabla \Phi(\theta) {\sigma}_{u, iT}^2 &  \tau^2(\nabla \Phi(\theta) \sigma_{u,i}^2 - \frac{2 \Phi(\theta)^2 \sigma_{u,i}^2 T }{T -1})
\end{bmatrix}
\end{equation}

\section{Additional Simulations}
\label{append:additional_simulations}

\subsection{Additional Comparison of Dynamic vs Nickell bias}
\label{append:additional_simulations_dynamic_vs_Nickell}

\begin{figure}[h!]
    \centering
    \includegraphics[scale=0.5]{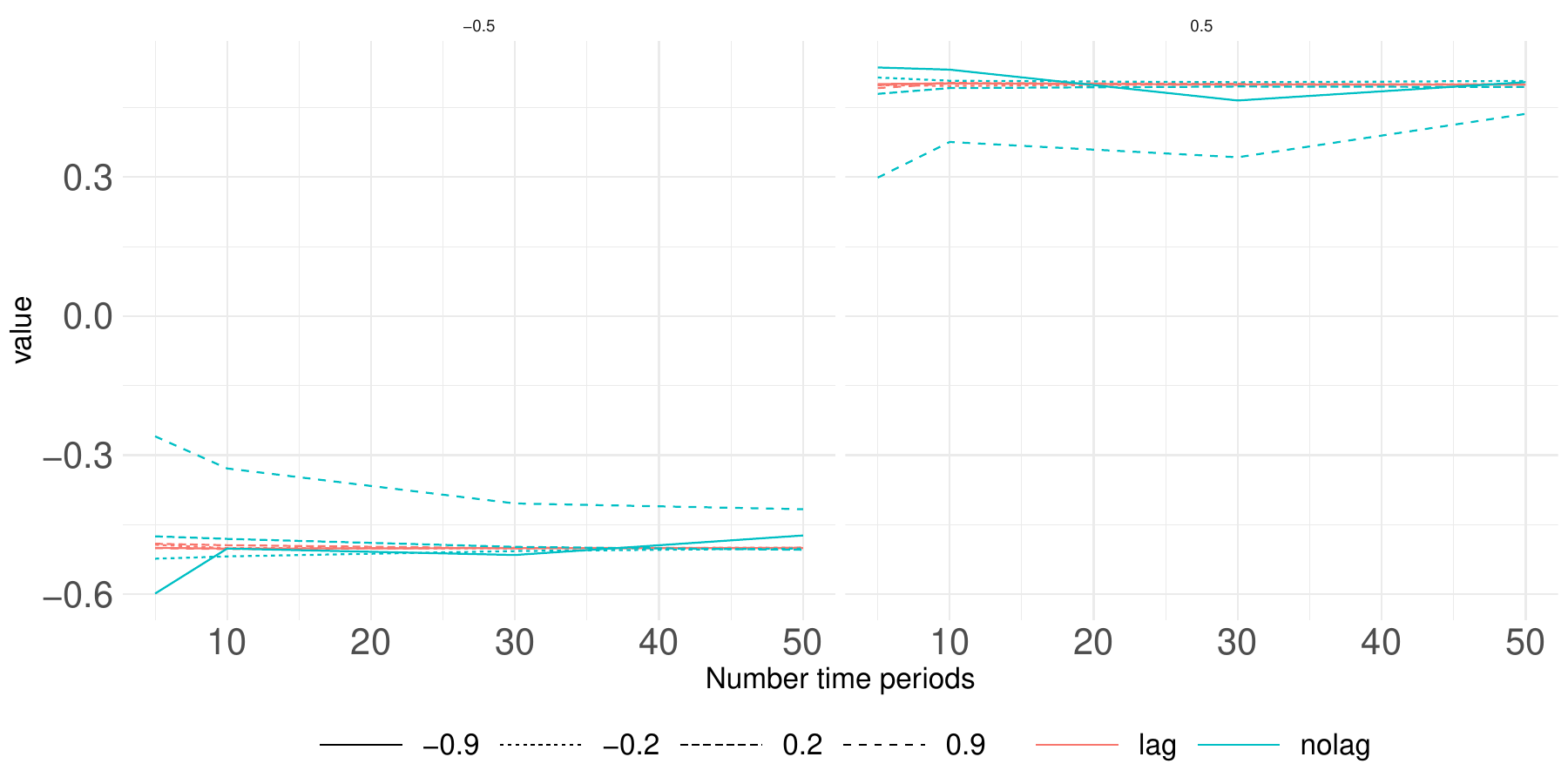} 
    \caption{Line type represents different values of $\rho_{10}$. Line color represents whether the model was estimates with an outcome lag or not. The left column contains results when the true $\tau_0 = -.5$ and the right column contains results when $\tau_0 = .5$.}
    \label{fig:more_nickell_vs_dynamic}
\end{figure}

Here I provide additional simulation results comparing dynamic and Nickell bias from Section \ref{sec:simulation_dynamic_vs_Nickell}. I plotted treatment estimates for a wide variety of DGP's in Figure \ref{fig:more_nickell_vs_dynamic}. 

\section{Additional Empirical Results}
\label{append:additional_empirics}

\subsection{Replicating \cite{dell2012temperature}}

I replicate column 2 of Table 2 in \cite{dell2012temperature} and obtain the same results.  My results are given in Table \ref{tab:combined_gdp_ben_og} column 3. In this Table \ref{tab:combined_gdp_ben_og}, I also show in column 4 how the treatment effect changes once you control for lagged GDP Growth. The change the coefficient in Temperature x Poor Country changes 9\% and is significantly different at the $p <.1$ level. I tested the difference in coefficients by using the clustered bootstrap. 

\begin{table}[!htbp] \centering 
  \caption{} 
  \label{tab:combined_gdp_ben_og} 
\begin{tabular}{@{\extracolsep{2pt}}lcccc} 
\\[-1.8ex]\hline 
\hline \\[-1.8ex] 
 & \multicolumn{2}{c}{\textit{Outcome variable:}} & \multicolumn{2}{c}{\textit{Outcome variable:}} \\ 
 & \multicolumn{2}{c}{GDP Level} & \multicolumn{2}{c}{GDP Growth} \\ 
\cline{2-5} 
\\[-1.8ex] & (1) & (2) & (3) & (4) \\ 
\hline \\[-1.8ex] 
 Temperature & 118.568 & 82.071$^{**}$ & 0.261 & 0.261 \\ 
  & (100.725) & (32.139) & (0.257) & (0.254) \\ 
  & & & & \\ 
 Temperature x Poor Country & 192.550 & $-$88.364$^{**}$ & $-$1.655$^{***}$ & $-$1.511$^{***}$ \\ 
  & (162.740) & (51.961) & (0.415) & (0.413) \\ 
  & & & & \\ 
 Outcome Lag &  & 0.922$^{***}$ &  & 0.192$^{***}$ \\ 
  &  & (0.005) &  & (0.015) \\ 
    & & & & \\ 
Controls & Yes & Yes & Yes & Yes \\ 
  & & & & \\ 
\hline \\[-1.8ex] 
Observations & 4,654 & 4,629 & 4,924 & 4,795 \\ 
R$^{2}$ & 0.941 & 0.994 & 0.223 & 0.251 \\ 
Adjusted R$^{2}$ & 0.935 & 0.993 & 0.150 & 0.179 \\ 
\hline 
\hline \\[-1.8ex] 
\textit{Note:}  & \multicolumn{4}{r}{$^{*}$p$<$0.1; $^{**}$p$<$0.05; $^{***}$p$<$0.01} \\ 
\end{tabular} 
\end{table}

\subsection{Using growth vs levels in outcomes}
\label{append:growth_vs_level}

Empirically, it is often observed that countries with lower initial GDP levels tend to have higher growth rates, which would suggest that GDP levels across countries should converge over time \citep{barro1992convergence}. By focusing on growth rates rather than levels, you might incorrectly suggest that these countries are diverging when, in fact, they could be converging in terms of absolute GDP levels. Using GDP growth as the dependent variable in regressions can lead to misleading interpretations about the relative economic performance of countries, potentially suggesting divergence when, in reality, countries might be converging in terms of GDP levels.

Still, in practice, it is common to study GDP growth. I formalize the econometric problems that arise. Consider following simple model for country GDP. This comes from \cite{solow1956contribution}. 

\begin{equation}
    Y_{i,t} = a_i + \rho_0 Y_{i,t-1} + \tau_0 D_{i,t} + \epsilon_{i,t}
\end{equation}

Where we say treatment $D_{i,t}$ is temperature, and it is random conditional on fixed effect. The error terms in both models are random idd shocks and $\epsilon_{i,t}$ and $u_{i,t}$. 

\begin{equation}
    D_{i,t} = a_i + u_{i,t}
\end{equation}

Let's say we want to run the regression using $\Delta Y_{i,t}$ as the outcome. 

\begin{equation}
      \Delta Y_{i,t} = Y_{i,t} - Y_{i,t-1} =  \rho_0 (Y_{i,t-1} - Y_{i,t-2})  + \tau_0(D_{i,t} - D_{i,t-1})  + (\epsilon_{i,t} - \epsilon_{i,t-1})
\end{equation}

What is the causal object of interest? The impact of current temperature on growth. 

\begin{equation}
\begin{aligned}
        \text{APD} &= \frac{\partial \Delta Y_{i,t}}{\partial D_{i,t}} \\
        &=\frac{\partial \rho_0 (Y_{i,t-1} - Y_{i,t-2})  + \tau_0(D_{i,t} - D_{i,t-1})  + (\epsilon_{i,t} - \epsilon_{i,t-1})}{\partial D_{i,t}} \\
        &= \tau_0
\end{aligned}
\end{equation}

So the treatment effect is $\tau_0$. 

Let's create an OLS model to estimate this effect. 

\begin{equation}
    \Delta Y_{i,t} = c_i + \beta D_{i,t} + e_{i,t}
\end{equation}

Does $\hat{\beta}$ provide an unbiased estimate for $\tau_0$? Recall that running OLS with fixed effects dummies is the same as running the within-trandformed regression. 

\begin{equation}
    \Delta \Tilde{Y}_{i,t} =  \beta \Tilde{D}_{i,t} + \Tilde{e}_{i,t}
\end{equation}

\begin{equation}
    \hat{\beta} = \frac{Cov(\Tilde{D}_{i,t}, \Delta \Tilde{Y}_{i,t} )}{Var(\Tilde{D}_{i,t})}
\end{equation}

Let us unpack this . 

\begin{equation}
\begin{aligned}
        \frac{Cov(\Tilde{D}_{i,t}, \Delta \Tilde{Y}_{i,t} )}{Var(\Tilde{D}_{i,t})} &= \frac{Cov(\Tilde{D}_{i,t}, \rho_0 (\Tilde{Y}_{i,t-1} - \Tilde{Y}_{i,t-2})  + \tau_0(\Tilde{D}_{i,t} - \Tilde{D}_{i,t-1})  + (\Tilde{\epsilon}_{i,t} - \Tilde{\epsilon}_{i,t-1}) )}{Var(\Tilde{D}_{i,t})} \\
        &= \frac{Cov(\Tilde{D}_{i,t}, \rho_0 (\Tilde{Y}_{i,t-1} - \Tilde{Y}_{i,t-2})  + \tau_0(\Tilde{D}_{i,t} - \Tilde{D}_{i,t-1}) )}{Var(\Tilde{D}_{i,t})} \\
        &= \tau_0 +  \underbrace{\frac{Cov(\Tilde{D}_{i,t}, \rho_0 (\Tilde{Y}_{i,t-1} - \Tilde{Y}_{i,t-2})  - \tau_0( \Tilde{D}_{i,t-1}) )}{Var(\Tilde{D}_{i,t})}}_{\text{bias}} \\
\end{aligned}
\end{equation}

So this bias is a type of dynamic bias. For now, I call this transformation bias, as it arises from transforming the outcome variable before running the fixed effects model. 

\subsection{Replicating \cite{annan2015federal}}

In this subsection, I replicate the findings of \cite{annan2015federal}, which examine the impact of federal crop insurance subsidies on agricultural outcomes. The results are given in Table \ref{table:schlenker}. The treatment variable of interest in this paper is ``frac:ddayHot''. This variable is the interaction of the insured fraction with exposure to hot days. In their original analysis, they used a static panel model that did not account for past outcomes. However, controlling for past outcomes is critical in settings where dynamic relationships are likely to exist, as previous outcomes can have a direct influence on current results.

In my replication, I modify their approach by introducing past outcomes into the model. The results show that the estimated treatment effects double when past outcomes are properly accounted for. This increase demonstrates the importance of controlling for dynamics in panel data settings. Failure to do so can lead to biased estimates.

\begin{table}[!htbp] \centering 
  \caption{} 
  \label{table:schlenker} 
\begin{tabular}{@{\extracolsep{5pt}}lcc} 
\\[-1.8ex]\hline 
\hline \\[-1.8ex] 
 & \multicolumn{2}{c}{\textit{Dependent variable:}} \\ 
\cline{2-3} 
\\[-1.8ex] & \multicolumn{2}{c}{yield\_log} \\ 
\\[-1.8ex] & (1) & (2)\\ 
\hline \\[-1.8ex] 
 frac & 0.0001 & 0.020 \\ 
  & (0.035) & (0.036) \\ 
  & & \\ 
 ddayMod & 0.430$^{***}$ & 0.420$^{***}$ \\ 
  & (0.017) & (0.017) \\ 
  & & \\ 
 ddayHot & $-$0.619$^{***}$ & $-$0.633$^{***}$ \\ 
  & (0.010) & (0.011) \\ 
  & & \\ 
 prec & 1.498$^{***}$ & 1.560$^{***}$ \\ 
  & (0.067) & (0.069) \\ 
  & & \\ 
 prec2 & $-$1.016$^{***}$ & $-$1.044$^{***}$ \\ 
  & (0.049) & (0.051) \\ 
  & & \\ 
 lag\_Y\_log &  & 0.063$^{***}$ \\ 
  &  & (0.004) \\ 
  & & \\ 
 frac:ddayMod & 0.008 & 0.004 \\ 
  & (0.013) & (0.013) \\ 
  & & \\ 
 frac:ddayHot & 0.046$^{***}$ & 0.070$^{***}$ \\ 
  & (0.015) & (0.015) \\ 
  & & \\ 
 frac:prec & $-$0.153 & $-$0.188$^{*}$ \\ 
  & (0.106) & (0.108) \\ 
  & & \\ 
 frac:prec2 & 0.142$^{*}$ & 0.154$^{*}$ \\ 
  & (0.079) & (0.080) \\ 
  & & \\ 
\hline \\[-1.8ex] 
Observations & 47,343 & 45,488 \\ 
R$^{2}$ & 0.736 & 0.738 \\ 
Adjusted R$^{2}$ & 0.725 & 0.727 \\ 
Residual Std. Error & 0.176 (df = 45447) & 0.175 (df = 43638) \\ 
\hline 
\hline \\[-1.8ex] 
\textit{Note:}  & \multicolumn{2}{r}{$^{*}$p$<$0.1; $^{**}$p$<$0.05; $^{***}$p$<$0.01} \\ 
\end{tabular} 
\end{table}

\subsection{Endogenous Treatment}
\label{sec:applied_endo_sims}
I repeat the same simulation exercise as in Section \ref{section:applied_motivation}, but update the true model. The true model is now given in Equation \eqref{eq:true_model_endo}. In this model I make treatment, $\text{Temp}_{i,t}$, a function of the past outcome, $\text{GDP}_{i,t-1}$.  Therefore our model has a new parameter $\rho_{20}$ which controls how much past GDP impacts temperature.\footnote{Most of the environmental literature does not think that past GDP impacts temperature, and I use setting mostly to illustrate my point on endogenous treatment. However, some papers discuss how economic growth, reflected in GDP, often correlates with increased industrial activity, energy consumption, and transportation. Historically, this has led to higher emissions of greenhouse gases (GHGs) such as carbon dioxide (CO2), which contribute to global warming \citep{nordhaus1992optimal}.} 

\begin{equation}
\label{eq:true_model_endo}
\begin{aligned}
     \text{True Model with Endogenous Treatment: }  \text{GDP}_{i,t} &= a_i + \tau_0 \text{Temp}_{i,t} +  \rho_{10} \text{GDP}_{i,t - 1} + \epsilon_{i,t}, \\
          \text{Temp}_{i,t} &= a_{i} + \rho_{20} \text{GDP}_{i,t-1} + u_{i,t}.
\end{aligned}
\end{equation}

\begin{figure}[h!]
    \centering
    \includegraphics[scale=0.5]{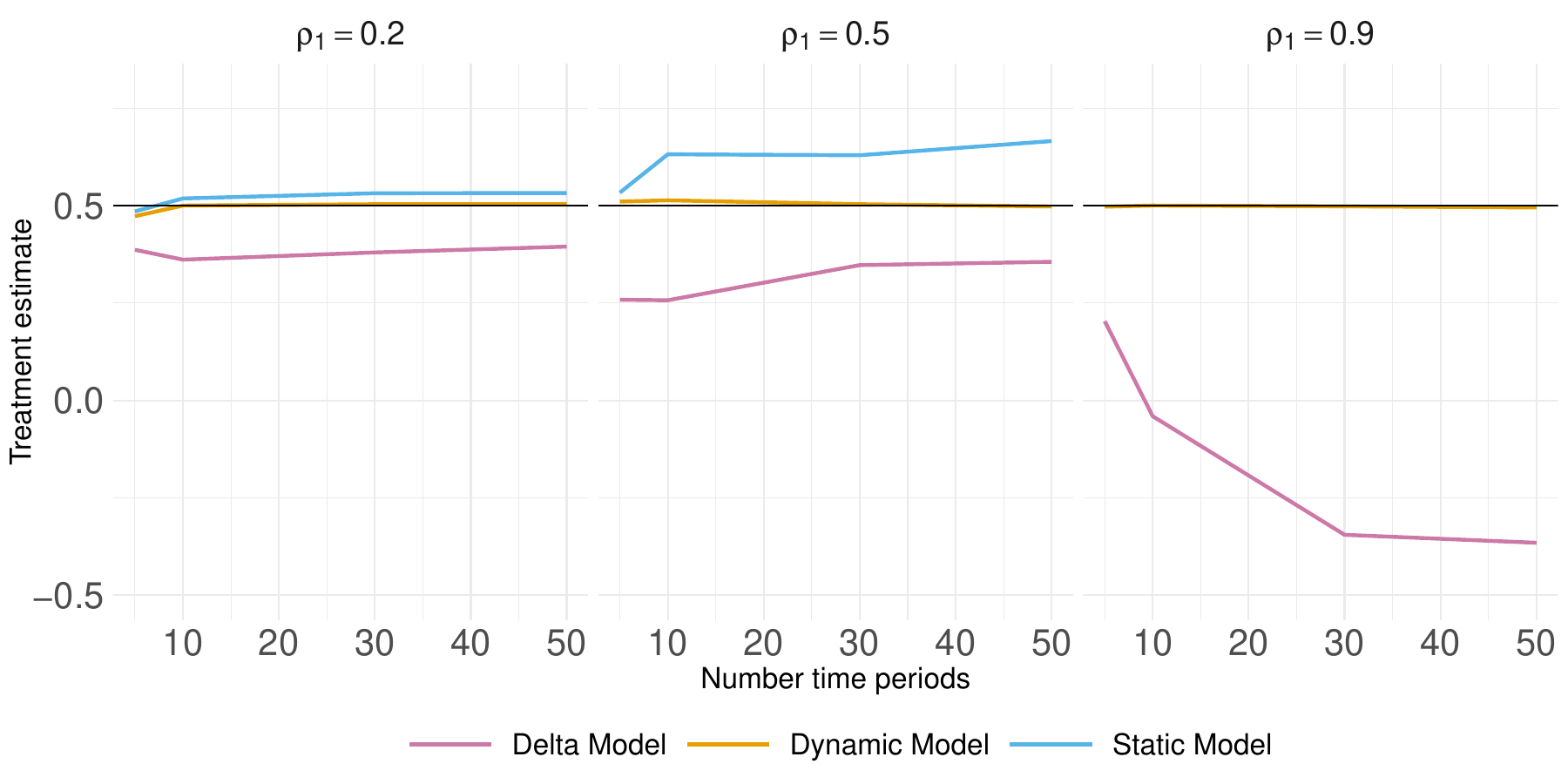}
    \caption{Bias of three different models.}
    \label{fig:growth_rate_into_endo}
\end{figure}

For a simulation, I add just a little bit of endogeneity and set $\rho_{20} = .1$.  The bias only gets larger the larger the absolute value of $\rho_{20}$ is.  Note in the plot, for the case of $\rho_{10} = .9$, the dynamic bias was so large it was omitted from the plot as it was much larger than all other biases.  The plot of the bias is given in Figure \ref{fig:growth_rate_into_endo}. 

\section{ML Extension}
\label{appendix:ML}

\subsection{Problem setup}

\paragraph{Data}
\begin{align}
    W_{i,t} &= (Y_{i,t}, Y_{i,t-1}, D_{i,t}, X_{i,t}) \\
    W_{i} &= (W_{1}, \cdots, W_{T}) \\
    \tilde{Y}_{i,t} &= Y_{i,t} - \frac{1}{T} \sum_{t} Y_{i,t}
\end{align}

\paragraph{Model setup}

\begin{equation}
\label{eq:CEF_func}
    Y_{i,t} = a_i + \theta_{0,1} Y_{i,t-1} + \theta_{0,2} D_{i,t} + \theta_{0, 3} X_{it} + \epsilon_{i,t}
\end{equation}
\begin{equation}
\label{eq:prop_func}
    D_{i,t} = c_i + \theta_{0, 4} Y_{i,t-1} + u_{i,t}
\end{equation}

Let $\theta_{0,3}$ represent a high-dimensional component, while the endogenous parameters $\theta_{0,1}, \theta_{0,2}, \theta_{0,4}$ are low-dimensional.

The true parameter vector is $\theta_0 := (\theta_{0,1}, \theta_{0,2}, \theta_{0,3}, \theta_{0,4})$.

The parameter of interest, $\theta_{0,2}$, corresponds to the treatment effect. The first equation (Equation \eqref{eq
}) represents the conditional expectation function (CEF), and the second equation (Equation \eqref{eq
}) represents the propensity function.

To implement Double Machine Learning (DML), I follow a two-stage process. In the first stage, I estimate the CEF and the propensity functions, which involves estimating the full parameter vector $\theta_0$. This step is critical, as it reduces the high-dimensional problem to manageable components. I rely on machine learning algorithms for flexible, consistent estimation of these nuisance parameters.

In the second stage, I use the first-stage estimates to construct a double-robust moment function. This moment is designed to be consistent, even if some of the first-stage estimates are slightly misspecified. The double-robust nature of DML ensures that the treatment effect $\theta_{0,2}$ can still be consistently estimated, provided that either the CEF or propensity function is estimated accurately.

Nickell bias prevents the consistent estimation of first stage using Lasso. To obtain consistent estimates, I use Nickell-bias-corrected moments with Regularized GMM (RGMM) instead of traditional Lasso. I outline the proof for the rates of RGMM in my setting to determine the rates of the first-stage function estimates.

\paragraph{Moment Conditions}

Note these are moments for one unit (average over time).

\begin{equation}
    g_1(W_i, \theta) = \frac{1}{T} \sum_{i = 1}^T \Tilde{Y}_{i, t-1} e(\theta) - e(\theta)^2\cdot C(\phi)
\end{equation}

\begin{equation}
    g_2(W_i, \theta) = \frac{1}{T} \sum_{i = 1}^T \Tilde{D}_{i, t} e(\theta) - e(\theta)^2 \cdot \theta_{0,4} \cdot C(\phi)
\end{equation}

\begin{equation}
    g_3(W_i, \theta) = \frac{1}{T} \sum_{i = 1}^T \Tilde{X}_{i, t} e(\theta) 
\end{equation}

\begin{equation}
    g_4(W_i, \theta) = \frac{1}{T} \sum_{i = 1}^T \Tilde{Y}_{i, t - 1} u(\theta) - u(\theta)^2 \cdot \theta_{0,2} \cdot C(\phi)
\end{equation}

\begin{equation}
    g(W_i, \theta) = [g_1(W_i, \theta) , g_2(W_i, \theta) , g_3(W_i, \theta), g_4(W_i, \theta)  ]
\end{equation}

Where 
\begin{equation}
    e(\theta) = (\Tilde{Y}_{i, t} - \theta_1 \Tilde{Y}_{i, t-1} - \theta_2 \Tilde{D}_{i, t} - \theta_3 \Tilde{X}_{i, t}  ) 
\end{equation}

\begin{equation}
    u(\theta) = (\Tilde{D}_{i, t} - \theta_4 \Tilde{Y}_{i, t-1} ) 
\end{equation}

\begin{equation}
    C(\phi) = \frac{1}{(1 - \phi)T}(1 - \frac{1 - \phi^T}{T(1 - \phi)})
\end{equation}

\begin{equation}
    \phi(\theta) = \theta_1 + \theta_2\cdot\theta_4
\end{equation}

\paragraph{Averages}

Define the empirical moment average and population expectations as. 

\begin{equation}
    \hat{g}(\theta) = \frac{1}{N} \sum_{i = 1}^N g(W_i, \theta)
\end{equation}

\begin{equation}
    {g}(\theta) = \E[ g(W_i, \theta)]
\end{equation}

\subsection{RGMM}

\begin{equation}
    \min_{\theta = \Theta} \| \theta \|_1 : \|\hat{g}(\theta) \|_{\infty} \leq \lambda
\end{equation}

Where $\lambda$ is our regularization parameter. I only regularize the high dimensional parameter $\theta_3$. 

\subsection{Rates for RGMM problem}

I require sufficiently fast rates for my first-stage estimation. In this stage, I aim to estimate $\theta_0$, as this allows me to construct consistent estimates of the conditional expectation function (CEF) and the propensity function.

My goal is to achieve a rate comparable to that of classical Lasso, but in the context of my Regularized GMM (RGMM) problem. \cite{belloni2018high} provide conditions and corresponding rates for RGMM when the moments possess an index structure. However, my moments do not naturally follow this structure.

To address this, I condition on a low-dimensional set of parameters, specifically $\theta_1, \theta_2$, and $\theta_4$. After this conditioning, the conditional moment exhibits the required index structure. This allows me to apply the results from \cite{belloni2018high} to determine the rate for my high-dimensional parameter, $\theta_3$, as a function of the low-dimensional parameters.




\subsection{Steps}

\begin{enumerate}
\item Create a grid with $K$ points over the parameters $\theta_1$, $\theta_2$, and $\theta_4$, which are the endogenous and low-dimensional parameters. For now, assume this grid is fixed (i.e., not increasing in size).

\item Select a point on the grid and fix the values of the parameters at this point. Denote this point as $(\theta_1^k, \theta_2^k, \theta_4^k)$.

\item Construct a new moment equation using these fixed values. This modified moment equation will still be a function of $\theta_3$. For example, consider the moment equation for $\theta_1$.

        \begin{equation}
        \label{eq:index_moment}
    g_1(W_i, \theta_3) = \frac{1}{T} \sum_{i = 1}^T \Tilde{Y}_{i, t-1} e(\theta) - e(\theta)^2\cdot \frac{1}{(1 - \phi)T}(1 - \frac{1 - \phi^T}{T(1 - \phi)})
\end{equation}

Where 
\begin{equation}
    e(\theta) = (\Tilde{Y}_{i, t} - \theta_1^k \Tilde{Y}_{i, t-1} - \theta_2^k \Tilde{D}_{i, t} - \theta_3 \Tilde{X}_{i, t}  ) 
\end{equation}
\begin{equation}
    \phi = \theta_1^k + \theta_2^k\cdot\theta_4^k
\end{equation}

\item Once $\theta_1^k$, $\theta_2^k$, and $\theta_4^k$ are fixed, the function $\phi$ becomes fixed and is treated as a constant, denoted $C_{\phi}^k$.

        \begin{equation}
    g_1(W_i, \theta_3) = \frac{1}{T} \sum_{i = 1}^T \Tilde{Y}_{i, t-1} e(\theta) - e(\theta)^2\cdot C_{\phi}^k
\end{equation}

\item  This allows me to express the moment equation in index form.


\begin{equation}
\begin{aligned}
        g_1(W_i, \theta_3) = \Tilde{m}(W_i, Z_{u(j)} (W)' v_{u(j)})
\end{aligned}
\end{equation}

\begin{equation}
    \Tilde{m}(W_{it}, Z_{u(j)} (W_{it})' v_{u(j)}) = \frac{1}{T} \sum_{i = 1}^T \Tilde{Y}_{i,t-1}Z_{u(j)} (W_{it})' v_{u(j)} - (Z_{u(j)} (W_{it})' v_{u(j)})^2 \cdot C_{\phi}^k
\end{equation}

\begin{equation}
    Z_{u(j)} (W_{it})' v_{u(j)}=  \Tilde{Y}_{i, t} - \theta_1^k \Tilde{Y}_{i, t-1} - \theta_2^k \Tilde{D}_{i, t} - \theta_3 \Tilde{X}_{i, t}  
\end{equation}

\begin{equation}
    Z_{u(j)} (W_{it})' = (\Tilde{Y}_{i,t}, \Tilde{Y}_{i,t-1}, \Tilde{D}_{i,t}, \Tilde{X}_{i,t} )
\end{equation}

\begin{equation}
    v_{u(j)} = (-1, \theta_1^k, \theta_2^k, \theta_3 )
\end{equation}

\item Now that the moment equation is in index form, I can apply the following relevant theorem from \cite{belloni2018high} to obtain $L_2$ rates. 

\begin{theorem}[Bounds on Empirical Error for Non-Linear RGMM]
\label{theorem:empirical_error_RGMM}
Consider the non-linear case and assume that Conditions L, DM, LID, NLID, and ENM are satisfied. Also, assume that \(\lambda\) is chosen so that \(\lambda \leq n^{-1/2}(\tilde{\ell}_n + \ell_n)\) and that the side condition \(n^{-1/2}(\tilde{\ell}_n + \ell_n) \leq \epsilon^*/2\) holds. Then with probability at least \(1 - \alpha - \delta_n\),

\[ \|\hat{\theta} - \theta_0\|_q \leq \frac{2(\tilde{\ell}_n + \ell_n)s^{1/q}}{\mu_n \sqrt{n}}, \quad q \in \{1, 2\}, \]

in the case of LID(a) (exactly sparse model); and, as long as \(a > 1\), and \(A > n^{-1/2} \ell_n (K + 1)\),

\[ \|\hat{\theta} - \theta_0\|_q \leq \frac{C_{a,q}(L_n + \mu_n)(\tilde{\ell}_n + \ell_n)s^{1/q}}{\mu_n \sqrt{n}}, \quad q \in \{1, 2\}, \]

in the case of LID(b) (approximately sparse model), where \(C_{a,q}\) is a constant depending only on \(a\) and \(q\).
\end{theorem}



\item Now, apply this bound to the problem at hand. Define $\hat{\theta}_3(\theta_1^k, \theta_2^k, \theta_4^k)$ as the parameter estimate obtained from the RGMM, using the moment equation with fixed values $(\theta_1^k, \theta_2^k, \theta_4^k)$ from the grid, as specified in Equation \eqref{eq:index_moment}.

I introduce the following notation: let $\hat{f}_3(\theta_1^k, \theta_2^k, \theta_4^k)$ represent a function that maps the points $(\theta_1^k, \theta_2^k, \theta_4^k)$ on the grid to the corresponding RGMM solution $\hat{\theta}_3(\theta_1^k, \theta_2^k, \theta_4^k)$. Let $f_{0,3}(\theta_1^k, \theta_2^k, \theta_4^k)$ denote the oracle version of this function. Note that this is distinct from $\theta_{0,3}$. Theorem 2.1 provides the bound for every point on the grid.

\begin{equation}
    \| \hat{f}_3(\theta_1^k, \theta_2^k, \theta_4^k) - f_{0,3}(\theta_1^k, \theta_2^k, \theta_4^k)\|_2 \leq \frac{2(\tilde{\ell}_n + \ell_n)s^{1/2}}{\mu_n \sqrt{n}} \forall k
\end{equation}

\item This holds for every point on the grid, so I take the maximum to establish the bounds.

\begin{equation}
    \max_{k = 1,2, ..., K } \|\hat{f}_3(\theta_1^k, \theta_2^k, \theta_4^k) - f_{0,3}(\theta_1^k, \theta_2^k, \theta_4^k)\|_2 \leq \max_{k = 1,2, ..., K } \frac{2(\tilde{\ell}_n + \ell_n)s^{1/q}}{\mu_n \sqrt{n}}
\end{equation}

From this maximum bound, I can calculate an $L_2$ rate for estimating the function $\hat{f}_3$. For now, assume that $\hat{f}_3$ is estimated at a rate of $O_p(n^{- \alpha})$.

\begin{equation}
\label{eq:rate_for_f}
    \|\hat{f}_3 - f_{0,3} \|_2 = O_p(n^{- \alpha})
\end{equation}

\item Now, I focus on the convergence of $\hat{\theta}_1$, $\hat{\theta}_2$, and $\hat{\theta}_4$, which can be solved through a low-dimensional GMM problem. Our moment conditions include the estimated $\hat{f}_3$ as an input. I introduce the following new moments.

        \begin{equation}
        \label{eq:low_dim_GMM}
    g(W_i, \theta_1, \theta_2, \theta_4, \hat{f}_3) = \frac{1}{T} \sum_{i = 1}^T \Tilde{Y}_{i, t-1} e(\theta) - e(\theta)^2\cdot \frac{1}{(1 - \phi)T}(1 - \frac{1 - \phi^T}{T(1 - \phi)})
\end{equation}

Where 
\begin{equation}
    e(\theta) = (\Tilde{Y}_{i, t} - \theta_1 \Tilde{Y}_{i, t-1} - \theta_2 \Tilde{D}_{i, t} - \hat{f}_3(\theta_1, \theta_2, \theta_4 ) \Tilde{X}_{i, t}  ) 
\end{equation}
\begin{equation}
    \phi = \theta_1 + \theta_2 \cdot\theta_4 
\end{equation}

Let 

\begin{equation}
    \hat{g}_n(\theta, f) = \frac{1}{N} \sum_{i=1}^n g(W_i, \theta, f)
\end{equation}

\begin{equation}
    {g}_0(\theta, f) = \E g(W_i, \theta, f)
\end{equation}

\paragraph{WTS:} 

\begin{equation}
\label{eq:wts_9}
        \| \hat{g}_n(\hat{\theta}, \hat{f}) -  g_0(\theta_0, {f}_0) \| = O_p(n^{-1/2} + n^{- \alpha})   
\end{equation}

\paragraph{Possible Argument}

\begin{enumerate}

\item 

I break the term on left hand side of Equation \eqref{eq:wts_9} into two parts. 

\begin{equation}
\begin{aligned}
       &=  \hat{g}_n(\hat{\theta}, \hat{f})   -   g_0(\theta_0, {f}_0) \\
       &= \hat{g}_n(\hat{\theta}, \hat{f}) \pm g_0(\hat{\theta}, \hat{f})  -   g_0(\theta_0, {f}_0) \\
      &=  \underbrace{\hat{g}_n(\hat{\theta}, \hat{f}) - g_0(\hat{\theta}, \hat{f})}_{\text{CF}}  -  \underbrace{ g_0(\hat{\theta}, \hat{f}) - g_0(\theta_0, {f}_0) }_{\text{Taylor}} 
\end{aligned}
\end{equation}

The CF term can be controlled with either cross fitting or empirical process theory. 

The ``Taylor'' term can be controlled by taking taylor expansions around $\theta_0$ and $f_0$.

    \item Taylor Term:

\begin{equation}
\label{eq:two_terms_moment}
\begin{aligned}
    = \underbrace{ g_0(\hat{\theta}, \hat{f}) - g_0(\theta_0, {f}_0) }_{\text{Taylor}} \\
    &=  g_0(\hat{\theta}, \hat{f}) \pm g_0(\hat{\theta}, f_0)  - g_0(\theta_0, {f}_0) \\
 &= \underbrace{ g_0(\hat{\theta}, \hat{f}) - g_0(\hat{\theta}, f_0)}_{Term 1}   +  \underbrace{ g_0(\hat{\theta}, f_0) - g_0(\theta_0, {f}_0)}_{Term 2} 
\end{aligned}
\end{equation}

\end{enumerate}

I perform a Taylor expansion of the first part of Term 1 around $f_0$, using a functional analog of a Taylor expansion (discussed on page 5 \href{https://www.reed.edu/physics/faculty/wheeler/documents/Classical%20Field%20Theory/Class%20Notes/Field%20Theory%20Chapter%205.pdf}{here}). Let $D_{f}(\hat{\theta}, f_0)$ represent the Jacobian matrix of the moment function $g_0$ with respect to the function $f$.

\begin{equation}
    \underbrace{g_0(\hat{\theta}, \hat{f}) - {g}_0(\hat{\theta}, {f}_0)}_{Term 1} \approx g_0(\hat{\theta}, f_0) + D_{f} g(\hat{\theta}, f_0)(f - f_0) - {g}_0(\hat{\theta}, {f}_0)
\end{equation}

Let $D_\theta \hat{g}(\theta_0, f_0)$ denote the Jacobian matrix of the moment function $g_0(\theta_0, f_0)$ with respect to the parameter vector $\theta$. 

\begin{equation}
     g_0(\hat{\theta}, f_0) \approx  g_0(\theta_0, f_0) +  D_{\theta}g_0(\theta_0, f_0)(\hat{\theta} - \theta_0)
\end{equation}

\begin{equation}
\underbrace{ g_0(\hat{\theta}, f_0) - g_0(\theta_0, {f}_0)}_{Term 2}  \approx   g_0(\theta_0, f_0) +  D_{\theta}g_0(\theta_0, f_0)(\hat{\theta} - \theta_0) - {g}_0(\theta_0, {f}_0)
\end{equation}

I rewrite Equation \eqref{eq:two_terms_moment} plugging in the expansions below.  

\begin{equation}
        \underbrace{ g_0(\hat{\theta}, \hat{f}) - g_0(\theta_0, {f}_0) }_{\text{Taylor}}  \approx D_{f} g(\hat{\theta}, f_0)(f - f_0) + D_{\theta}g_0(\theta_0, f_0)(\hat{\theta} - \theta_0)
\end{equation}

I determine the rate of the LHS by substituting the rates of the two terms on the RHS. From Equation \eqref{eq
}, we have $|\hat{f} - f_0 |_2 = O_p(n^{- \alpha})$. Since ${g}_0(\hat{\theta}, {f}0)$ is smooth and $D{f}\hat{g}(\hat{\theta}, {f}0)$ is a bounded linear operator, it follows that $|D{f}{g}_0(\hat{\theta}, f_0)(\hat{f} - f_0) |_2 = O_p(n^{- \alpha})$.


Conclude with Equation \eqref{eq:rate_moments}.

\begin{equation}
\label{eq:rate_moments}
        \| \hat{g}_n(\hat{\theta}, \hat{f}) -  g_0(\theta_0, {f}_0) \| = O_p(n^{-1/2} + n^{- \alpha})     
\end{equation}


\item Conclude with a convergence rate for the parameter estimates. 

\paragraph{WTS:} 

\begin{equation}
    \| (\hat{\theta}_1, \hat{\theta}_2, \hat{\theta}_3, \hat{\theta}_4) - ({\theta}_{0,1}, {\theta}_{0,2}, {\theta}_{0,3}, {\theta}_{0,4}) \|_2 = O_p(n^{-{\alpha}})
\end{equation}

where 

\begin{equation}
   \hat{\theta}_3 = \hat{f}_3(\hat{\theta}_1, \hat{\theta}_2, \hat{\theta}_4 )
\end{equation}

Given the smoothness of the moment conditions and the invertibility of the Jacobian, the goal is to apply the Delta method to transfer the rate of convergence from the sample moments to the parameter estimates.

\end{enumerate}

\section{Absorbing Treatment}
\label{sec:bias_corr_formula_absorbing}

\begin{equation}
    Y_{i,t} = a_i  + \tau (D_{i} \times 1_{t > T_{start}})  + \epsilon_{i,t}
\end{equation}

\begin{equation}
    D_{i} = a_i +  \rho_2 Y_{i,T_{start} -1} + u_{i}
\end{equation}

I label the bias in this model $b_4(\theta_0)$. I follow similar steps as above. To do this I need to calculate the formula for $\widebar{D_{i} \times 1_{t > T_{start}}}$.

\begin{equation}
\begin{aligned}
        \widebar{D_{i} \times 1_{t > T_{start}}} &= \frac{1}{T} \sum_{t = 1}^T (a_i +  \rho_2 Y_{i,T_{start} -1} + u_{i})(1_{t > T_{start}}) \\
        &= \frac{1}{T} \sum_{t = T_{start}}^T (a_i +  \rho_2 Y_{i,T_{start} -1} + u_{i}) \\
        &= \frac{T - T_{start}}{T} (a_i +  \rho_2 Y_{i,T_{start} -1} + u_{i}) \\
                &= \frac{T - T_{start}}{T} (a_i +  \rho_2 (a_ i + \epsilon_{T_{start} - 1} ) + u_{i})
\end{aligned}
\end{equation}

\begin{align}
 b_4(\theta_0) 
 &= \plim_{N \rightarrow \infty} \bigg(  \frac{1}{N} \sum_{i = 1}^N \sum_{t = 1}^T \widetilde{D_{i} \times 1_{t > T_{start}}}  \Tilde{\epsilon}_{i,t} \bigg) \\
 &= - \plim_{N \rightarrow \infty} \bigg(\frac{T}{N} \sum_{i = 1}^{N} \widebar{D_{i} \times 1_{t > T_{start}}}\Bar{\epsilon}_{i} \bigg)  \\
 &= - \plim_{N \rightarrow \infty} \bigg(\frac{T}{N} \sum_{i = 1}^{N} \frac{T - T_{start}}{T} (a_i +  \rho_2 (a_ i + \epsilon_{T_{start} - 1} ) + u_{i}) \Bar{\epsilon}_{i} \bigg)  \\ 
        &= - \frac{1}{T} (T - T_{start} -1) \rho_2 \sigma^2_{\epsilon}       
\end{align}

This formula is very intuitive. It is only error in the time period $T_{start} -1$ that is causing correlation in the errors. So the Nickell bias is much smaller in this case.

This is particularly interesting because past outcomes are typically viewed as time-varying covariates that are not absorbed by the individual fixed effect. However, in the case where the treatment is absorbed, the past outcome used for selection becomes fixed. As a result, the history remains constant over time and is controlled for by the fixed effect. Although there is still bias due to violation of strict exogenity, it is not the standard OVB (omitted variable bias) type.

\end{document}